\documentclass[useAMS]{mn2e}
\usepackage{graphicx,amssymb,amsmath}
\usepackage{hyperref,xcolor}
\usepackage{subfigure}
\usepackage{pdflscape}
\usepackage{mathrsfs}
\newcommand{\scs}{\scriptsize}

\title[NGC 1342, NGC 1662, NGC 1912, NGC 2354 and NGC 2447]
  {Comprehensive abundance analysis of red giants in the open clusters NGC 1342, 1662,
    1912, 2354 and 2447}
\author[A. B. S. Reddy, S. Giridhar and D. L. Lambert ]
  { Arumalla B. S. Reddy$^{1,2}$\thanks{E-mail: sudha@iiap.res.in (ABSR);
giridhar@iiap.res.in (SG); dll@astro.as.utexas.edu (DLL)},
    Sunetra Giridhar$^2$ and David L. Lambert$^1$  \\
   $^1$W.J. McDonald Observatory and Department of Astronomy, The University of Texas at Austin, Austin, TX 78712, USA \\
   $^2$Indian Institute of Astrophysics, Bangalore 560034, India }

\begin{document}

\pagerange{\pageref{firstpage}--\pageref{lastpage}} \pubyear{2014}

\maketitle

\label{firstpage}

\begin{abstract}

We have observed high-dispersion echelle spectra of red giant members in the five open clusters NGC 1342, NGC 1662, NGC 1912, NGC 2354 and NGC 2447 and determined their radial velocities and chemical compositions. These are the first chemical abundance measurements for all but NGC 2447. 
We combined our clusters from this and previous papers with a sample drawn from the literature for which we remeasured the chemical abundances to establish a common abundance scale. With this homogeneous sample of open clusters, we study the relative elemental abundances of stars in open clusters in comparison with field stars as a function of age and metallicity. We find a range of mild enrichment of heavy (Ba$-$Eu) elements in young open cluster giants over field stars of the same metallicity. Our analysis succinct that the youngest stellar generations in cluster might be under-represented by the solar neighborhood field stars.

\end{abstract}


\section{Introduction} 

The first generation stars formed in galaxies are composed almost entirely of hydrogen and helium. When these stars evolve and return their nuclear-processed interiors to the interstellar medium (ISM), the enriched gas then present will be incorporated into future generation of stars. The amount of the chemical elements observed today and the timescales over which the ISM is being enriched with heavy metals are then, a function of many processes: the star formation rate (SFR), the initial mass function (IMF), the rate of element production and eventual return to the ISM via mass-loss and thresholds on the gas density for the star formation to proceed. The synthesis of the chemical elements and their return to the ISM are functions of lifetime and mass of stars. As the physical conditions such as the surface density of gas in the galactic disk and the SFR vary throughout many galaxies, the observed/derived abundances are a function of position as well. Therefore, the precise measurement of the variation of chemical elements as a function of radius in the galactic disk (i.e. the radial abundance gradient) and the gradient's temporal variation over the disk's lifetime are essential to develop a complete picture of Galactic evolution.

Stars numbering hundreds to thousands are born in rich open clusters (OCs) (see for example, De Silva et al. 2009). Many high-dispersion spectroscopic abundance analyses support the presence of chemical homogeneity among cluster members with a typical star-to-star abundance scatter of about 0.01 to 0.05 over many elements (Carretta et al. 2005; De Silva et al. 2006 \& 2007; Pancino et al. 2010, Reddy et al. 2012, hereafter Paper {\scs I} and Reddy et al. 2013, hereafter Paper {\scs II}). This implies that the OCs harbor coeval group of stars all formed in a single burst of star formation from a well mixed proto-cloud. An OC facilitates the measurement of basic parameters like age, distance, kinematics and metallicity more accurately than for field stars. The broad coverage of ages and Galactocentric distances of these OCs make them powerful tracers to map the structure, kinematics, and chemistry of the Galactic disk with respect to Galactic coordinates and its evolution with time.

\begin{table*}
 \centering
\caption{Target clusters and their properties from the literature.}
 \label{tab1}
\begin{tabular}{lcccccccc}   \hline
\multicolumn{1}{l}{Cluster}&  \multicolumn{1}{c}{$\ell$}& \multicolumn{1}{c}{$b$}& \multicolumn{1}{c}{Age}& 
\multicolumn{1}{c}{[Fe/H]$_{\rm phot.}$}& \multicolumn{1}{c}{R$_{\rm gc}$}& \multicolumn{1}{c}{(m-M)$_{V}$}&
\multicolumn{1}{c}{E(B-V)}& \multicolumn{1}{c}{[Fe/H]$_{\rm ref}$} \\
\multicolumn{1}{c}{}& \multicolumn{1}{c}{(deg.)}& \multicolumn{1}{c}{(deg.)}& \multicolumn{1}{c}{(Gyr)}&
\multicolumn{1}{c}{(dex.)}& \multicolumn{1}{c}{(kpc)}& \multicolumn{1}{c}{(mag.)}& \multicolumn{1}{c}{(mag.)}&
\multicolumn{1}{l}{ } \\
\hline

NGC 1342 & 154.95 & $-$15.34 & 0.45 & $-$0.16 &  8.6 & 10.10 & 0.32 & Gratton (2000) \\
NGC 1662 & 187.69 & $-$21.11 & 0.42 & $-$0.09 &  8.4 & 09.14 & 0.30 & Twarog et al. (1997) \\
NGC 1912 & 172.25 & $+$00.69 & 0.29 & $-$0.11 &  9.1 & 10.91 & 0.25 & Lyng\r{a} (1987)  \\
NGC 2354 & 238.37 & $-$06.79 & 0.13 & $-$0.30 &  8.8 & 14.01 & 0.31 & Claria et al. (1999) \\
NGC 2447 & 240.04 & $+$00.13 & 0.39 & $-$0.09 &  8.6 & 10.22 & 0.05 & Claria et al. (2005) \\

\hline
\end{tabular}
\end{table*}

\begin{table*}
 \centering
 \caption{The journal of the observations for each of the cluster members analysed in this paper.} 
 \label{tab2}
\begin{tabular}{lcccccccccl}   \hline
\multicolumn{1}{l}{Cluster}& \multicolumn{1}{c}{Star ID} & \multicolumn{1}{c}{$\alpha(2000.0)$}& \multicolumn{1}{c}{$\delta(2000.0)$}&  
\multicolumn{1}{c}{V}& \multicolumn{1}{c}{B-V} & \multicolumn{1}{c}{V-K$_{\rm s}$} &  \multicolumn{1}{c}{J-K$_{\rm s}$} &
\multicolumn{1}{c}{$RV_{\rm helio}$} & \multicolumn{1}{l}{S/N at} & \multicolumn{1}{l}{Date of} \\
\multicolumn{1}{c}{}& \multicolumn{1}{c}{}& \multicolumn{1}{c}{(hh mm s)}& \multicolumn{1}{c}{($\degr$ $\arcsec$ $\arcmin$)}&
\multicolumn{1}{c}{(mag.)}& \multicolumn{1}{c}{(mag.)} & \multicolumn{1}{c}{(mag.)}& \multicolumn{1}{c}{(mag.)} & 
\multicolumn{1}{c}{(km s$^{-1}$)} & \multicolumn{1}{l}{6000 \AA } & \multicolumn{1}{l}{observation} \\
\hline

NGC 1342 &  4  & 03 32 11.23 & +37 22 55.43  & 09.26 &$+$1.31 &$+$3.26 &$+$0.78 &$-$10.9$\pm$0.2 & 180 & 29-11-2012 \\
         &  6  & 03 31 26.98 & +37 21 28.62  & 09.65 &$+$1.18 &$+$3.02 &$+$0.66 &$-$10.3$\pm$0.2 & 170 & 29-11-2012 \\
         &  7  & 03 32 02.46 & +37 21 21.50  & 09.98 &$+$1.22 &$+$2.89 &$+$0.64 &$-$10.8$\pm$0.2 & 170 & 30-11-2012 \\
NGC 1662 &  1  & 04 48 29.51 & +10 55 48.27  & 08.34 &$+$1.18 &$+$2.99 &$+$0.71 &$-$13.6$\pm$0.2 & 180 & 29-11-2012 \\
         &  2  & 04 48 32.08 & +10 57 59.02  & 08.87 &$+$1.16 &$+$3.02 &$+$0.69 &$-$12.9$\pm$0.2 & 180 & 29-11-2012 \\
NGC 1912 &   3 & 05 28 44.05 & +35 49 52.77  & 09.85 &$+$1.19 &$+$2.85 &$+$0.66 &$-$00.2$\pm$0.2 & 140 & 18-11-2011 \\
         &  70 & 05 29 08.37 & +35 51 29.78  & 10.04 &$+$1.10 &$+$3.13 &$+$0.66 &$+$00.6$\pm$0.2 & 170 & 18-11-2011 \\
NGC 2354 & 183 & 07 13 51.93 &$-$25 44 24.30 & 11.41 &$+$1.25 &$+$2.79 &$+$0.70 &$+$35.6$\pm$0.4 & 100 & 06-03-2013 \\
         & 205 & 07 13 59.21 &$-$25 45 50.31 & 11.13 &$+$1.20 &$+$2.50 &$+$0.71 &$+$35.0$\pm$0.3 & 110 & 06-03-2013 \\
NGC 2447 &  28 & 07 44 50.25 &$-$23 52 27.14 & 09.96 &$+$0.82 &$+$2.33 &$+$0.56 &$+$21.1$\pm$0.3 & 100 & 04-03-2013 \\
         &  34 & 07 44 33.67 &$-$23 51 42.20 & 10.15 &$+$0.90 &$+$2.21 &$+$0.57 &$+$22.7$\pm$0.3 & 110 & 04-03-2013 \\
         &  41 & 07 44 25.73 &$-$23 49 52.95 & 10.16 &$+$0.89 &$+$2.28 &$+$0.52 &$+$22.2$\pm$0.2 & 110 & 04-03-2013 \\
    
\hline
\end{tabular}
\end{table*}

For this reason, independent groups have attempted to derive two fundamental relations, the age-metallicity relation and the radial metallicity gradient in the disk, using OC elemental abundances. But, such studies are presently limited by the lack of large and homogeneous datasets. Several recent attempts have been made to construct a homogeneous sets of precise metallicity measurements for OCs (see for example, Heiter et al. 2014). Despite this effort to provide a homogeneous set of metallicities, many clusters lack spectroscopically determined chemical compositions. This series is intended to examine some of these clusters.

This is our third paper reporting a comprehensive abundance measurements for red giants in OCs lacking detailed information on their chemical composition. In our previous papers (Paper {\scs I} \& {\scs II}) we have presented  abundance measurements for eleven OCs whose Galactocentric distances (R$_{\rm gc}$) lie between 8.3 and 11.3 kpc with ages between 0.2 to 4.3 Gyr. Here, we add five OCs, four of which have not been previously analysed in detail, in the Galactic anticentre direction. Our overall goal is to improve our understanding of Galactic chemical evolution (GCE). We combine our newly observed OCs with clusters in the literature for which high-resolution optical spectra have been published. Useful data on chemical composition is thus provided for an additional 53 clusters.

The layout of the paper is as follows: In Section 2 we describe the target selection and observations, and Section 3 is devoted to the data reduction and radial velocity measurements. Section 4 is devoted to the abundance analysis and in Section 5 we merge our sample OCs with those from the literature and describe the method adopted to establish a common abundance scale followed by a discussion on the assignment of OCs to Galactic populations. We discuss in Section 6 the relative abundances of open clusters in comparison with field star chemical compositions with respect to age and metallicity. In Section 7 we discuss the spread in chemical composition from cluster to cluster in the context of exploring the elements suitable for chemical tagging. Finally, Section 8 provides the conclusions.
  
\section{Target selection and Observations}

The new sample of OCs not yet subjected to abundance analysis, except NGC 2447, via high-resolution optical spectroscopy was chosen from the {\it New catalogue of optically visible open clusters and candidates\footnote{http://www.astro.iag.usp.br/~wilton/}} (Dias et al. 2002). Selection of the red giant members of an OC increases the distance over which the OC sample may be drawn but eliminates the very youngest clusters. Red giants provide spectra favourable for abundance determination: sharp lines with strengths from weak to strong for elements sampling the major processes of stellar nucleosynthesis.

We have made use of the {\small WEBDA}\footnote{\url{http://www.univie.ac.at/webda/}} database for the selection of suitable cluster members and cross-checked their astrometric and photometric measurements with the {\small SIMBAD}\footnote{\url{http://simbad.u-strasbg.fr/simbad/}} astronomical database and more recent measurements are adopted. Table \ref{tab1} summarizes the basic properties of the target clusters: references to the adopted photometric [Fe/H] are also given; all quantities are from the databases while the Galactocentric distance, R$_{\rm gc}$, was calculated assuming a distance of the Sun from the Galactic centre of 8.0$\pm$0.6 kpc (Ghez et al. 2008). The present sample of OCs covers a galactocentric distance of 8.4 to 9.1 kpc and an age range of 0.1 to 0.4 Gyr. 

Observations were carried out during observing runs in 2011 May \& November, 2012 November, and 2013 March using the Robert G. Tull echelle spectrograph (Tull et al. 1995) at the coud\'{e} focus of the 2.7-m Harlan J. Smith telescope located at the McDonald observatory. On all occasions we employed a 2048$\times$2048 24 $\mu$m pixel, backside illuminated, anti-reflection coated CCD as a detector and the 52.67 grooves mm$^{-1}$ echelle grating with exposures centred at 5060 \AA. 

Each night's observing run included 5-10 zero second exposures (bias frames), 10-15 quartz lamp exposures (flat frames), and 2-3 comparison Th-Ar spectra to provide the reference wavelength scale. We also obtained on each observing night the spectrum of hot, rapidly-rotating stars to monitor the presence of telluric lines. 

A total of twelve stars spread across the five OCs were observed. For each target, we obtained two to three exposures, each lasting for 20-30 min to minimize the influence of cosmic rays and to acquire a good signal-to-noise (S/N) ratio. All the spectra correspond to a resolving power of $R$ $\gtrsim$ 55,000 ($<$ 6 km s$^{-1}$) as measured by the FWHM of Th {\scs I} lines in comparison spectra, except for the members of NGC 2354 whose spectra was taken at the lower resolution of 30,000. The spectral coverage in a single exposure from 3700 \AA\ to 9800 \AA\ across various orders was complete but for the inter-order gaps which begin to appear longward of 5600 \AA.

\section{Data reduction and radial velocities}

\begin{figure}
\begin{center}
\includegraphics[trim=0.6cm 0.6cm 0.5cm 4.5cm, clip=true,height=0.30\textheight]{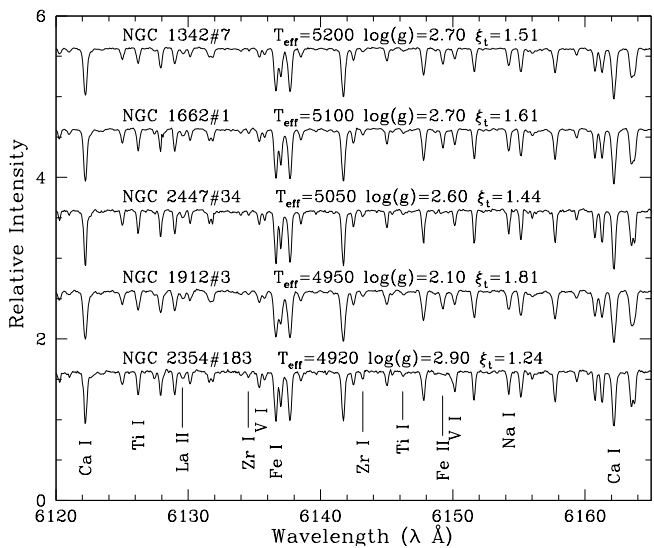}
\caption{Normalized spectra of red giant members of the five OCs as described in Table \ref{tab2} are presented in descending order of temperature (top to bottom) in the 6120$-$6165 \AA\ region. All the spectra are on the same scale but have been displaced vertically from each other for clarity and better visibility.}
\label{spectra} 
\end{center}
\end{figure}

The spectroscopic data reductions were done exactly as in Paper {\scs I} using various routines available within the \textit{imred} and \textit{echelle} packages of the standard spectral reduction software {\scs IRAF}\footnote{IRAF is a general purpose software system for the reduction and analysis of astronomical data distributed by NOAO, which is operated by the Association of Universities for Research in Astronomy, Inc. under cooperative agreement with the National Science Foundation.}. Briefly, the two-dimensional target star frames were de-trended by subtracting bias level, correcting for scattered light and then divided by the normalized flat field. The individual echelle orders were traced, extracted to one-dimensional spectra and then wavelength calibrated using Th-Ar spectra as a reference. 

Multiple spectra were combined to acquire a single, high S/N spectra for each star. The combined spectra have S/N ratios of about 100-180 as measured around 6000 \AA\ region, while shortward of 5000 \AA\ the S/N ratio decreases with wavelength and reaches a value of about 25 around 3700 \AA\ region. The noisy ends of each echelle order were trimmed to reduce the edge effects before continuum fitting. The spectrum was normalized interactively to unity with the available cursor commands in \textit{splot} task of {\scs IRAF} by marking continuum regions, those regions which are uneffected by the presence of spectral lines, on each aperture which were then fitted by a slowly varying function such as cubic spline of appropriate order for better normalization. We referred our spectra to high-resolution spectrum of Arcturus and Sun (Hinkle et al. 2000) to identify the continuum regions. Spectra of a representative region are shown in Figure \ref{spectra} for one star from each of the five OCs.

We have verified the membership of target stars in each of the OCs using their radial velocities (RVs). RV of each star was calculated from its normalized spectrum by measuring the shift in the central wavelength from their laboratorary values of well defined line cores. The absence of systematic RV shift for the spectral lines selected from the blue and red wavelength regions strengthens the accuracy of our dispersion solution. The observed RVs were corrected for solar motion using {\scs IRAF}'s {\it rvcorrect} routine. Our mean RV measurements for all OCs save NGC 1912 are in fair agreement with the previous RV measurements for the red giants in OCs (Mermilliod et al. 2008). 

\begin{figure*}
\begin{center}
\includegraphics[trim=1.8cm 5.5cm 3.5cm 5.2cm, clip=true,height=0.45\textheight]{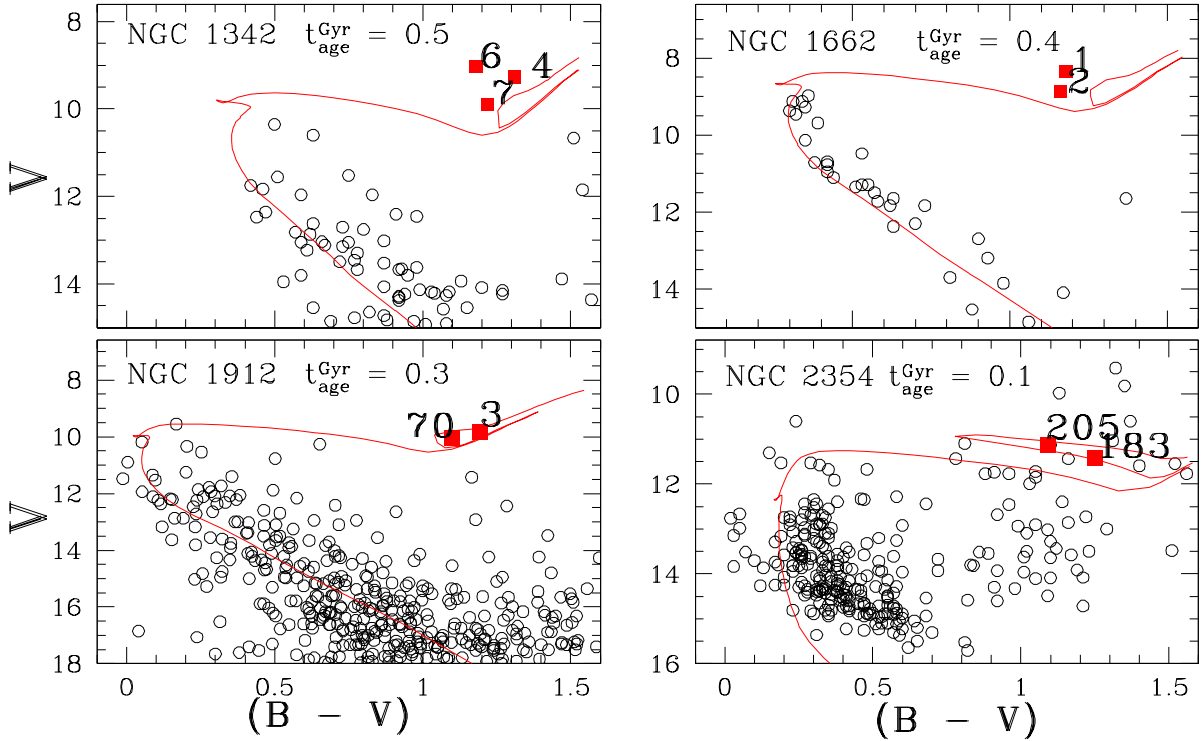}
\caption[]{The fitting of Padova isochrones by Marigo et al. (2008) to the BV color-magnitude diagram of the clusters NGC 1342 (photometry from Hoag et al. 1961), NGC 1662 (photometry from Hoag et al. 1961), NGC 1912 (photometry from Subramaniam \& Sagar 1999) \& NGC 2354 (photometry from D\"{u}rbeck 1960) using the distance, reddening and age information from Dias et al. (2002) catalogue. The open circles (black) represent the photometric cluster members while the filled squares (red) for the position of the program stars for which we derived abundances.} 
\label{cmd}
\end{center}
\end{figure*}

For NGC 1912, Mermilliod et al. (2008) derived a RV of $-$45.0$\pm$0.1 km s$^{-1}$ using a single star ($\#$405) which as noted by proper motion data of Mills (1967) is a cluster non-member. We have observed two stars ($\#$3 \& $\#$70) from this cluster, whose individual proper motions are consistent with the mean proper motions of cluster, implying that they are cluster members. Moreover, the measured RVs (Table \ref{tab2}) and positions of the stars in a (B-V) colour-magnitude diagram (see Figure \ref{cmd}) are also consistent with membership. Based on a cross-correlation technique, Glushkova \& Rastorguev (1991) derived a radial velocity of $-$1.0$\pm$0.58 km s$^{-1}$ (one star), in a good agreement with our measured value of $\langle$RV$\rangle$ = +0.2$\pm$0.3 km s$^{-1}$ (two stars). Hence, our radial velocity estimate for this OC would appear to be the first measurement available in the literature using high dispersion echelle spectra of two potential cluster members. Though the brighter giants in NGC 2447 are cluster members, we avoided them for observations as they have cool atmospheres with T$_{\rm eff}$ $\lesssim$ 3800 K

The journal of the observations for each of the cluster members is given in Table \ref{tab2} together with the identifications, J2000 coordinates, available optical and 2MASS\footnote{\url{http://irsa.ipac.caltech.edu/applications/Gator}} photometry (Cutri et al. 2003)\footnote{Originally published in University of Massachusetts and Infrared Processing and Analysis Center (IPAC)/ California Institute of Technology.}, computed heliocentric RVs, and S/N of the spectra around 6000 \AA\ for each of the stars.    

\begin{figure}
\begin{center}
\includegraphics[trim=1.9cm 8.1cm 4.5cm 5.3cm, clip=true,height=0.28\textheight]{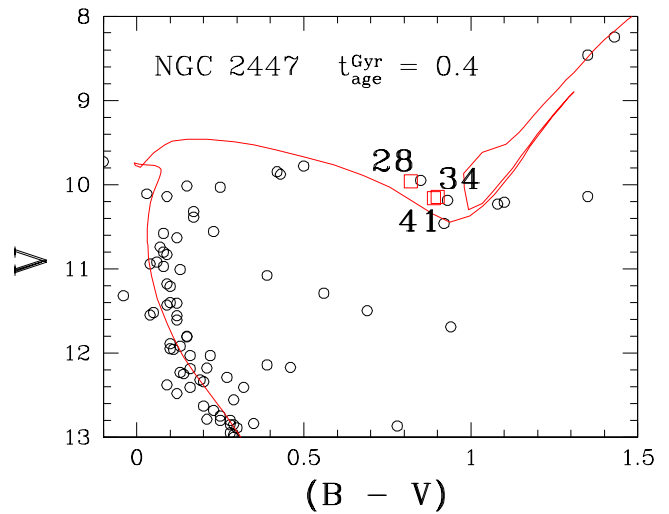}
\caption[]{The fitting of Padova isochrones to the BV color-magnitude diagram of NGC 2447 (photometry from Becker et al. 1976) using the distance, reddening and age information from Dias et al. (2002) catalogue. We marked our program stars with open squares (red).}  
\label{cmd2447}
\end{center}
\end{figure}

\begin{table*}
{\fontsize{7}{8}\selectfont
\caption{The linelist for all cluster members analysed in this paper.}
\label{EWmeasurement}
\begin{tabular}{llllrrr|rrrrrrrrr} \hline
\multicolumn{4}{l}{ } & \multicolumn{3}{c}{NGC 1342} & \multicolumn{2}{c}{NGC 1662} & \multicolumn{2}{c}{NGC 1912} & \multicolumn{2}{c}{NGC 2354} & \multicolumn{3}{c}{NGC 2447}  \\ 
\multicolumn{1}{l}{$\lambda$(\AA)} & \multicolumn{1}{c}{Species$^{a}$ } & \multicolumn{1}{c}{LEP$^{b}$} & \multicolumn{1}{c}{$\log~{gf}$} & \multicolumn{1}{c}{\#4} & \multicolumn{1}{c}{\#6} & \multicolumn{1}{c}{\#7} & \multicolumn{1}{c}{\#1} &\multicolumn{1}{c}{\#2} & \multicolumn{1}{c}{\#3} & \multicolumn{1}{c}{\#70} & \multicolumn{1}{c}{\#183} & \multicolumn{1}{c}{\#205} & \multicolumn{1}{c}{\#28} & \multicolumn{1}{c}{\#34} & \multicolumn{1}{c}{\#41} \\ \hline
   
 4668.56 & 11.0 & 2.10 & -1.31 &  90.8 &  82.3 &  73.0 &  79.0 &  73.1 &$\ldots$&$\ldots$&$\ldots$&$\ldots$&$\ldots$&$\ldots$&$\ldots$  \\
 4982.82 & 11.0 & 2.10 & -0.91 & 116.4 & 104.9 &$\ldots$&$\ldots$&$\ldots$&$\ldots$&$\ldots$& 97.1 &$\ldots$&  101.5 & 101.8 &  96.2    \\
 5688.22 & 11.0 & 2.10 & -0.45 & 149.3 & 141.9 & 128.8 & 143.1 & 134.0 & 158.1 & 156.5 &$\ldots$&$\ldots$& 129.3 &$\ldots$& 130.4       \\
 6154.23 & 11.0 & 2.10 & -1.55 &  76.4 &  66.9 &  60.9 &  71.9 &  67.7 &$\ldots$& 96.3 &  67.2 & 66.0 &   66.2 &  62.6 &  60.3          \\
 6160.75 & 11.0 & 2.10 & -1.27 &  96.7 &  88.1 &  75.4 &  91.9 &  83.7 & 101.7 & 105.8 &  80.1 & 87.8 &   80.9 &  80.5 &$\ldots$        \\
 
\hline
\end{tabular} }

\flushleft  
$^{a}$ The integer part of the 'Species' indicates the atomic number, and the decimal component indicates the ionization state \\ 
\hspace{0.2cm} (0 = neutral, 1 = singly ionized). \\
$^{b}$ All the lines are arranged in the order of their increasing Lower Excitation Potential (LEP). \\ 
Note. Only a portion of this table is shown here for guidance regarding its form and content. A machine-readable version of the full table is available as Supporting Information with the online version of this Paper. 
\end{table*}

\section{Abundance analysis} 

\subsection{\small Line selection and Equivalent widths} \label{line_selection}

The method of abundance analysis has been described in our Paper {\scs I}, so we refer the reader to that paper for more information on the adopted line list and the source of references to atomic data. We employed a strict criteria to select clean, unblended, isolated and symmetric spectral lines due to weak and moderately strong lines of various atomic/ionic species avoiding the wavelength regions affected by telluric contamination and heavy line crowding. The line equivalent widths (EWs) were measured manually using the cursor commands in {\scs IRAF}'s \textit{splot} package by fitting often a Gaussian profile and for a few lines a direct integration was performed for the best measure of EW.     

Our final line list has about 300 absorption lines covering 24 elements in the spectral range $\sim$ 4300 $-$ 8850 \AA. Our selection criteria renders, on an average across the sample of 12 stars, a list of 100 Fe {\scs I} lines with lower excitation potentials (LEPs) ranging from 0.9 to 5.0 eV and EWs of up to 180 m\AA\ and 15 Fe {\scs II} lines with LEPs of 2.8 to 3.9 eV and EWs from $\simeq$ 25 to 130 m\AA. Chemical abundances for almost all elements are based on lines weaker than 100 m\AA\ and strong lines were employed only for species represented by a few lines (for example, Mg {\scs I}). 

The accuracy of the derived abundances to a great extent depend on the quality of the $gf$-values which varies from line to line of a given element. 
The $gf$-values of high accuracy (5 to 10\%) are available for a large fraction of iron lines (Fe {\scs I} \& Fe {\scs II}) from recent critical reviews (F\"{u}hr \& Wiese 2006). This gave us a freedom to select a large sample of absorption lines due to iron thanks to the presence of numerous iron lines in the optical region of spectra. As the overall goal is to have the best possible abundance determination, for non-Fe elements represented by fewer lines we have identified a set of self-consistent absorption lines having high-quality relative $gf$-values.  
We provide in Table \ref{EWmeasurement}, the atomic data adopted for each spectral line and the EW measurements for all stars analysed.

We derived the solar abundances using solar EWs, measured off the solar integrated disk spectrum (Kurucz et al. 1984), and the ATLAS9 model atmosphere for T$_{\rm eff},_{\odot}$ = 5777 K, log~{g}$_{\odot}$ = 4.44 cm s$^{-2}$ to establish a reference abundance scale and to reduce the systematic errors in final abundances. We found a microturbulence velocity of $\xi_{t}$ = 0.95 km s$^{-1}$ using Fe {\scs II} lines. The derived solar abundances that agree well with the published values of Asplund et al. (2009) are reported in Table 4 of Paper {\scs I}. We refer to our solar abundances when determining the stellar abundances, [X/H] and [X/Fe], i.e., our analysis is essentially a differential one relative to the Sun.   

\begin{table*}
\centering
\begin{minipage}{165mm}
\caption{Basic photometric and spectroscopic atmospheric parameters for the stars in each cluster.}
\label{tab3}
\begin{tabular}{lcccccccccc}  \hline
\multicolumn{1}{l}{Cluster}& \multicolumn{1}{c}{Star ID} & \multicolumn{3}{c}{T$^{\rm phot}_{\rm eff}$ (K)}&
\multicolumn{1}{c}{$\log g^{(V-K)}_{\rm phot}$}& \multicolumn{1}{c}{T$^{\rm spec}_{\rm eff}$}& 
\multicolumn{1}{c}{$\log g_{\rm spec}$}& \multicolumn{1}{c}{$\xi_{\rm spec}$}&
\multicolumn{1}{c}{$\log(L/L_\odot)$} & \multicolumn{1}{l}{$\log(L/L_\odot)$} \\  \cline{3-5} 
\multicolumn{1}{c}{}& \multicolumn{1}{c}{} & \multicolumn{1}{c}{(B-V)}& (V-K)& (J-K)& 
\multicolumn{1}{c}{(cm s$^{-2}$)}& \multicolumn{1}{c}{(K)} & \multicolumn{1}{c}{(cm s$^{-2}$)}& 
\multicolumn{1}{c}{(km sec$^{-1}$)}& \multicolumn{1}{l}{spectroscopy} & \multicolumn{1}{l}{photometry} \\
\hline

NGC 1342 &  4 & 4766 & 4667 & 4614 & 2.21 & 5100 & 2.40 & 1.70 & 2.31 & 2.34 \\
         &  6 & 5035 & 4908 & 5022 & 2.50 & 5200 & 2.80 & 1.66 & 1.95 & 2.14 \\
         &  7 & 4948 & 5058 & 5120 & 2.68 & 5200 & 2.70 & 1.51 & 2.05 & 2.03 \\
NGC 1662 &  1 & 5023 & 4894 & 4800 & 2.32 & 5100 & 2.70 & 1.61 & 1.97 & 2.28 \\
         &  2 & 5067 & 4862 & 4884 & 2.53 & 5200 & 2.85 & 1.49 & 1.85 & 2.06 \\
NGC 1912 &  3 & 4852 & 4875 & 4883 & 2.27 & 4950 & 2.10 & 1.81 & 2.61 & 2.42 \\
         & 70 & 5044 & 4604 & 4869 & 2.28 & 4950 & 2.00 & 1.70 & 2.71 & 2.30 \\
NGC 2354 &183 & 4822 & 5132 & 4854 & 2.43 & 4920 & 2.90 & 1.24 & 1.80 & 2.35 \\
         &205 & 4926 & 5534 & 4829 & 2.47 & 4850 & 2.80 & 1.29 & 1.88 & 2.44 \\
NGC 2447 & 28 & 5257 & 4849 & 4837 & 2.58 & 5050 & 2.70 & 1.42 & 1.97 & 2.02 \\
         & 34 & 5069 & 4976 & 4802 & 2.64 & 5050 & 2.60 & 1.44 & 2.05 & 1.98 \\
         & 41 & 5091 & 4898 & 5001 & 2.64 & 5100 & 2.80 & 1.59 & 1.88 & 1.97 \\
\hline
\end{tabular}
\end{minipage}
\end{table*}

\subsection{Determination of atmospheric parameters} \label{Dap}

As the EW of a spectral line is affected by the physical conditions in the stellar atmosphere and number density of absorbers, it is necessary to predetermine the atmospheric parameters to estimate the stellar abundances. We obtained our preliminary estimates from dereddened\footnote{The adopted interstellar extinctions are (A$_{V}$, A$_{K}$, E(V-K), E(J-K))= (3.1, 0.28, 2.75, 0.54)*E(B-V), where E(B-V) is taken from {\scs WEBDA}} optical and 2MASS photometry, following the precepts discussed in our Paper {\scs I}. Effective temperature of each star was calculated using the empirically calibrated color$-$temperature relations\footnote{Colours selected included are (B-V), (V-K) and (J-K).} of Alonso et al. (1999). 

The surface gravities were computed by incorporating the known distance to the OCs, derived effective temperature T$^{\rm phot}_{\rm eff_\star}$, bolometric correction $BC_{V}$, cluster turn-off mass $M_{\star}$ and solar values of T$_{\rm eff},_{\odot}$= 5777 K and log~{g}$_{\odot}$= 4.44 cm s$^{-2}$ into the log~$g$\,$-$\,T$_{\rm eff}$ relation given by Allende Prieto et al. (1999). The $BC_{V}$s were derived from estimated effective temperatures and photometric metallicities (Table \ref{tab1}) using Alonso et al's (1999) calibration. 

The turn-off mass of giants has been estimated by fitting the cluster CMD with Padova stellar evolutionary tracks of Marigo et al. (2008): the adopted turn-off masses are 3.1 $M_{\odot}$ for NGC 1342, 2.8 $M_{\odot}$ for NGC 1662, 3.5 $M_{\odot}$ for NGC 1912, 3.5 $M_{\odot}$ for NGC 2354, 2.9 $M_{\odot}$ for NGC 2447. For all clusters, the fundamental cluster parameters (age, distance and reddening) of Dias et al. (2002) catalogue were adopted.

We refined our photometric estimates of atmospheric parameters using spectroscopy. The line list and model atmospheres were then used as inputs to the local thermodynamical equilibrium (LTE) line analysis and spectrum synthesis code {\scs \bf MOOG}\footnote{{\scs \bf MOOG} was developed and updated by Chris Sneden and originally described in Sneden (1973)} for an abundance analysis. The one-dimensional, line-blanketed plane-parallel uniform model atmospheres by assumption of LTE, hydrostatic equilibrium and flux conservation were interpolated linearly from the ATLAS9 model atmosphere grid of Castelli \& Kurucz (2003).     
   
We performed a differential abundance analysis relative to the Sun by running the {\scs \bf MOOG} in {\it abfind} mode. Starting from a model with photometric estimates of atmospheric parameters and the measured EWs, the individual line abundances were force-fitted to match the computed EWs to the observed ones while satisfying the following three constraints simultaneously: First, the microturbulence, $\xi_{t}$, assumed to be isotropic and depth independent was determined by the requirement that the Fe abundance from Fe {\scs II} lines be independent of a line's EW. Fe {\scs II} lines are preferred over Fe {\scs I} as they cover a small range in LEP so that the derived abundance will be free of non-LTE effects, and cover a good range in EWs so that $\xi_{t}$ will be measured accurately. A check on the $\xi_{t}$ is provided by lines of Sc {\scs I}, Ti {\scs I}, Ti {\scs II}, V {\scs I}, Cr {\scs I} and Cr {\scs II} species. Second, the effective temperature, T$_{\rm eff}$, is estimated by imposing the requirement that the Fe abundance from Fe I lines (as they cover a good range in LEP $\sim$ 0.0 to 5.0 eV) be independent of a line's LEP (excitation equilibrium). This condition is also verified with the lines of other species like Ti {\scs I} and Ni {\scs I}. Third, the surface gravity is adjusted until the Fe abundance derived by Fe I and Fe II lines matched within 0.02 dex (i.e. maintaining the ionization equilibrium between the neutral and ionized species) for the derived T$_{\rm eff}$ and $\xi_{t}$. A check on this condition is also performed by Sc, Ti, V, and Cr species as they provide both neutral and ionized lines. As all these atmospheric parameters are interdependent, several iterations are needed to choose a suitable model from the grid so that all the spectral lines in observed spectra are readily reproduced.

The error analysis on the derived spectroscopic atmospheric parameters was performed in the same way as described in Paper {\scs II}. The typical errors considered in our analysis are 100 K in T$_{\rm eff}$, 0.25 cm s$^{-2}$ in log~$g$ and 0.20 km s$^{-1}$ in $\xi_{t}$. The derived stellar parameters for program stars in each of the cluster are given in Table \ref{tab3}.

With few exceptions, our spectroscopic estimates are in good agreement with photometric ones. The uncertainties affecting the photometric estimates are discussed in Paper {\scs II}, and were ascribed to adopted colors and reddening estimates. The mean difference in photometric temperatures estimated using (B-V) and (V-K) is $+$53 $\pm$ 279 K and using (V-K) and (J-K) is $+$62 $\pm$ 236 K. The corresponding mean differences in T$^{(B-V)}_{\rm eff}$, T$^{(V-K)}_{\rm eff}$ and T$^{(J-K)}_{\rm eff}$ with respect to spectroscopic T$^{spec}_{\rm eff}$'s are $-$64 $\pm$ 145 K, $-$117 $\pm$ 290 K and $-$179 $\pm$ 132 K respectively. Mean differences in log~$g$ and log(L/L$_{\odot}$) across the 12 stars are $-0.15\pm0.22$ cm-s$^{-2}$ and $+0.10\pm0.27$ cm-s$^{-2}$ respectively. The corresponding comparison of the spectroscopic [Fe/H] with photometric ones in Table \ref{tab1} also illustrates fair agreement: $\Delta$[Fe/H] = $-$0.02 (NGC 1342), $-$0.01 (NGC 1662), 0.00 (NGC 1912), $+$0.11 (NGC 2354), and $-$0.04 (NGC 2447). 

\begin{table}
{\fontsize{8}{8}\selectfont
\begin{minipage}{90mm}
\caption{Sensitivity of abundances to the uncertainties in the model parameters for the star 
with ID 6 in NGC 1342 with T$_{\rm eff}$= 5200 K, $\log{g}$= 2.80 cm s$^{-2}$,and $\xi_{t}$= 1.66 km s$^{-1}$.}
\label{sensitivity}
\begin{tabular}{lllll}   \hline
\multicolumn{1}{l}{ }&\multicolumn{1}{l}{T$_{\rm eff}\pm$100 K}&\multicolumn{1}{l}{$\log~{g}\pm$ 0.25} &
\multicolumn{1}{l}{$\xi_{t}\pm$ 0.20} & \multicolumn{1}{l}{ } \\ \cline{2-5}
\multicolumn{1}{l}{Species}&\multicolumn{1}{c}{$\sigma_{T_{\rm eff}}$}&\multicolumn{1}{c}{$\sigma_{log{g}}$} & 
\multicolumn{1}{c}{$\sigma_{\xi_{t}}$} &\multicolumn{1}{c}{$\sigma_{2}$}  \\ \hline

Na {\scs I} & $+0.08/-0.07$ &$ 0.00/+0.03$  &$-0.05/+0.05$   & 0.05   \\
Mg {\scs I} & $+0.07/-0.05$ &$-0.01/+0.05$  &$-0.04/+0.04$   & 0.05   \\
Al {\scs I} & $+0.04/-0.04$ &$-0.01/+0.01$  &$-0.02/+0.02$   & 0.03   \\
Si {\scs I} & $+0.02/+0.01$ &$+0.05/-0.01$  &$-0.02/+0.02$   & 0.02   \\
Ca {\scs I} & $+0.10/-0.09$ &$-0.01/+0.04$  &$-0.02/+0.11$   & 0.07   \\
Sc {\scs I} & $+0.11/-0.12$ &$-0.01/+0.01$  &$-0.01/ 0.00$   & 0.07   \\
Sc {\scs II}&$ 0.00/+0.01$ &$+0.12/-0.11$  &$-0.06/+0.06$   & 0.07   \\ 
Ti {\scs I} & $+0.12/-0.11$ &$ 0.00/+0.02$  &$-0.02/+0.03$   & 0.07   \\
Ti {\scs II}&$+0.01/+0.02$ &$+0.13/-0.09$  &$-0.06/+0.07$   & 0.07   \\
V  {\scs I} & $+0.12/-0.14$ &$-0.01/+0.01$  &$-0.01/+0.01$   & 0.08   \\
Cr {\scs I} & $+0.11/-0.10$ &$+0.01/+0.02$  &$-0.07/+0.08$   & 0.08   \\ 
Cr {\scs II}&$-0.01/+0.06$ &$+0.15/-0.07$  &$-0.05/+0.07$   & 0.08   \\ 
Mn {\scs I} & $+0.10/-0.09$ &$+0.01/-0.02$  &$-0.09/+0.11$   & 0.08   \\
Fe {\scs I} & $+0.09/-0.08$ &$+0.01/+0.01$  &$-0.08/+0.10$   & 0.07   \\
Fe {\scs II}&$-0.02/+0.06$ &$+0.14/-0.10$  &$-0.07/+0.08$   & 0.08   \\ 
Co {\scs I} & $+0.08/-0.08$ &$+0.02/-0.01$  &$-0.03/+0.02$   & 0.05   \\ 
Ni {\scs I} & $+0.06/-0.06$ &$+0.03/-0.02$  &$-0.05/+0.05$   & 0.05   \\ 
Cu {\scs I} & $+0.08/-0.06$ &$+0.03/-0.01$  &$-0.05/+0.07$   & 0.05   \\
Zn {\scs I} & $+0.01/+0.02$ &$+0.09/-0.06$  &$-0.13/+0.14$   & 0.09   \\
Y  {\scs II}& $+0.01/ 0.00$ &$+0.12/-0.10$  &$-0.05/+0.05$   & 0.07   \\
Zr {\scs I} & $+0.14/-0.15$ &$ 0.00/+0.01$  &$-0.01/+0.01$   & 0.08   \\
Zr {\scs II}&$+0.01/+0.02$ &$+0.13/-0.09$  &$-0.01/+0.02$   & 0.06   \\
Ba {\scs II}&$+0.03/-0.03$ &$+0.08/-0.08$  &$-0.17/+0.15$   & 0.10   \\
La {\scs II}&$+0.03/-0.01$ &$+0.12/-0.10$  &$-0.02/+0.02$   & 0.07   \\
Ce {\scs II}&$+0.02/-0.01$ &$+0.12/-0.10$  &$-0.05/+0.06$   & 0.07   \\
Nd {\scs II}&$+0.03/-0.01$ &$+0.12/-0.09$  &$-0.01/+0.03$   & 0.06   \\
Sm {\scs II}&$+0.03/-0.01$ &$+0.13/-0.09$  &$-0.02/+0.04$   & 0.07   \\
Eu {\scs II}&$ 0.00/+0.01$ &$+0.11/-0.11$  &$-0.02/+0.02$   & 0.06   \\

\hline
\end{tabular}
\end{minipage} } \vspace{-0.4cm}
\end{table} 

\subsection{Abundances and error estimation}

A complete abundance analysis was conducted by running the {\it abfind} driver of {\scs \bf MOOG} with the atmospheric parameters determined from the Fe {\scs I} and Fe {\scs II} lines as described in previous section. In most cases, the abundances are derived from the measured EWs but a few lines were analysed with synthetic spectra. 

\begin{table*}
\caption{Elemental abundance ratios [X/Fe] for elements from Na to Eu for NGC 1342, 1662, 1912, 2354 and 2447 from this study. Abundances calculated by synthesis are presented in bold typeface.}
\label{mean_abundance}
\begin{tabular}{lccccc}   \hline
\multicolumn{1}{c}{Species}  & \multicolumn{1}{c}{NGC 1342} & \multicolumn{1}{c}{NGC 1662} &  
\multicolumn{1}{c}{NGC 1912} & \multicolumn{1}{c}{NGC 2354} & \multicolumn{1}{c}{NGC 2447}   \\ \hline

$[$Na I/Fe$]$  &$+0.28\pm0.04$ &$+0.22\pm0.04$ &$+0.33\pm0.04$ &$+0.12\pm0.04$ &$+0.12\pm0.04$      \\
$[$Mg I/Fe$]$  &$ 0.00\pm0.04$ &$-0.06\pm0.03$ &$+0.03\pm0.02$ &$-0.17\pm0.04$ &$-0.02\pm0.04$     \\
$[$Al I/Fe$]$  &$-0.05\pm0.02$ &$-0.03\pm0.03$ &$+0.06\pm0.02$ &$-0.11\pm0.03$ &$-0.14\pm0.03$     \\
$[$Si I/Fe$]$  &$+0.11\pm0.02$ &$+0.16\pm0.03$ &$+0.23\pm0.02$ &$+0.16\pm0.03$ &$+0.11\pm0.03$     \\
$[$Ca I/Fe$]$  &$+0.07\pm0.05$ &$+0.11\pm0.06$ &$+0.14\pm0.03$ &$-0.07\pm0.06$ &$+0.02\pm0.05$     \\
$[$Sc I/Fe$]$  &$+0.04\pm0.05$ &$+0.02\pm0.10$  &$\ldots   $   &$\ldots   $    &$+0.04\pm0.09$      \\
$[$Sc II/Fe$]$ &$\ldots   $    &$+0.11\pm0.08$  &$\bf+0.10$    &$+0.05\pm0.10$ &$+0.10\pm0.05$     \\ 
$[$Ti I/Fe$]$  &$+0.02\pm0.05$ &$+0.06\pm0.05$ &$-0.07\pm0.03$ &$+0.01\pm0.07$ &$-0.04\pm0.05$     \\
$[$Ti II/Fe$]$ &$-0.04\pm0.05$ &$+0.05\pm0.07$ &$+0.03\pm0.06$ &$-0.06\pm0.06$ &$-0.05\pm0.06$     \\
$[$V I/Fe$]$   &$+0.01\pm0.06$ &$+0.03\pm0.06$ &$-0.07\pm0.04$ &$+0.04\pm0.06$ &$-0.02\pm0.06$     \\
$[$Cr I/Fe$]$  &$+0.01\pm0.06$ &$ 0.00\pm0.04$ &$+0.01\pm0.06$ &$-0.03\pm0.05$ &$-0.04\pm0.04$     \\ 
$[$Cr II/Fe$]$ &$+0.03\pm0.06$ &$+0.07\pm0.07$ &$+0.05\pm0.05$ &$-0.03\pm0.06$ &$+0.02\pm0.07$     \\ 
$[$Mn I/Fe$]$  &  $\bf-0.12 $  &   $\bf-0.05$  &$\bf-0.12$     &$\bf-0.05$     &   $\bf-0.07$          \\
$[$Fe I/H$]$   &$-0.14\pm0.05$ &$-0.10\pm0.06$ &$-0.11\pm0.05$ &$-0.19\pm0.04$ &$-0.13\pm0.05$      \\
$[$Fe II/H$]$  &$-0.13\pm0.06$ &$-0.11\pm0.07$ &$-0.09\pm0.06$ &$-0.16\pm0.08$ &$-0.11\pm0.08$      \\ 
$[$Co I/Fe$]$  &$-0.03\pm0.04$ &$ 0.00\pm0.04$ &$-0.10\pm0.02$ &$+0.07\pm0.04$ &$-0.04\pm0.04$      \\ 
$[$Ni I/Fe$]$  &$-0.06\pm0.04$ &$-0.02\pm0.04$ &$-0.02\pm0.05$ &$ 0.00\pm0.06$ &$-0.07\pm0.04$      \\ 
$[$Cu I/Fe$]$  &  $\bf-0.29$   &  $\bf-0.24$   &$\bf-0.30$     & $\bf-0.12$    &  $\bf-0.28$          \\
$[$Zn I/Fe$]$  &  $\bf-0.29$   &  $\bf-0.13$   &$\bf+0.10$     & $\bf-0.31$    &  $\bf-0.38$         \\  
$[$Rb I/Fe$]$  &  $\bf-0.04$   &  $\bf-0.14$   &$\bf-0.30$     & $\bf-0.17$    &  $\bf-0.18$         \\ 
$[$Y II/Fe$]$  &$+0.12\pm0.05$ &$+0.15\pm0.05$ &$+0.06\pm0.04$ &$+0.14\pm0.05$ &$+0.03\pm0.06$      \\
$[$Zr I/Fe$]$  &$+0.18\pm0.06$ &$+0.25\pm0.07$ &$+0.10\pm0.04$ &$+0.13\pm0.08$ &$+0.13\pm0.07$      \\
$[$Zr II/Fe$]$ &$+0.25\pm0.04$ &$+0.31\pm0.03$ &$\ldots      $ &$\ldots   $    &$+0.16\pm0.05$      \\
$[$Ba II/Fe$]$ & $\bf+0.32$    &   $\bf+0.54$  &$\bf+0.70$     &$\bf+0.17$     &   $\bf+0.23$        \\
$[$La II/Fe$]$ &$+0.16\pm0.05$ &$+0.22\pm0.05$ &$+0.14\pm0.04$ &$+0.23\pm0.08$ &$+0.13\pm0.05$      \\
$[$Ce II/Fe$]$ &$+0.36\pm0.05$ &$+0.37\pm0.06$ &$+0.23\pm0.04$ &$+0.38\pm0.05$ &$+0.32\pm0.06$      \\  
$[$Nd II/Fe$]$ &$+0.29\pm0.04$ &$+0.26\pm0.05$ &$+0.13\pm0.04$ &$+0.33\pm0.05$ &$+0.22\pm0.05$      \\
$[$Sm II/Fe$]$ &$+0.24\pm0.05$ &$+0.22\pm0.05$ &$+0.04\pm0.04$ &$+0.24\pm0.05$ &$+0.19\pm0.05$     \\
$[$Eu~II/Fe$]$ &$\bf+0.22$     & $\bf+0.20$    &$\bf+0.07$     &$\bf+0.16$     &$\bf+0.22$          \\

\hline
\end{tabular}
\end{table*}

We computed synthetic spectra for species affected by hyperfine structure (hfs) and isotopic splitting and/or affected by blends and matched them to the observed ones by adjustment of abundances. The features analysed by spectrum synthesis, the adopted hfs data and isotopic ratios are same as those listed in our Paper {\scs II}. We have tested our linelists extensively to reproduce the solar and Arcturus spectra and measured the solar abundances to establish a reference abundance scale. The spectrum synthesis was carried out by running the MOOG in '\textit{synth}' mode. 

Abundance results for the individual stars averaged over all available lines of given species are presented in Tables \ref{abu_1342}$-$\ref{abu_2447}, relative to solar abundances derived from the adopted $gf$-values. The Tables give the average [Fe/H] and [X/Fe] for all elements, standard deviation and the number of lines used in calculating the abundance of that element. Abundances calculated by synthesis are presented in bold typeface. Inspection of the Tables \ref{abu_1342}$-$\ref{abu_2447} shows that, in general, all stars in a given cluster have very similar chemical compositions [X/Fe] for almost all the elements and are identical to within the (similar) standard deviations computed for an individual star. Exceptions tend to occur for species represented by one or a few lines, as expected when the uncertainty in measuring equivalent widths is a contributor to the total uncertainty. From the spread in the abundances for the stars of a given cluster we obtain the standard deviation $\sigma_1$ in the Tables \ref{abu_1342}$-$\ref{abu_2447} in the column headed `average'. 

The sensitivity of the derived abundances to the variations in adopted atmospheric parameters were estimated following the procedure described in Paper {\scs II}. The changes in abundances caused by varying atmospheric parameters by 100 K, 0.25 cm s$^{-2}$ and 0.2 km s$^{-1}$ with respect to the chosen model atmosphere are summarized in Table \ref{sensitivity}. The quadratic sum of all the three contributors are represented by $\sigma_{2}$. The total error $\sigma_{tot}$ for each of the element is the quadratic sum of $\sigma_{1}$ and $\sigma_{2}$. The final OC mean abundances along with the $\sigma_{tot}$ from this study are presented in Table \ref{mean_abundance}.  

\section{Comparison with literature} \label{comp_lit}

We merged our sample of sixteen OCs with the available high-quality results in the literature to enlarge the dataset. As the abundances are collected from the literature, the data are liable to systematic errors concerning the atomic data (not only the quality of $gf$-values but also the choice of specific lines), model atmosphere grids employed, reference solar abundances adopted, abundance analysis programs and type of stars considered by different authors in their analysis. 

\begin{table}
\centering
\caption{Comparison of our elemental abundance ratios [X/Fe] for OC NGC 2447 with those from Hamdani et al. (2000).}
\label{compare_2447}
\begin{tabular}{lcc}   \hline
\multicolumn{1}{c}{Species}  & \multicolumn{1}{c}{This work} & \multicolumn{1}{c}{Hamdani}  \\
\hline

$[$Na {\scs I}/Fe$]$  &$+0.12\pm0.02$ & $+0.18\pm0.03$ \\
$[$Mg {\scs I}/Fe$]$  &$-0.02\pm0.01$ & $+0.02\pm0.04$ \\
$[$Si {\scs I}/Fe$]$  &$+0.11\pm0.02$ & $ 0.00\pm0.04$ \\
$[$Ca {\scs I}/Fe$]$  &$+0.02\pm0.03$ & $+0.02\pm0.00$ \\
$[$Sc {\scs II}/Fe$]$ &$+0.10\pm0.01$ & $-0.02\pm0.03$ \\
$[$Ti {\scs I}/Fe$]$  &$-0.04\pm0.03$ & $+0.09\pm0.03$ \\
$[$V {\scs I}/Fe$]$   &$-0.02\pm0.02$ & $+0.08\pm0.02$ \\
$[$Cr {\scs I}/Fe$]$  &$-0.04\pm0.02$ & $+0.07\pm0.01$ \\ 
$[$Cr {\scs II}/Fe$]$ &$+0.02\pm0.02$ & $+0.11\pm0.03$ \\
$[$Fe {\scs I}/H$]$   &$-0.13\pm0.02$ & $+0.03\pm0.03$ \\
$[$Co {\scs I}/Fe$]$  &$-0.04\pm0.03$ & $+0.12\pm0.03$ \\ 
$[$Ni {\scs I}/Fe$]$  &$-0.07\pm0.02$ & $-0.11\pm0.02$ \\ 
$[$Y {\scs II}/Fe$]$  &$+0.03\pm0.02$ & $-0.05\pm0.02$ \\
$[$Ce {\scs II}/Fe$]$ &$+0.32\pm0.01$ & $+0.06\pm0.02$ \\  
$[$Eu {\scs II}/Fe$]$ &$\bf+0.22 $ & $\bf+0.01$ \\

\hline
\end{tabular} \vspace{-0.5cm}
\end{table}

Many researchers have noticed a systematic offset when combining abundance results of OCs from different authors to obtain a complete picture of chemical evolution of the Galaxy. Among such investigators are Friel et al. (2010), Yong et al. (2012), and more recently the robust analysis by Heiter et al. (2014). In Table 13 of our Paper {\scs II}, we offered a comparison of our abundance estimates with literature studies for the OC NGC 2682, and spotted a systematic offset of $\pm$0.15 dex or smaller in [X/Fe] ratios for almost all elements.  
Our present sample also includes a OC NGC 2447 for which high-resolution spectroscopic abundance analyses have been reported by Hamdani et al. (2000) for all stars in common with our sample. Using their Table 4, we calculated the mean abundances for all the elements and those values are presented in Table \ref{compare_2447} in comparison with our results. While calculating their mean abundances we have not included the measurement uncertainties of the order of $0.15$ dex or even larger for a few elements. We notice from the Table \ref{compare_2447} that the mean difference in [X/Fe] between ours and Hamdani et al. (2000) analyses are $\pm$0.15 dex or smaller for almost all elements in common: exceptions include Fe, Co, Ni, Ce and Eu. Therefore, we emphasize that a simple merger of all cluster abundances from various resources can easily wash out or mask subtle abundance trends unless it is done with extreme care\footnote{Note, however that Mishenina et al. (2015) found a good agreement of their results with ours for one star in OC NGC 2506 in common with ours (see Table 13 in their paper), while for other OCs they too noticed a systematic offset in the results obtained by different authors.}

To cancel systematic errors we have followed a different procedure. The only observed quantity that has to be extracted from spectra for abundance estimates is the EW of a spectral line. On the assumption that published EWs have been measured reliably, we reestimated the cluster abundances from the EWs using our models, linelists and reference solar abundances. This approach should minimize, or even cancel out, all sorts of systematics mentioned earlier and place results on a common abundance scale. A wide wavelength coverage of echelle data employed by various authors made it easy to pick up common lines for Na, Al, $\alpha$- elements, Cr {\scs I}, Fe {\scs I}, Fe {\scs II} and Ni {\scs I}, but note that for s- and r- process elements measurements are often lacking in the literature, thereby making it impossible to homogenize the data for heavy elements, but still we can measure abundances for [Y/Fe] for 13 OCs, [Zr/Fe] for 8 OCs and [Ce/Fe] for 5 OCs in the literature sample. The abundances derived for literature sample of OCs are shown in Tables \ref{abu_lit} - \ref{lit_heavy}.  

This process provides a list of 69 OCs drawn from our studies and from the literature, covering a range of $\sim$ $-$0.5 to 0.3 dex in metallicity and an age of few Myr to 9 Gyr. All these OCs, except the OC data from Villanova et al. (2010) and Jacobson et al. (2007 \& 2011), have a high-quality (S/N $>$ 50) and high-dispersion echelle spectra ($R$ $>$ 25,000) observed with various telescopes and spectrographs. The OC data from Villanova et al. (2010), Jacobson et al. (2007) have high quality spectra (S/N $>$ 50) observed at $R$ $\sim$ 17,000 and 15,000 while the OC data from Jacobson et al. (2011) was observed at two different resolutions $R$ $\sim$ 18,000 and 21,000. Here, we emphasize that our sample contribute about 23\% of the total OCs explored so far with high quality and high-dispersion echelle spectroscopy. Moreover, we have done a homogeneous and comprehensive abundance analysis extending up to s- and r-process elements. Before we use this homogeneous sample to look at trends with age and metallicity in the disk, it is worth while to know the cluster's kinematics in order to assign them to the thin or the thick disk. 

\subsection{Assignment of OCs to Galactic populations}  \label{mem_prob}

The separation between thin disk, thick disk and halo populations is made by invoking the kinematic criteria used in studies of local field stars by Bensby et al. (2005, 2014), Mishenina et al. (2004), and Reddy et al. (2006). The method of assigning OCs either to the thin disk, the thick disk or the halo relies on the assumption that an observed sample is a mixture of three stellar populations with their respective Galactic space velocity U$_{LSR}$, V$_{LSR}$, W$_{LSR}$ components represented by Gaussian distributions, with given mean values and dispersions $\sigma_{U}$, $\sigma_{V}$, $\sigma_{W}$. The remaining constraints are the relative densities of thin, thick and halo stars in the solar vicinity. The values for the velocity dispersions, the asymmetric drift and population fractions are same as those given in Table 1 of Reddy et al. (2006).

We compute the membership probabilities P$_{thin}$, P$_{thick}$, P$_{halo}$ and the associated uncertainties using the equations and recipes described in Reddy et al.(2006). Our program successfully reproduces the probabilities given in Reddy et al. (2006). We associate the clusters to thin disk population as those that have probabilities greater than 75\% and at least twice the probability of belonging to the thin disk over other populations (and likewise for other populations). We assume that placing a lower limit of 75\% in probability ensures a minimum contamination of each subsample from the rest of the stellar populations. The clusters that do not satisfy these criteria but equally probable to belong to either thin and thick disks or thick disk and halo are assigned as respective transition objects. On account of these  probabilities, NGC 2266 (one out of sixteen OCs studied by us) turned out to be a thick disk member and the remaining OCs in our sample display probabilities typical of the thin disk, not surprisingly, none belongs to the halo. Table \ref{membership} lists the computed space and GSR velocity components of OCs along with their membership probabilities. 

\begin{table*} 
{\fontsize{5}{6}\selectfont
\caption{Mean elemental abundance ratios, [X/Fe], for Na to Eu for our and literature sample of OCs, the thin disk mean abundances from Luck \& Heiter (2007) for giants, and Luck \& Heiter (2005 \& 2006) and Bensby et al. (2014) for dwarfs. Abundances calculated by synthesis are presented in bold typeface.} 
\label{ocmean_intra}
\begin{tabular}{lc|cc|cc|cc|cc}   \hline
\multicolumn{1}{l}{Species} & \multicolumn{1}{c}{Our sample} & \multicolumn{2}{c}{Literature} & \multicolumn{2}{c}{Luck \& Heiter (2007)} & \multicolumn{2}{c}{Luck \& Heiter (2005 \& 2006) } & \multicolumn{2}{c}{Bensby et al. (2014)} \\ 
\multicolumn{1}{l}{ } & \multicolumn{3}{c}{thin disk OCs} & \multicolumn{2}{c}{thin disk giants} & \multicolumn{4}{c}{thin disk dwarfs} \\ \cline{2-4}\cline{5-6}\cline{7-10} 
\multicolumn{1}{l}{$[$Fe/H$]$ $\sim$} & \multicolumn{1}{c}{($-$0.20 to 0.0)} & \multicolumn{1}{c}{(-0.20 to 0.0)} & \multicolumn{1}{c}{(0.0 to 0.20)} & \multicolumn{1}{c}{(-0.20 to 0.0)} & \multicolumn{1}{c}{(0.0 to 0.20)} & \multicolumn{1}{c}{(-0.20 to 0.0)} & \multicolumn{1}{c}{(0.0 to 0.20)} & \multicolumn{1}{c}{(-0.20 to 0.0)} & \multicolumn{1}{c}{(0.0 to 0.20)} \\ 
\hline

$[$Na /Fe$]$  &$+0.23\pm0.07$ & +0.18$\pm$0.11 & +0.16$\pm$0.13 & $+0.10\pm0.06$ & +0.13$\pm$0.06 & +0.02$\pm$0.08 & +0.04$\pm$0.11 & ~0.00$\pm$0.05 & +0.02$\pm$0.06  \\
$[$Mg /Fe$]$  &$+0.04\pm0.08$ & +0.06$\pm$0.10 & +0.03$\pm$0.09 & $+0.08\pm0.10$ & +0.08$\pm$0.10 & +0.16$\pm$0.09 & +0.13$\pm$0.13 & +0.07$\pm$0.11 &$-$0.06$\pm$0.10  \\
$[$Al /Fe$]$  &$+0.03\pm0.09$ & +0.14$\pm$0.06 & +0.19$\pm$0.10 & $+0.09\pm0.05$ & +0.08$\pm$0.05 & +0.04$\pm$0.10 & +0.02$\pm$0.09 & +0.05$\pm$0.09 & +0.03$\pm$0.05  \\
$[$Si /Fe$]$  &$+0.15\pm0.06$ & +0.11$\pm$0.08 & +0.10$\pm$0.10 & $+0.12\pm0.04$ & +0.12$\pm$0.05 & +0.05$\pm$0.06 & +0.02$\pm$0.06 & +0.03$\pm$0.05 & +0.02$\pm$0.03  \\
$[$Ca /Fe$]$  &$+0.07\pm0.06$ & +0.08$\pm$0.09 & +0.04$\pm$0.07 & $-0.04\pm0.05$ &$-$0.07$\pm$0.06 & +0.03$\pm$0.04 & +0.01$\pm$0.07 & +0.05$\pm$0.04 & +0.01$\pm$0.03  \\
$[$Sc /Fe$]$  &$+0.07\pm0.07$ & $\ldots$      &   $\ldots$    & $-0.08\pm0.06$ &$-$0.14$\pm$0.06 & +0.02$\pm$0.11 & ~0.00$\pm$0.09 &  $\ldots$     &  $\ldots$      \\
$[$Ti /Fe$]$  &$+0.02\pm0.07$ & +0.04$\pm$0.06 & +0.01$\pm$0.08 & ~0.00$\pm$0.03 & +0.07$\pm$0.04 & +0.05$\pm$0.07 & +0.05$\pm$0.09 & +0.04$\pm$0.06 & +0.01$\pm$0.04  \\
$[$V /Fe$]$   &$+0.05\pm0.07$ & $\ldots$       &  $\ldots$  & $-0.09\pm0.07$ &$-$0.04$\pm$0.07 & +0.02$\pm$0.09 & ~0.00$\pm$0.10 &   $\ldots$    &   $\ldots$     \\
$[$Cr /Fe$]$  &$+0.02\pm0.05$ & +0.09$\pm$0.11  & +0.03$\pm$0.07 & $+0.01\pm0.05$ & +0.03$\pm$0.03 & +0.03$\pm$0.05 & +0.03$\pm$0.06 & ~0.00$\pm$0.04 & ~0.00$\pm$0.02  \\ 
$[$Mn /Fe$]$  &$\bf-0.10\pm0.07$ & $\ldots$    &   $\ldots$    & $+0.06\pm0.07$ & +0.17$\pm$0.11 &$-$0.05$\pm$0.12 & ~0.00$\pm$0.11 &   $\ldots$ &   $\ldots$     \\
$[$Fe /H$]$ &$\bf-0.12\pm0.05$ &$-0.08\pm0.05$ & +0.10$\pm$0.07 & $-0.08\pm0.05$ & +0.09$\pm$0.05 &$-$0.08$\pm$0.06 & +0.08$\pm$0.05 &$-$0.09$\pm$0.06 & +0.10$\pm$0.05  \\
$[$Co /Fe$]$  &$+0.04\pm0.07$  & $\ldots$      &   $\ldots$    & $+0.06\pm0.08$ & +0.10$\pm$0.09 & +0.03$\pm$0.10 & +0.04$\pm$0.07 &   $\ldots$    &   $\ldots$     \\ 
$[$Ni /Fe$]$  &$ 0.00\pm0.05$ & $-0.01\pm0.05$ & ~0.00$\pm$0.06 & ~0.00$\pm$0.03 & +0.02$\pm$0.04 & +0.00$\pm$0.04 & +0.00$\pm$0.03 &$-$0.03$\pm$0.03 & ~0.00$\pm$0.04  \\ 
$[$Cu /Fe$]$  &$\bf-0.19\pm0.07$ & $\ldots$    &   $\ldots$    & $+0.01\pm0.13$ & +0.03$\pm$0.12 &$-$0.12$\pm$0.19 &$-$0.06$\pm$0.17 &   $\ldots$    &   $\ldots$  \\
$[$Y /Fe$]$  &$+0.09\pm0.05$ & $\ldots$    &   $\ldots$    & $+0.07\pm0.15$ & +0.03$\pm$0.08 &$-$0.07$\pm$0.14 &$-$0.06$\pm$0.16 &$-$0.01$\pm$0.15 &$-$0.02$\pm$0.08  \\
$[$Ba /Fe$]$ & $\bf+0.21\pm0.20$ & $\ldots$   &  $\ldots$    & $+0.04\pm0.16$ & $-0.10\pm0.17$ & +0.02$\pm$0.12 &$-$0.03$\pm$0.10 & +0.07$\pm$0.18 & +0.01$\pm$0.11 \\
$[$La /Fe$]$ &$+0.17\pm0.08$ & $\ldots$       &   $\ldots$    & $-0.12\pm0.11$ & $-0.10\pm0.11$ & +0.13$\pm$0.09 & +0.00$\pm$0.08 &   $\ldots$    &   $\ldots$     \\
$[$Ce /Fe$]$ &$\bf+0.22\pm0.11$ & $\ldots$    &   $\ldots$   & $+0.05\pm0.09$ & $-0.02\pm0.12$ & +0.09$\pm$0.12 & +0.04$\pm$0.13 &   $\ldots$    &   $\ldots$  \\  
$[$Nd /Fe$]$ &$+0.18\pm0.10$  & $\ldots$     &   $\ldots$    & $-0.01\pm0.07$ &$-0.05\pm0.06$  & +0.02$\pm$0.11 & +0.03$\pm$0.17  &   $\ldots$    &   $\ldots$     \\
$[$Sm /Fe$]$ &$+0.16\pm0.09$ & $\ldots$      &   $\ldots$    & $\ldots$  &   $\ldots$         &  $-$0.10$\pm$0.18 & $-$0.13$\pm$0.22 &   $\ldots$    &   $\ldots$    \\
$[$Eu /Fe$]$ &$\bf+0.13\pm0.07$ & $\ldots$   &   $\ldots$    & $+0.08\pm0.06$ & +0.04$\pm$0.08   & +0.15$\pm$0.10  & +0.12$\pm$0.08 &   $\ldots$    &   $\ldots$     \\

\hline
\end{tabular} } 
\flushleft
Note: Although, we measured abundance ratios, [Y/Fe] (13 OCs), [Zr/Fe] (8 OCs) and [Ce/Fe] (5 OCs), for literature sample of OCs they were not tabulated because of the scarcity of large sample OCs with measured heavy elements.
\vspace{-0.4cm}
\end{table*}

\section{The cluster and field star populations}

OCs form from dense molecular clouds. Stars in an OC have the composition of the natal cloud. It is assumed that stars within a cluster are of a common age and composition and that gas remaining with the cluster at birth is lost before it can be contaminated by the ejecta from the cluster's stars. Such ejecta are presumed to be lost to the cluster and not to contaminate the remaining stars. Stars are slowly lost from the cluster and join the field star population. Small and poorly populated OCs dissolve primarily from internal dynamical effects in less than 10$^{8}$ Myr. Clusters with intermediate masses ($\sim$ 500 to 1000 $M_\odot$) can survive several Gyr, if they are located in the external regions of the disk (Pavani \& Bica 2007). One expects a close, if not exact, correspondence between the compositions of OCs and field stars. This expectation is searched for first through a global comparison of the runs of [X/Fe] with [Fe/H]  and age with [Fe/H] for field and cluster stars. Then, in the next section, intracluster abundance variations, the basis of chemical tagging (Freeman \& Bland-Hawthorn 2002), are discussed. Our primary sources for data on field stars are Luck \& Heiter's (2007) survey of local disk giants and Luck \& Heiter (2005 \& 2006) and Bensby et al. (2014) for nearby FG dwarfs. 

Intrinsic variations in [X/Fe], as distinct from those arising from systematic errors in abundance analyses, are in the main relatable to the principal sites of stellar nucleosynthesis which left their imprint on a cluster's natal cloud. For the elements considered here, the principal sites include Type II  supernovae from massive stars, Type Ia supernovae from white dwarfs which exceed the Chandrasekhar mass limit,  AGB stars and possibly a merging pair of neutron stars. The composition of stars in a cluster will be determined by the cumulative effect on the natal cloud of these principal sites within prior generations of stars. For a cluster, the principal sites will seed other clouds and, thus, a future generation of clusters. In contrast to the case of globular clusters, it seems improbable that nucleosynthetically-enriched ejecta from cluster members will determine the composition of the cluster except in rare cases, e.g., the creation of Barium stars in a binary star with an $s$-process enriched AGB star donating mass to a less-evolved companion.  

\subsection{Relative abundances and stellar ages}

A global representation of cluster and field star compositions is provided by Table \ref{ocmean_intra}. Fifteen of our 16 clusters have [Fe/H] from $-0.2$ to $0$ for a mean [Fe/H] of $-0.12$ with the remaining cluster having [Fe/H] $= -0.45$. With the exception of the latter, the clusters have been assigned to the thin disk. About half of the additional OCs from the literature have [Fe/H] $\geq 0.0$ and all of these belong to the thin disk. Table \ref{ocmean_intra} provides [X/Fe] for the following nine samples: our 15 thin disk clusters with a mean [Fe/H] = -0.1, clusters from the literature in the intervals $-0.20$ to $0.0$ (15) and 0.0 to $+0.20$ (20), thin disk giants from Luck \& Heiter (2007) in the intervals $-0.20$ to 0.0 and 0.0 to $+0.20$, and thin disk dwarfs from both Luck \& Heiter (2005 \& 2006) and Bensby et al. (2014) for the intervals $-0.20$ to 0.0 and 0.0 to $+0.20$. Table entries come from similar but not identical analyses of spectra of comparable quality and, thus, small differences in [X/Fe] for the same element X are to be expected. Perhaps, the largest systematic differences are present between dwarfs and giants where the model atmospheres have major differences in T$_{\rm eff}$ and $log~{g}$ and, perhaps, systematic departures from the true atmospheres and certainly differences are likely in corrections for non-LTE effects and stellar granulation (i.e., 3D effects). In addition, the convective envelope of a giant brings nuclear processed material into the atmosphere but, except for Na, our analyses do not discuss elements expected to be affected such as Li, C, N, and O.

Inspection of Table \ref{ocmean_intra} shows that the [X/Fe] around the solar abundance [Fe/H] = 0 are similar in their mean values and $\sigma$ for field and OC stars for almost all elements X. For the purposes of discussion, elements in Table \ref{ocmean_intra} are grouped together beginning with Na and Al and ending with the heavy elements Y to Eu among which are enriched in some OCs. The remarks below refer to thin and thick disk clusters and stars.

\subsection{Na and Al}

\begin{figure} 
\begin{center}
\includegraphics[trim=0.5cm 3.5cm 10.2cm 8.0cm, clip=true,height=0.29\textheight]{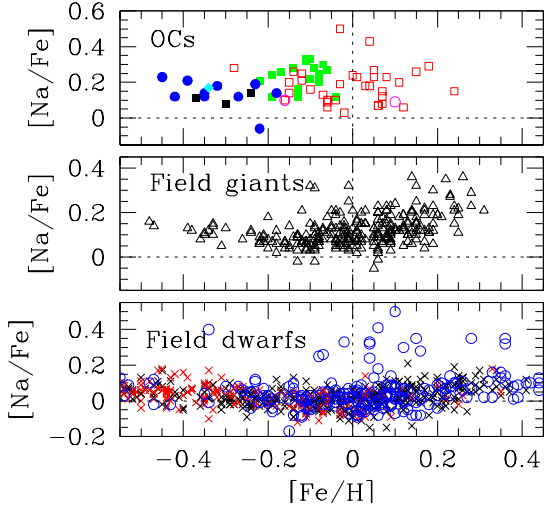}
\caption[]{The relative abundance ratios of [Na/Fe] vs. [Fe/H] for OCs (top panel), field giants (middle panel) and field dwarfs (bottom panel).
Our sample of OCs are shown as green filled squares. Clusters from the literature are presented as red open squares (thin disk), blue filled circles (thick disk) and black filled squares (halo). Intermediate stellar populations are designated as magenta open circles (thin$-$thick disk) and cyan filled diamond (thick disk$-$halo). Luck \& Heiter's (2007) sample of field giants are marked as black open triangles. Bensby et al. (2014) sample of field dwarfs are shown as black crosses (thin disk) and red crosses (thick disk). Luck \& Heiter (2005 \& 2006) sample of field dwarfs are presented as blue open circles.} 
\label{na_fe} 
\end{center} \vspace{-0.5cm}
\end{figure}

Entries for Na and Al in Table \ref{ocmean_intra} are expanded in Figure \ref{na_fe}. For Na, Figure \ref{na_fe} shows, as indicated in Table \ref{ocmean_intra}, that [Na/Fe] for our OCs and to a smaller degree for the literature sample are slightly larger than provided by Luck \& Heiter (2007) for field giants. Both the OC and field giant samples appear to yield slightly more positive [Na/Fe] than the samples of dwarfs from Bensby et al. (2014) and Luck \& Heiter (2005 \& 2006) dwarfs is not confirmed by Bensby et al. (2014). 

That giants and dwarfs give slightly different [Na/Fe] values may be due to larger non-LTE effects for giants augmented by mild Na enrichment by the first dredge-up (see also Smiljanic 2012). Non-LTE effects for the dwarfs are negligible -- see Bensby et al. (2014, their Figure 7). For typical red giants, Alexeeva et al. (2014) found non-LTE Na overabundances from our lines to be lowered from their LTE values by (say) about 0.1 dex, a correction sufficient to render Na in the OC giants very similar to the values seen in the dwarfs. Non-LTE effects for Fe {\scs I} and Fe {\scs II} lines have been estimated by Lind et al. (2012) for giants. At the metallicity of our cluster's giants (0.0 to $-$0.2 dex), the non-LTE corrections for both the Fe {\scs I} and Fe {\scs II} lines are smaller than 0.02 dex. Therefore, these corrections to our LTE Fe abundance do not affect the global behaviour of [Na/Fe] (or any [X/Fe]) with [Fe/H]. Note that rise in [Na/Fe] with [Fe/H] for [Fe/H] $>$ 0.2 which occurs for both dwarfs and giants may have been reported first by Feltzing \& Gustafsson (1998). The first dredge-up increases the [Na/Fe] by a mere 0.01 dex at 1$M_\odot$, and 0.1 dex at 2$M_\odot$, to approximately 0.2 dex at 3$M_\odot$ for [Fe/H] $\sim 0.0$ (Karakas \& Lattanzio 2014).  

\begin{figure} 
\begin{center}
\includegraphics[trim=0.5cm 3.5cm 10.2cm 8.0cm, clip=true,height=0.29\textheight]{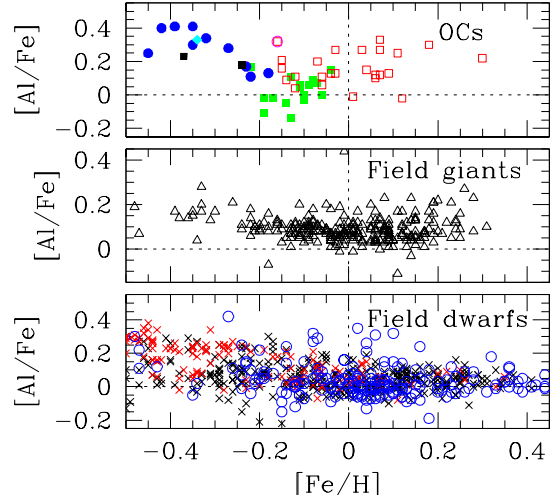}
\caption[]{Same as Figure \ref{na_fe} but for [Al/Fe].} 
\label{al_fe} 
\end{center} \vspace{-0.5cm}
\end{figure}

In the case of Al, the samples provide a similar mean [Al/Fe] with the exception that the clusters drawn from the literature may have a higher mean [Al/Fe] (Figure \ref{al_fe}). This is not regarded as a real effect because the scatter across the sample is large.

For Na and Al, our conclusion is that [Na/Fe] and [Al/Fe] for the thin and thick disk OCs are in fair agreement with respect to mean values and scatter across the sampled [Fe/H] range with their values for field dwarfs and giants.

\subsection{The $\alpha$-elements}

\begin{figure} 
\begin{center}
\includegraphics[trim=0.5cm 3.5cm 10.2cm 8.0cm, clip=true,height=0.29\textheight]{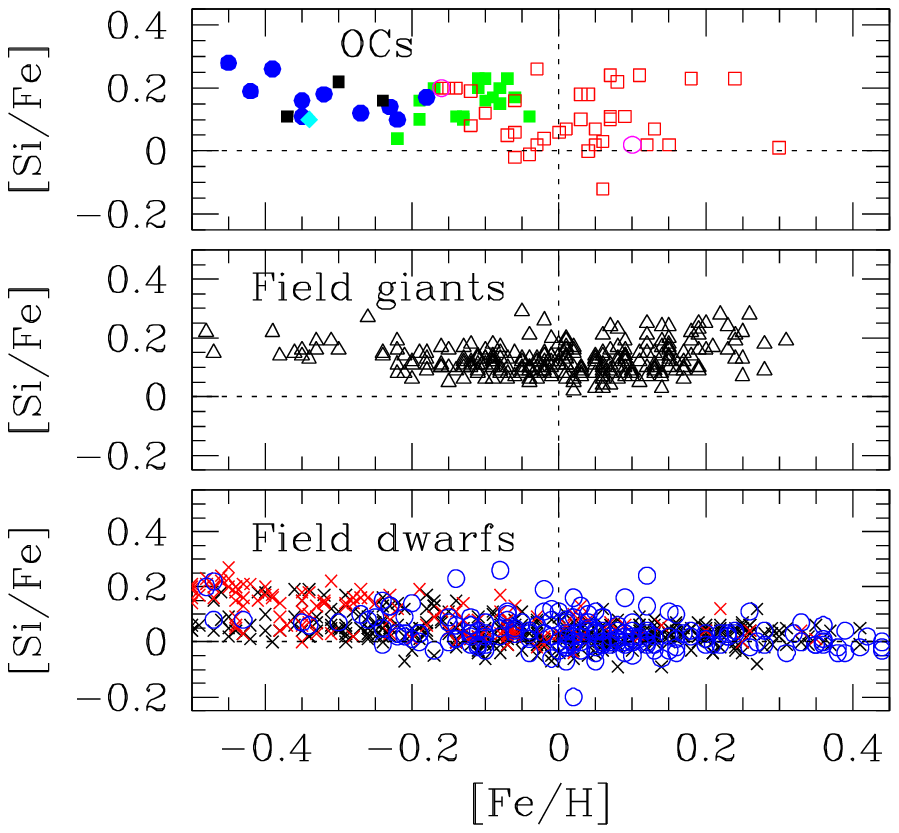}
\caption[]{Same as Figure \ref{na_fe} but for the alpha element Si.}
\label{si_fe} 
\end{center} \vspace{-0.5cm}
\end{figure}

\begin{figure} 
\begin{center}
\includegraphics[trim=0.5cm 3.5cm 10.2cm 8.0cm, clip=true,height=0.29\textheight]{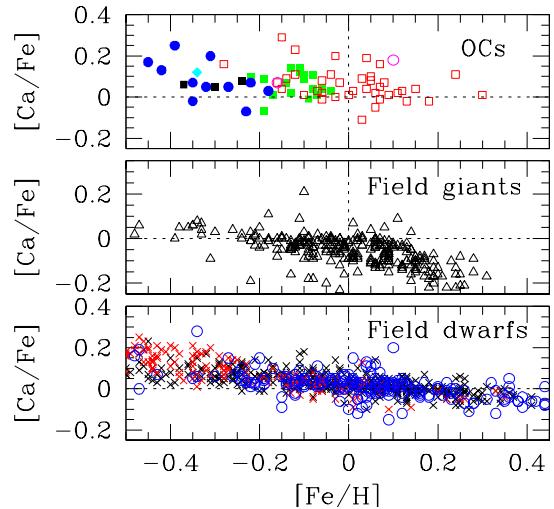}
\caption[]{Same as Figure \ref{na_fe} but for the alpha element Ca.}
\label{ca_fe} 
\end{center} \vspace{-0.7cm}
\end{figure}

Here, the $\alpha$-elements comprise Mg, Si, Ca and Ti. Inspection of Table \ref{ocmean_intra} shows that the samples of cluster and field giants and dwarfs generally share a common [X/Fe]. A graphical comparison is provided in Figures \ref{si_fe} and \ref{ca_fe} for two $\alpha$-elements well represented by lines. For the giants in the field and clusters, [Si/Fe] appears to be about 0.1 dex greater than for the dwarfs. For [Ca/Fe], the cluster giants and field dwarfs share a very similar mean value and the sole apparent exception is that Luck \& Heiter (2007)'s giants have a slightly lower value. It is of interest to note that the gradients of [Ca/Fe] with [Fe/H] provided by Luck \& Heiter (2007)'s giants and Bensby et al.'s dwarfs are similar: a slight increase of [Ca/Fe] with decreasing [Fe/H]. [X/Fe] for $\alpha$-elements in thin and thick disk OCs closely match their values for thin and thick disk field dwarfs and giants. There is a slight indication that the spread in the OCs [X/Fe] around [Fe/H] $\sim 0.0$ is larger than for field giants.

A similar behaviour of $\alpha$-elements Si and Ca is anticipated because both (and Mg) are synthesized primarily in Type II supernovae. Although the predicted yield of Ti from these supernovae is too small, Ti is very often considered an $\alpha$-element because the observed run of [Ti/Fe] with [Fe/H] in local thin and thick disc stars resembles those of Mg, Si and Ca.

\subsection{The iron group} 

\begin{figure}
\begin{center}
\includegraphics[trim=0.6cm 5.8cm 3.0cm 8.0cm, clip=true,width=0.60\textheight,height=0.25\textheight]{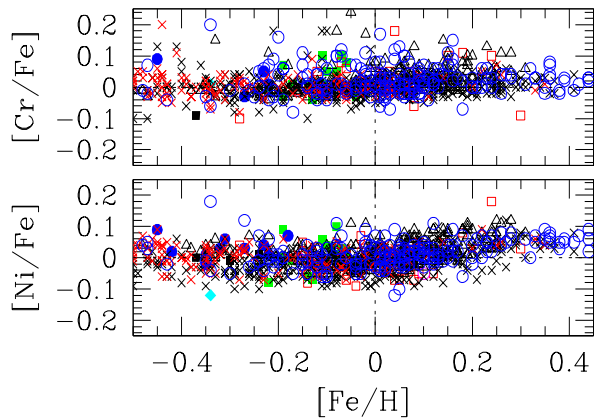} 
\caption[]{The relative abundance ratios [X/Fe] vs. [Fe/H] for the iron peak elements Cr and Ni. All the symbols have their usual meaning as in Figure \ref{na_fe} except that the OCs, field giants \& field dwarfs are all plotted in a single panel.}
\label{iron_fe} 
\end{center} \vspace{-0.5cm}
\end{figure}

Within the run from Sc to Cu, Table \ref{ocmean_intra} shows almost complete agreement for [X/Fe] from all samples. Figure \ref{iron_fe} shows a typical result. Minor exceptions are noted for Mn and Cu where our OCs appear to have lower [X/Fe] than reported by Luck \& Heiter for field giants and perhaps too for field dwarfs. A major part and even all of these exceptions may be attributable to neglect of hyperfine and isotopic splitting affecting Mn {\scs I} and Cu {\scs I} lines (R. E. Luck, private communication).

\begin{figure}
\begin{center}
\includegraphics[trim=0.6cm 3.5cm 10.5cm 8.0cm, clip=true,height=0.28\textheight]{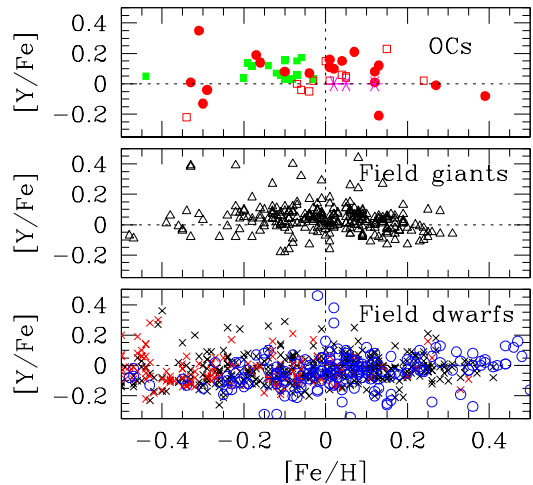}
\caption[]{The relative abundance ratios [X/Fe] versus [Fe/H] for the $s$-process element Y. All the symbols have their usual meaning as in Figure \ref{na_fe}, while the red filled dots are for abundances of OCs from Maiorca et al. (2011) and the magenta asterisks are abundances of nearby young associations from D'Orazi et al. (2012). }
\label{y_fe} 
\end{center}  \vspace{-0.5cm}
\end{figure}

\subsection{Heavy elements} \label{he}

Even casual perusal of Table \ref{ocmean_intra} shows that differences between [X/Fe] for the OCs and for field stars occur for the heavy elements. In particular, Ba, La, Ce, Nd, Sm and possibly Eu give mean [X/Fe] values that are distinctly positive with $\sigma$s which are greater than the corresponding $\sigma$s for individual clusters (see, for example, Table \ref{mean_abundance}). This behaviour for Ba-Eu does not seem to apply to the lighter elements Y and Zr. This exceptional pattern for heavy elements has been noticed previously, see, for example, De Silva et al.'s (2009, Table 2) compilation of [X/Fe] and $\sigma$ for 24 clusters of which only one (M 67) is included our sample featured in Table \ref{ocmean_intra}. (Heavy element abundances were available for a minority of the 24 clusters.) An obvious inference is that OCs clearly differ in [X/Fe] for heavy elements; across our sample, NGC 2354 (Table \ref{abu_2354}) is among the most enriched in Ba-Sm and NGC 2682 the least enriched with the range of about 0.35 dex far exceeding the star-to-star $\sigma$ of 0.05 dex or less. Moreover, stars like those in NGC 2682 seem not to have their counterparts in the samples of field dwarfs and giants selected for Table \ref{ocmean_intra}. These field-cluster differences have to be related to the neutron-capture processes which synthesize these heavy elements.  

\begin{figure}
\begin{center}
\includegraphics[trim=0.5cm 1.0cm 10.2cm 8.5cm, clip=true,width=0.35\textheight,height=0.38\textheight]{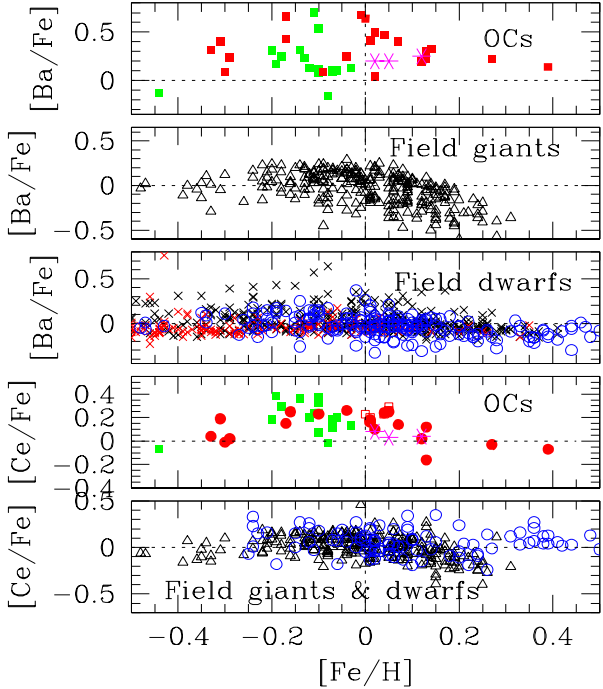}
\caption[]{The relative abundance ratios [X/Fe] versus [Fe/H] for the $s$-process elements Ba and Ce. All the symbols have their usual meaning as in  Figure \ref{y_fe}, while the red filled squares are the Ba abundances of OCs from D'Orazi et al. (2009).}
\label{bace_fe} 
\end{center} \vspace{-0.5cm}
\end{figure}

Heavy elements from Y to Eu sample the neutron-capture processes of nucleosynthesis. The weak $s$-process occuring in massive stars contributes greatly to synthesis of Y and Zr. The $r$-process in Type II SN from massive stars and/or in merging neutron stars is the major contributor to Eu and a serious contributor to La to Sm. The main $s$-process from AGB stars is the dominant contributor to Ba and an appreciable contributor to La to Sm. The $s$-process contribution to solar abundances is 85\% for Ba, 75\% for La, 81\% for Ce, 45\% for Nd and 3\% for Eu (Burris et al. 2000).

Examination of Table \ref{ocmean_intra} suggests that Y has similar [X/Fe] in cluster and field stars (see Figures \ref{y_fe}). Although Zr has yet to be measured in field dwarfs and giants, it would appear that Zr in OCs is consistent with the claim that Y in OCs is unchanged from its abundance in field dwarfs and giants, i.e., [Zr/Y] $\simeq 0.0$. The larger scatter in [Y/Fe] from Luck \& Heiter's sample of dwarfs is not confirmed by Bensby et al. (2014).

\begin{figure}
\begin{center}
\includegraphics[trim=0.5cm 3.5cm 10.2cm 8.5cm, clip=true,height=0.28\textheight]{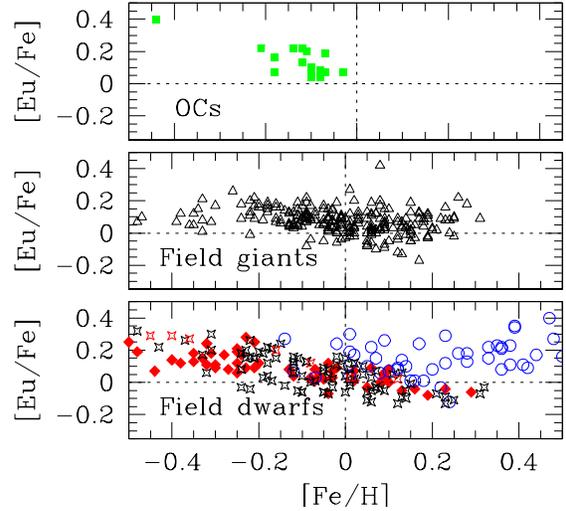}
\caption[]{Same as Figure \ref{y_fe} but for the $r$-process element Eu, while the red filled diamonds are for field dwarfs drawn from Koch \& Edvardsson (2002), and the red open stars (thick disk) and black open stars (thin disk) are for field dwarfs from Gorbaneva et al. (2012). }
\label{eu_fe} 
\end{center}  \vspace{-0.5cm}
\end{figure}

In contrast to Y (and Zr), the mean [X/Fe] for Ba-Sm  for the OCs are greater than in field stars (Table \ref{ocmean_intra} and Figure \ref{bace_fe}) and, as we show in Section 7, there is a clear variation in [X/Fe] among the OCs even for the same [Fe/H]. Although this variation is smaller for Eu (Figure \ref{eu_fe}). For Luck \& Heiter's sample of field dwarf stars, there may be an offset in [Ce/Fe] and [Eu/Fe]. 

\section{Intracluster abundance variations}

The proposal that Galactic field stars began life in open clusters implies a close match between the compositions of field and cluster stars. In the previous section, we showed that the [X/Fe] for field and cluster stars over the [Fe/H] range sampled by the OCs were similar with regards to mean values of [X/Fe] and their $\sigma$ except for heavy elements from Ba to Eu; [X/Fe] for the latter elements span a range exceeding that expected purely from measurement uncertainties.  For our OC sample, it is apparent too that the $\sigma$ for [X/Fe] for each element is noticeably larger than the $\sigma$ derived from the several stars in a given cluster with the latter $\sigma$ being comparable to that expected from the measurement uncertainties (see Table \ref{sensitivity}). These two features suggest that there are measurable differences in some [X/Fe] values amongst clusters.

Composition differences between clusters offer the opportunity to exploit a technique known as `chemical tagging' (Freeman \& Bland-Hawthorn 2002). On the assumption that all stars in a given cluster have the same composition, the composition is a tag or label identifying cluster members even after the cluster has dissolved. The efficacy of chemical tagging depends - obviously - on there being measurable differences in composition from cluster to cluster. 

\begin{figure}
\begin{center}
\includegraphics[trim=0.5cm 6.0cm 11.0cm 8.5cm, clip=true,height=0.18\textheight]{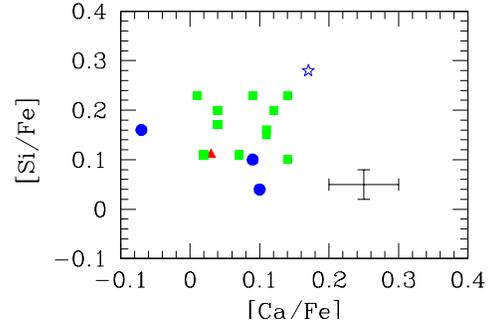}
\caption[ ]{The relative abundance ratios [Si/Fe] vs. [Ca/Fe] for our sample of sixteen OCs. The clusters with mean [Fe/H] $= 0.00\pm0.05$ (one), $= -0.10\pm0.05$ (eleven) and $= -0.20\pm0.05$ (three) are shown as a red filled triangle, green filled squares and blue filled circles respectively. The most metal-poor ([Fe/H] $= -0.45\pm0.04$) cluster NGC 2266 is designated as a blue open star.}
\label{ca_si} 
\end{center} \vspace{-0.5cm}
\end{figure}

Our sample of sixteen clusters offers an excellent opportunity to assess the role which chemical tagging might play in Galactic forensics (i.e., in searching for solar twins and siblings, tying stars in moving groups, superclusters and stellar streams to potential parent open clusters). Stars in our sample are similar red giants analysed identically and, thus, systematic errors affecting the abundances and abundance ratios [X/Fe] should be consistent (and small) across the sample which spans a small range in metallicity [Fe/H]. 

Composition differences which might be the basis for chemical tagging are most likely to be found among elements whose synthesis samples different nucleosynthetic processes. Differences should be minimized for elements sampling a single or very similar nucleosynthetic processes.  Our assessment begins with elements provided by Type II and Type Ia supernovae, i.e., Mg to Ni. A representative illustration from this set of elements is in Figure \ref{ca_si} where [Si/Fe] versus [Ca/Fe] is plotted for our sample. Silicon is primarily a product of Type II supernovae and Fe at the typical [Fe/H] ($\simeq -0.1$) of our sample is primarily a product of Type Ia supernovae and Ca is a sample of both Type II and Ia supernovae. For the subset of our OCs with [Fe/H] within $-0.1\pm0.05$, both [Si/Fe] and [Ca/Fe] cover a range of 0.13 dex when the $\sigma$ for each cluster is $\pm0.03$ and $\pm0.05$ dex, respectively for Si and Ca. In Figure \ref{ca_si}, each cluster is represented by mean values but results for the individual stars of a cluster are closely arranged about the mean values for [Si/Fe] and [Ca/Fe] and such clusters of points are generally well separated from the mean values for other clusters.

The ratio of spread to $\sigma$ for [Si/Fe] may suggest intrinsic differences across the OC sample and, thus, different proportions of Type II to Type Ia products in the clusters.  These differences are less obvious for [Ca/Fe] because of  the larger measurement uncertainty for [Ca/Fe] and the mixed sites for Ca synthesis. A cautionary note is required here: the range in [Ni/Fe], elements with similar expected relative yields, is 0.17 dex for the OCs with [Fe/H] = $-0.10\pm0.05$ but a measurement $\sigma$ of typically $\pm0.03$. 

\begin{figure}
\begin{center}
\includegraphics[trim=0.5cm 6.0cm 11.0cm 8.5cm, clip=true,height=0.18\textheight]{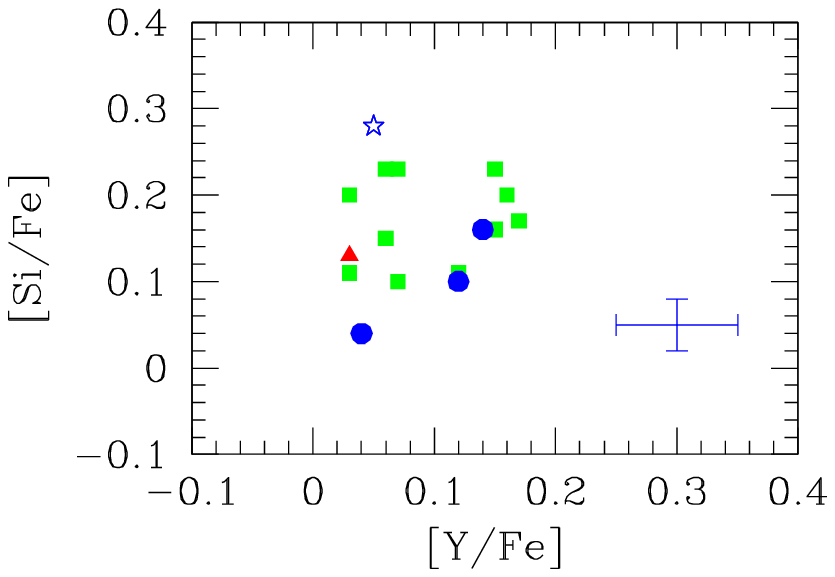}
\caption[ ]{Same as Figure \ref{ca_si}, but for [Si/Fe] vs. [Y/Fe]. }
\label{si_y} 
\end{center}  \vspace{-0.5cm}
\end{figure}

\begin{figure}
\begin{center}
\includegraphics[trim=0.5cm 6.0cm 11.0cm 8.5cm, clip=true,height=0.18\textheight]{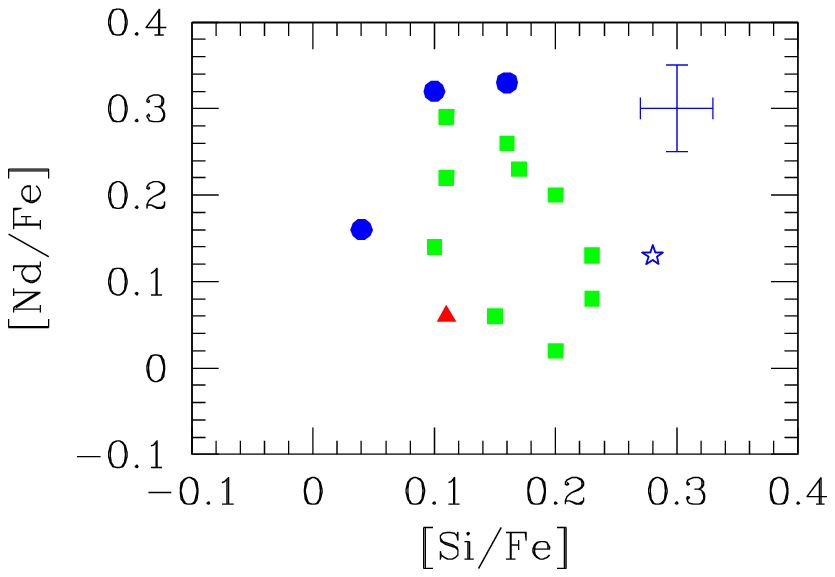}
\caption[ ]{Same as Figure \ref{ca_si}, but for [Nd/Fe] vs. [Si/Fe]. }
\label{si_nd} 
\end{center}  \vspace{-0.5cm}
\end{figure}

Examination of Table \ref{ocmean_intra} listing the [X/Fe] for field and cluster stars points very clearly to a difference in [X/Fe] for heavy elements. Among these heavy elements, the cluster-to-cluster scatter is small for Y and Zr, larger for Ba to Sm and somewhat smaller for Eu. Synthesis of Eu is primarily attribtable to the $r$-process occurring in either Type II supernovae and/or neutron star mergers. The $r$-process makes a contribution to the other sampled heavy elements (see \S \ref{he}). AGB stars and their dredge-up (and mass loss) are expected to be the dominant source of $s$-process products via what is known as the main $s$-process. Massive stars via the weak $s$-process may contribute to Y and Zr and lighter post-Fe elements. 

The cluster-to-cluster scatter among the Y abundances is not appreciably greater than that expected from measurement errors. In Figure \ref{si_y}, [Si/Fe] versus [Y/Fe] is shown for our clusters. With a typical error bar (see figure) of $\pm0.05$ and $\pm0.03$ for Y and Si, respectively, the points in Figure \ref{si_y} for the 12 clusters with [Fe/H] $= -0.10\pm0.05$ suggest a possible mild intrinsic spread in [Si/Fe] but no detectable intrinsic spread in [Y/Fe]. 

\begin{figure*}
\begin{center}
\includegraphics[trim=1cm 9.3cm 1.0cm 5cm, clip=true,height=0.2\textheight]{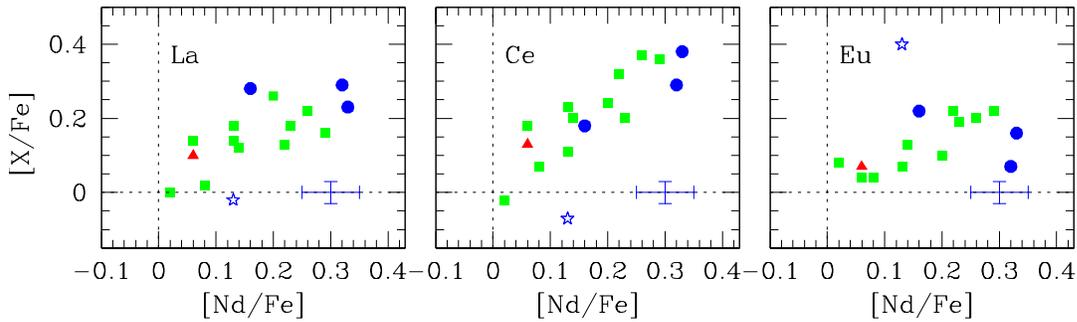}
\caption[ ]{Same as Figure \ref{ca_si}, but for mean cluster abundances for [La/Fe] (left panel), [Ce/Fe] (middle panel) and [Eu/Fe] (right panel) against [Nd/Fe].}
\label{laceeu_nd} 
\end{center}  \vspace{-0.5cm}
\end{figure*}

In contrast to the Y and Zr abundances, heavier elements -- Ba to Eu -- show cluster-to-cluster variations larger than the measurement uncertainties. Figure \ref{si_nd} shows the [Nd/Fe] results as a function of [Si/Fe]. The [Nd/Fe] values range over 0.3 dex for our OCs with [Fe/H]$= -0.10\pm0.05$ dex with no obvious dependence on [Si/Fe]; this range is almost ten times larger than the $\sigma$ for [Nd/Fe] from a given cluster which is only about $\pm0.04$ dex. 

A striking feature of the Ba to Eu variations is that they are not clearly related to either the $s$- or the $r$-process alone. Figure \ref{laceeu_nd} shows the relation between [Nd/Fe] and [La/Fe], [Ce/Fe] and [Eu/Fe] for the [Fe/H]$= -0.10\pm0.05$ clusters. (Similar relations are found for [Ba/Fe] with considerable scatter and [Sm/Fe].) With the exception of NGC 2682 (M 67) with an age of 4.1 Gyr, the clusters with [Fe/H]$=-0.1\pm0.05$ have similar young ages (0.3-0.6 Gyr). Qualitatively, these relations suggest that either the $r$- and $s$-process mix for Ba-Eu across the birthplaces of these clusters is quite similar to a solar mix or they result from variations in the Fe abundance from - presumably - Type Ia supernovae. If an $s$-process contribution dominated the spread, the [Eu/Fe] dependence on [Nd/Fe] would be much weaker; the $s$-process contribution to solar abundances is just 3\% for Eu against 45\% for Nd. Similarly, the La and Ce trends would be steeper with solar $s$-process contributions of 75\% and 81\% for La and Ce, respectively. By a similar but inverse argument, the $r$-process cannot be said to dominate the cluster-to-cluster spread. Variations in the Fe abundance for a given heavy element abundance would superficially account for the approximately unit slope in the three panels of Figure \ref{laceeu_nd}. 

Enhanced interest in heavy-element abundances in OCs has been stimulated by D'Orazi et al. (2009) who analysed by standard LTE procedures the Ba {\scs II} 5854\AA\ and 6496\AA\ lines in dwarf stars from 20 OCs with ages from 0.04 Gyr to 8.4 Gyr to find that the [Ba/Fe] values decreased with age: [Ba/Fe] $\simeq 0.0$ at ages greater than about a billion years in OCs and field stars and rose to [Ba/Fe] $\simeq +0.6$ in some youngest clusters. 

\begin{figure}
\begin{center}
\includegraphics[trim=1.1cm 1.4cm 10.1cm 8.4cm, clip=true,width=0.30\textwidth,height=0.50\textheight,angle=-90]{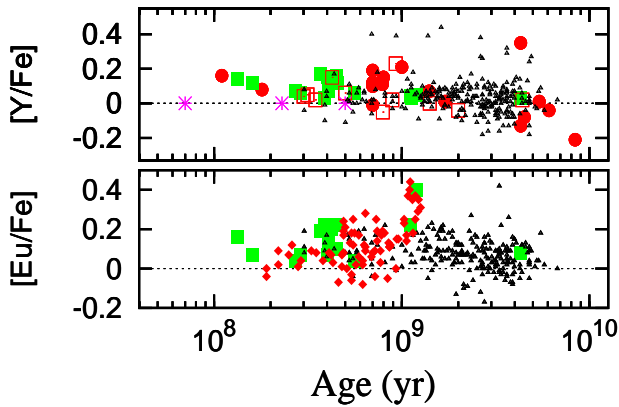}
\caption[]{The relative abundance ratios [X/Fe] versus age (yr) for the $s$-process element Y and the $r$-process element Eu. All the symbols have their usual meaning as in Figures \ref{y_fe} and \ref{eu_fe}.}
\label{yeu_age} 
\end{center}  \vspace{-0.5cm}
\end{figure}

Although the definitive study of elements across the Y-Eu interval for a large sample of OCs is awaited, our results generally confirm reports in the literature. For Y, our small spread in [Y/Fe] is confirmed  - see, for example, Maiorca et al. (2011) and Mishenina et al. (2013, 2015). Figure \ref{yeu_age} shows [Y/Fe] versus age for our and other OCs. In addition, field giants from Luck \& Heiter (2007) are included where we have assigned stellar ages using the relation between turn-off mass to obtain a red giant's age (Takeda et al. 2008):

\begin{equation}
 \log\, age\,\mbox{(yr)} \backsimeq 10.74 - 1.04\,(M/M_{\odot}) + 0.0999\,(M/M_{\odot})^{2} 
\end{equation}

The figure shows the OC [Y/Fe] merging smoothly with the results for the local field giants and the spread in the two samples is comparable.

\begin{figure}
\begin{center}
\includegraphics[trim=1.1cm 1.4cm 10.1cm 8.4cm, clip=true,width=0.30\textwidth,height=0.50\textheight,angle=-90]{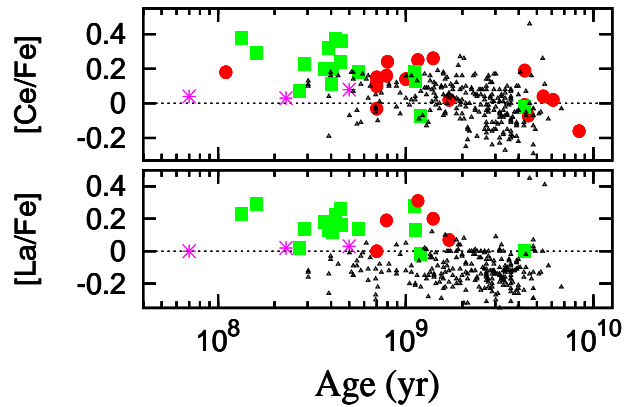}
\caption[]{Same as Figure \ref{yeu_age} but for the $s$-process elements La and Ce.}
\label{lace_age} 
\end{center}   \vspace{-0.5cm}
\end{figure}

Reported results for [Ba/Fe] versus [Fe/H] and age show a large dispersion at a given [Fe/H] or age - see, for example, Mishenina et al. (2013, their Figure 8) even after non-LTE effects on the Ba {\scs II} lines are taken into account. In contrast, elements such as La and Ce show tighter
relations for [La/Fe] and [Ce/Fe] as in Figure \ref{lace_age} (see also Jacobson \& Friel 2013). It is difficult to ignore a suspicion that Ba presents a special problem because the Ba {\scs II} lines are strong whereas the line selection for La {\scs II} and Ce {\scs II} includes several weak lines. However, some authors close to the abundance analyses have invoked a long-forgotten neutron-capture process to account for the high Ba abundances. Mishenina et al. (2015) reintroduce the intermediate neutron-capture $i$-process discussed by Cowan \& Rose (1977) in which the neutron density operating the process is approximately 10$^{15}$ cm$^{-3}$. i.e., intermediate between the low densities for the $s$-process and the much higher densities adopted for the $r$-process. However, abundance predictions illustrated by Mishenina at al. (2015) appear not to fit the OC heavy element abundances: [Ba/La] $\simeq 1.2$ with [Eu/La] $\simeq -0.6$ are predicted but with a single exception our results are [Ba/La] $\sim 0.0$ with [Eu/La] near $-0.1$.

\begin{figure*}
\begin{center}
\includegraphics[trim=0.5cm 5.5cm 1.0cm 5cm, clip=true,width=0.6\textheight,height=0.25\textheight]{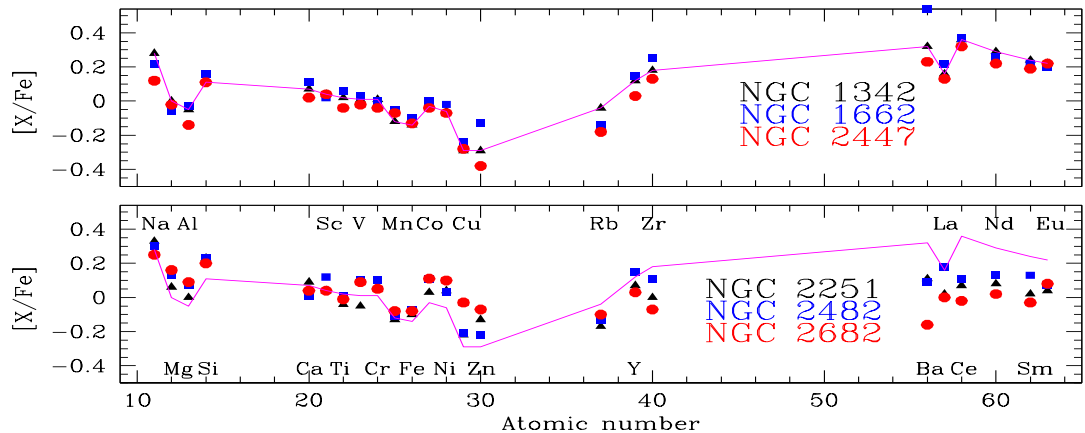}
\caption[]{[X/Fe] versus atomic number for elements Na-Eu for the three most Ce-enriched clusters (top panel: NGC 1662, 1342 and 2447) and the three least Ce-enriched clusters (bottom panel: NGC 2682, 2251 and 2482). The magenta line shown in both panels connects the elements for Ce-enriched cluster NGC 1342. }
\label{ce_enrich} 
\end{center}
\end{figure*}

Intracluster abundance variations offer the opportunity for chemically tagging a cluster's stars after the cluster has dissolved and contributed its stars to the field population. As shown above, the most significant abundance variations are seen among the heavy elements Ba-Eu. Abundances of elements Na-Zn and even Y and Zr display smaller variations and although such variations of some elements appear to exceed their measurement uncertainties, these variations are not obviously correlated with the tightly coupled Ba-Eu variations. In Figure \ref{ce_enrich}, we show [X/Fe] for elements Na-Eu for the three most Ce-enriched clusters (top panel: NGC 1662, 1342 and 2447) and the three least Ce-enriched clusters (bottom panel: NGC 2682, 2251 and 2482). The line shown in both panels connects the elements for Ce-enriched cluster NGC 1342. Relative to the most Ce-enriched clusters, the least Ce-enriched appear to be richer in Mg, Si, Cr, Co and Ni but by only about 0.1 dex Other elements exhibit even smaller differences across the spectrum of Ba-Eu abundance enrichment. These conclusions about intracluster variations for Ba-Eu and much smaller variations for lighter elements confirm comments advanced by De Silva et al. (2009) from a cluster sample drawn from the literature. The present conclusions are based on our sample of OCs observed and analysed in a homogeneous way. Of the 16 OCs in the sample, the primary conclusiuons are drawn from 12 OCs covering a narrow [Fe/H] interval ($-0.10\pm0.05$ dex), a limited range in Galactocentric distnce (8.3 to 9.9 kpc) and all with young ages but for a single cluster.

\section{Concluding remarks}

One aim of our studies of the chemical compositions of red giants is to provide insights into Galactic chemical evolution (GCE). The principal GCE-related result presented here is that among our sample of young clusters with slightly sub-solar [Fe/H] there is a striking spread in the overabundances of elements Ba to Eu in stars with a range from [X/Fe] $\simeq 0.0$, a value exhibited by field stars which are generally older, to
approximately [X/Fe] = 0.3. (Ba is an exceptional elements with a higher maximum [Ba/Fe] but the great strength of the Ba\,{\sc II} lines may give an erroneous Ba abundance,) Such a spread in heavy element  overabundances from cluster to cluster has been reported previously from young associations and clusters but our analysis is the first to demonstrate clearly that the overabundances arise from contributions of both an $s$-process and an $r$-process. Thus, heavy elements from Ba to Eu are the outstanding prospects for chemical tagging. 

Inspection of the [X/Fe] suggests that the [X/Fe] do not exactly mirror the solar abundance distribution, i.e., the combination of $s$-process and $r$-process is not a uniform amplification of the neutron-capture processes which gave the solar abundances. Additionally, the fact that the high [X/Fe] for Ba-Eu appears in young OCs  and not older OCs and field stars suggests that the processes are long delayed in making their contributions to GCE. And the existence of spread from OC to OC in the heavy elements [X/Fe] suggests that there is incomplete mixing among the OCs' birthplaces. Such incompleteness may be due to the relative rarity of the sites for these processes and/or a long timescale for mixing.

In the spirit of speculation to encourage further spectroscopic analyses, the suggestion is made that the GCE is now influenced by $r$-process contributions from the mergers of neutron star binaries where at earlier times Type II supernovae were the dominant source of $r$-process products. For the $s$-process, if it is shown that a distinctly non-solar contribution is required, low mass AGB stars may now be more important contributors
than intermediate mass AGB stars at earlier times. To test such speculations, it would be useful to determine the abundances of as many elements as possible for stars most enriched in Ba-Eu and then to compare the abundance distribution with that for stars least enriched in Ba-Eu.

\vskip1ex 
{\bf Acknowledgements:}
 
We are grateful to the McDonald Observatory's Time Allocation Committee for granting us observing time for this project. DLL wishes to thank the Robert A. Welch Foundation of Houston, Texas for support through grant F-634. We also thank the anonymous referee for useful comments and suggestions which led to improvement of the Paper.

This research has made use of the WEBDA database, operated at the Institute for Astronomy of the University of Vienna and the NASA ADS, USA. This research has also made use of Aladin. This publication makes use of data products from the Two Micron All Sky Survey, which is a joint project of the University of Massachusetts and the Infrared Processing and Analysis Center/California Institute of Technology, funded by the National Aeronautics and Space Administration (NASA) and the National Science Foundation (NSF).

\begin{table*}
{\fontsize{8}{8}\selectfont
\caption[Elemental abundances for stars in the OC NGC 1342]{Elemental abundances for stars in the OC NGC 1342. The abundances calculated by synthesis are presented in bold typeface while the remaining elemental abundances were calculated using the line EWs. Numbers in the parentheses indicate the number of lines used in calculating the abundance of that element. } \vspace{0.2cm}
\label{abu_1342}
\begin{tabular}{lccccc}   \hline
\multicolumn{1}{c}{Species}& \multicolumn{1}{c}{\#4}& \multicolumn{1}{c}{ \#6}& \multicolumn{1}{c}{\#7} & \multicolumn{1}{c}{NGC 1342$_{\mbox{Avg.}}$} \\ \hline

$[$Na I/Fe$]$ &$+0.35\pm0.03$(5) &$+0.27\pm0.03$(5) &$+0.21\pm0.03$(4) &$+0.28\pm0.03$   \\
$[$Mg I/Fe$]$ &$+0.02\pm0.03$(5) &$-0.04\pm0.05$(6) &$+0.02\pm0.05$(6) &$ 0.00\pm0.03$  \\
$[$Al I/Fe$]$ &$-0.06\pm0.02$(6) &$-0.06\pm0.01$(3) &$-0.02\pm0.03$(3) &$-0.05\pm0.01$  \\
$[$Si I/Fe$]$ &$+0.12\pm0.03$(15)&$+0.11\pm0.05$(12)&$+0.09\pm0.04$(12)&$+0.11\pm0.02$   \\
$[$Ca I/Fe$]$ &$+0.02\pm0.04$(12)&$+0.09\pm0.03$(14)&$+0.11\pm0.04$(10) &$+0.07\pm0.02$  \\
$[$Sc I/Fe$]$ &$-0.01\pm0.06$(6) &$-0.01\pm0.07$(5) &$+0.15\pm0.05$(2)  &$+0.04\pm0.03$    \\
$[$Ti I /Fe$]$&$-0.03\pm0.04$(12) &$+0.04\pm0.04$(20)&$+0.04\pm0.03$(15)&$+0.02\pm0.02$    \\
$[$Ti II/Fe$]$&$-0.06\pm0.03$6) &$-0.01\pm0.04$(8) &$-0.06\pm0.02$(5)   &$-0.04\pm0.02$    \\
$[$V I /Fe$]$ &$-0.07\pm0.04$(14)&$+0.04\pm0.03$(13)&$+0.06\pm0.02$(9)  &$+0.01\pm0.02$    \\
$[$Cr I/Fe$]$ &$-0.03\pm0.04$(12) &$+0.03\pm0.02$(12)&$+0.03\pm0.04$(8) &$+0.01\pm0.02$    \\
$[$Cr II/Fe$]$&$+0.02\pm0.03$(6)  &$+0.04\pm0.03$(7)&$+0.04\pm0.02$(6) &$+0.03\pm0.02$     \\
$[$Mn I/Fe$]$ &$\bf -0.09 $          &$\bf-0.10 $    & $\bf-0.18 $      &$\bf-0.12$        \\
$[$Fe I/H $]$ &$-0.12\pm0.04$(126)&$-0.12\pm0.03$(106)&$-0.17\pm0.04$(107)&$-0.14\pm0.02$  \\
$[$Fe II/H$]$ &$-0.10\pm0.04$(19)&$-0.11\pm0.02$(19)&$-0.17\pm0.04$(14) &$-0.13\pm0.02$     \\ 
$[$Co I/Fe$]$ &$-0.06\pm0.04$(9) &$ 0.00\pm0.05$(8) &$-0.04\pm0.06$(6)  &$-0.03\pm0.03$     \\
$[$Ni I/Fe$]$ &$-0.10\pm0.04$(33)&$-0.03\pm0.04$(32)&$-0.06\pm0.04$(32) &$-0.06\pm0.02$     \\  
$[$Cu I/Fe$]$ &   $\bf-0.26$        &  $\bf-0.30$       &$\bf-0.30$      &$\bf-0.29$          \\
$[$Zn I/Fe$]$ &   $\bf-0.33$        &  $\bf-0.27$       &$\bf-0.27$      &$\bf-0.29$          \\ 
$[$Rb I/Fe$]$ &   $\bf-0.06$        &  $\bf-0.08$       &  $\bf+0.02$    &$\bf-0.04$          \\
$[$Y II/Fe$]$ &$+0.12\pm0.03$(6) &$+0.13\pm0.04$(8) &$+0.12\pm0.04$(6)  &$+0.12\pm0.02$      \\
$[$ Zr I/Fe$]$&$+0.14\pm0.04$(5) &$+0.18\pm0.04$(5) &$+0.21\pm0.04$(5)  &$+0.18\pm0.02$     \\
$[$ Zr II/Fe$]$&$+0.24\pm0.02$(3) &$+0.22\pm0.05$(3) &$+0.29\pm0.00$(2)  &$+0.25\pm0.02$     \\
$[$Ba II/Fe$]$&  $\bf+0.36 $       &    $\bf+0.32$   &$\bf+0.27$         &$\bf+0.32$          \\
$[$La II/Fe$]$&$+0.17\pm0.04$(4) &$+0.15\pm0.05$(7) &$+0.16\pm0.04$(5) &$+0.16\pm0.03$       \\
$[$Ce II/Fe$]$&$+0.30\pm0.04$(5) &$+0.43\pm0.02$(6)  &$+0.36\pm0.00$(2) &$+0.36\pm0.01$      \\
$[$Nd II/Fe$]$&$+0.27\pm0.03$(12) &$+0.33\pm0.02$(15) &$+0.26\pm0.03$(9)&$+0.29\pm0.02$      \\
$[$Sm II/Fe$]$&$+0.17\pm0.03$(7)  &$+0.27\pm0.05$(8)  &$+0.27\pm0.03$(7) &$+0.24\pm0.02$     \\
$[$Eu II/Fe$]$&  $\bf+0.21 $     & $\bf+0.22$          &$\bf+0.22$       &$\bf+0.22$          \\

\hline
\end{tabular}
}
\end{table*}

\begin{table*} 
\caption[Elemental abundances for stars in the OCs NGC 1662 \& 1912]{ Same as Table \ref{abu_1342} but for stars in clusters NGC 1662 \& 1912 } 
\vspace{0.2cm}
\label{abu_1912}
\begin{tabular}{lcccccc}   \hline
\multicolumn{1}{l}{Species} & \multicolumn{1}{c}{NGC 1662$\#$1} & \multicolumn{1}{c}{NGC 1662$\#$2} &\multicolumn{1}{c}{NGC 1662$_{\mbox{Avg.}}$} \vline & \multicolumn{1}{c}{NGC 1912$\#$3} & \multicolumn{1}{c}{NGC 1912$\#$70} & \multicolumn{1}{c}{NGC 1912$_{\mbox{Avg.}}$}  \\ \hline

$[$Na I/Fe$]$  &$+0.23\pm0.03$(4) &$+0.21\pm0.04$(4) &$+0.22\pm0.03$   &$+0.33\pm0.05$(4) &$+0.34\pm0.02$(4) &$+0.33\pm0.04$    \\
$[$Mg I/Fe$]$  &$-0.07\pm0.02$(6) &$-0.05\pm0.01$(4) &$-0.06\pm0.01$   &$-0.01\pm0.05$(4) &$+0.08\pm0.03$(5) &$+0.03\pm0.04$   \\
$[$Al I/Fe$]$  &$-0.02\pm0.03$(3) &$-0.05\pm0.04$(6) &$-0.03\pm0.03$   &$+0.05\pm0.05$(4) &$+0.07\pm0.01$(2) &$+0.06\pm0.04$   \\
$[$Si I/Fe$]$  &$+0.17\pm0.04$(13)&$+0.15\pm0.04$(12)&$+0.16\pm0.03$   &$+0.20\pm0.05$(9)&$+0.26\pm0.05$(7)  &$+0.23\pm0.05$   \\
$[$Ca I/Fe$]$  &$+0.12\pm0.03$(13) &$+0.11\pm0.04$(13)&$+0.11\pm0.03$  &$+0.11\pm0.04$(9) &$+0.18\pm0.04$(7) &$+0.14\pm0.04$   \\
$[$Sc I/Fe$]$  &$+0.02\pm0.11$(4) &$\ldots$           &$+0.02\pm0.11$  & $\ldots$        &  $\ldots$         &  $\ldots$      \\
$[$Sc II/Fe$]$ &$+0.15\pm0.08$(5) &$+0.07\pm0.08$(5)  &$+0.11\pm0.06$  &$\bf+0.10$        &$\bf+0.10$            &$\bf+0.10$   \\
$[$Ti I/Fe$]$  &$+0.07\pm0.03$(16) &$+0.05\pm0.04$(13) &$+0.06\pm0.03$ &$-0.08\pm0.05$(11) &$-0.06\pm0.04$(9) &$-0.07\pm0.04$   \\
$[$Ti II/Fe$]$ &$+0.03\pm0.04$(9) &$+0.06\pm0.03$(6)  &$+0.05\pm0.03$  &$-0.07\pm0.03$(3) &$+0.01\pm0.04$(4)  &$+0.03\pm0.03$  \\
$[$V I/Fe$]$   &$+0.04\pm0.03$(11)&$+0.02\pm0.04$(10) &$+0.03\pm0.03$  &$-0.13\pm0.03$(8) &$-0.01\pm0.05$(7) &$-0.07\pm0.04$   \\
$[$Cr I/Fe$]$  &$ 0.00\pm0.03$(15)&$+0.01\pm0.03$(10)  &$ 0.00\pm0.02$ &$-0.01\pm0.05$(6) &$+0.04\pm0.03$(4)  &$+0.01\pm0.04$  \\
$[$Cr II/Fe$]$ &$+0.10\pm0.04$(7) &$+0.04\pm0.04$(7)  &$+0.07\pm0.03$  &$+0.05\pm0.05$(6) &$+0.06\pm0.00$(2)  &$+0.05\pm0.03$  \\
$[$Mn I/Fe$]$  &$ -0.05 $         &$-0.04 $           &$-0.05$         &$\bf-0.15 $       &$\bf-0.10 $        &$\bf-0.12$      \\
$[$Fe I/H$]$   &$-0.09\pm0.03$(112)&$-0.11\pm0.04$(107) &$-0.10\pm0.03$&$-0.12\pm0.05$(49)&$-0.10\pm0.04$(45) &$-0.11\pm0.04$  \\
$[$Fe II/H$]$  &$-0.10\pm0.03$(15)&$-0.12\pm0.04$(16) &$-0.11\pm0.03$  &$-0.09\pm0.05$(13)&$-0.10\pm0.04$(12) &$-0.09\pm0.04$  \\
$[$Co I/Fe$]$  &$+0.01\pm0.04$(9) &$ 0.00\pm0.04$(8)  &$ 0.00\pm0.03$  &$-0.10\pm0.01$(4) &$-0.11\pm0.05$(8)  &$-0.10\pm0.03$  \\
$[$Ni I/Fe$]$  &$ 0.00\pm0.03$(31)&$-0.04\pm0.04$(28) &$-0.02\pm0.03$  &$-0.01\pm0.05$(16)&$-0.04\pm0.04$(11) &$-0.02\pm0.04$  \\
$[$Cu I/Fe$]$  &   $-0.27$        &  $-0.21$          &$-0.24$         &   $\bf-0.30$        &  $\bf-0.30$    &$\bf-0.30$      \\
$[$Zn I/Fe$]$  &   $-0.14$        &  $-0.13$          &$-0.13$         &   $\bf+0.10$        &  $\bf+0.10$    &$\bf+0.10$      \\
$[$Rb I/Fe$]$  &   $-0.15$        &  $-0.14$          &$-0.14$         &   $\bf-0.30$        &  $\bf-0.30$    &$\bf-0.30$      \\
$[$Y II/Fe$]$  &$+0.18\pm0.04$(6) &$+0.13\pm0.04$(5)  &$+0.15\pm0.03$  &$+0.04\pm0.01$(4) &$+0.09\pm0.03$(5)  &$+0.06\pm0.02$  \\
$[$Zr I/Fe$]$  &$+0.26\pm0.02$(5) &$+0.25\pm0.04$(4)  &$+0.25\pm0.02$  &$+0.12\pm0.02$(3) &$+0.09\pm0.05$(4)  &$+0.10\pm0.03$  \\
$[$Zr II/Fe$]$ &$+0.30\pm0.04$(4) &$+0.32\pm0.05$(4) &$+0.31\pm0.03$   & $\ldots$        &  $\ldots$         &  $\ldots$      \\
$[$Ba II/Fe$]$ &  $ +0.55 $       &    $+0.54$        &$+0.54$          &  $\bf +0.70 $       &   $\bf+0.70$   &$\bf+0.70$      \\
$[$La II/Fe$]$ &$+0.23\pm0.02$(4) &$+0.21\pm0.04$(5)  &$+0.22\pm0.02$  &$+0.13\pm0.02$(4) &$+0.16\pm0.03$(3)  &$+0.14\pm0.02$  \\
$[$Ce II/Fe$]$ &$+0.37\pm0.04$(5) &$+0.38\pm0.02$(6)  &$+0.37\pm0.02$  &$+0.24\pm0.02$(3) &$+0.23\pm0.03$(3)  &$+0.23\pm0.02$  \\
$[$Nd II/Fe$]$ &$+0.25\pm0.03$(13) &$+0.28\pm0.03$(13) &$+0.26\pm0.02$ &$+0.14\pm0.03$(6) &$+0.13\pm0.04$(9) &$+0.13\pm0.03$   \\
$[$Sm II/Fe$]$ &$+0.24\pm0.03$(7) &$+0.20\pm0.04$(8)  &$+0.22\pm0.03$  &$+0.06\pm0.04$(4) &$+0.03\pm0.03$(5)  &$+0.04\pm0.03$   \\
$[$Eu II/Fe$]$ &  $+0.20$         & $+0.20$           &$+0.20$         &  $\bf+0.10$         & $\bf+0.05$     &$\bf+0.07$      \\

\hline
\end{tabular}
\end{table*}

\begin{table*}
{\fontsize{8}{8}\selectfont
 \begin{minipage}{12cm}
\caption[Elemental abundances for stars in the 2354]{Same as Table \ref{abu_1342} but for stars in the OC NGC 2354 }
\vspace{0.2cm}
\label{abu_2354}
\begin{tabular}{lccc}   \hline
\multicolumn{1}{l}{Species} & \multicolumn{1}{c}{NGC 2354$\#$183} & \multicolumn{1}{c}{NGC 2354$\#$205} & \multicolumn{1}{c}{NGC 2354$_{\mbox{Avg.}}$}  \\ \hline

$[$Na I/Fe$]$  &$+0.14\pm0.04$(3) &$+0.11\pm0.03$(2) &$+0.12\pm0.03$  \\
$[$Mg I/Fe$]$  &$-0.18\pm0.04$(3) &$-0.16\pm0.04$(5) &$-0.17\pm0.03$  \\
$[$Al I/Fe$]$  &$-0.14\pm0.05$(5) &$-0.09\pm0.03$(3) &$-0.11\pm0.03$   \\
$[$Si I/Fe$]$  &$+0.15\pm0.03$(8) &$+0.17\pm0.03$(7) &$+0.16\pm0.02$   \\
$[$Ca I/Fe$]$  &$-0.07\pm0.05$(9) &$-0.07\pm0.04$(5) &$-0.07\pm0.03$  \\
$[$Sc II/Fe$]$ &$+0.05\pm0.10$(3) &  $\ldots $          &$+0.05\pm0.10$  \\
$[$Ti I/Fe$]$  &$-0.07\pm0.04$(12) &$+0.09\pm0.04$(16)&$+0.01\pm0.03$ \\
$[$Ti II/Fe$]$ &$-0.12\pm0.03$(3) &$ 0.00\pm0.03$(4)  &$-0.06\pm0.02$ \\
$[$V I/Fe$]$   &$ 0.00\pm0.05$(8) &$+0.08\pm0.05$(7)  &$+0.04\pm0.04$ \\
$[$Cr I/Fe$]$  &$-0.07\pm0.04$(5) &$+0.01\pm0.04$(6)  &$-0.03\pm0.03$ \\
$[$Cr II/Fe$]$ &$-0.03\pm0.03$(2) &$-0.03\pm0.04$(2)  &$-0.03\pm0.03$ \\
$[$Mn I/Fe$]$  &$ -0.01 $         &$-0.09 $           &$-0.05$        \\
$[$Fe I/H$]$   &$-0.20\pm0.03$(71)&$-0.18\pm0.03$(67) &$-0.19\pm0.02$ \\
$[$Fe II/H$]$  &$-0.18\pm0.02$(5) &$-0.15\pm0.03$(8)  &$-0.16\pm0.02$ \\
$[$Co I/Fe$]$  &$ 0.00\pm0.03$(5) &$+0.14\pm0.01$(3)  &$+0.07\pm0.02$ \\
$[$Ni I/Fe$]$  &$-0.01\pm0.04$(18)&$+0.01\pm0.04$(13) &$ 0.00\pm0.03$ \\
$[$Cu I/Fe$]$  &   $-0.13$        &  $-0.11$          &$-0.12$        \\
$[$Zn I/Fe$]$  &   $-0.30$        &  $-0.33$          &$-0.31$        \\
$[$Rb I/Fe$]$  &   $-0.21$        &  $-0.14$          &$-0.17$        \\
$[$Y II/Fe$]$  &$+0.16\pm0.03$(3) &$+0.12\pm0.03$(3)  &$+0.14\pm0.02$ \\
$[$Zr I/Fe$]$  &$+0.11\pm0.04$(3) &$+0.16\pm0.01$(3)  &$+0.13\pm0.02$ \\
$[$Ba II/Fe$]$ &  $ +0.19 $       &    $+0.16$        &$+0.17$       \\
$[$La II/Fe$]$ &$+0.25\pm0.04$(4) &$+0.22\pm0.04$(4)  &$+0.23\pm0.03$ \\
$[$Ce II/Fe$]$ &$+0.34\pm0.03$(4) &$+0.43\pm0.03$(4)  &$+0.38\pm0.02$ \\
$[$Nd II/Fe$]$ &$+0.35\pm0.04$(4) &$+0.32\pm0.02$(8)  &$+0.33\pm0.02$ \\
$[$Sm II/Fe$]$ &$+0.26\pm0.02$(3) &$+0.22\pm0.02$(6)  &$+0.24\pm0.01$ \\
$[$Eu II/Fe$]$ &  $\ldots $        & $+0.16$           &$+0.16$        \\

\hline
\end{tabular} 
\end{minipage}
}
\end{table*}

\begin{table*}
{\fontsize{9}{9}\selectfont
 \begin{minipage}{12cm}
\caption[Elemental abundances for stars in the OC NGC 2447]{Same as Table \ref{abu_1342} but for stars in the OC NGC 2447.}
\vspace{0.2cm}
\label{abu_2447}
\begin{tabular}{lcccc}   \hline
\multicolumn{1}{l}{Species} & \multicolumn{1}{c}{NGC 2447$\#$28} & \multicolumn{1}{c}{NGC 2447$\#$34} & \multicolumn{1}{c}{NGC 2447$\#$41} & \multicolumn{1}{c}{NGC 2447$_{\mbox{Avg.}}$}  \\ \hline

$[$Na I/Fe$]$  &$+0.13\pm0.04$(5) &$+0.13\pm0.04$(3) &$+0.09\pm0.02$(4) &$+0.12\pm0.02$    \\
$[$Mg I/Fe$]$  &$-0.08\pm0.02$(5) &$+0.01\pm0.02$(4) &$ 0.00\pm0.02$(4) &$-0.02\pm0.01$    \\
$[$Al I/Fe$]$  &$-0.16\pm0.03$(5) &$-0.13\pm0.03$(7) &$-0.12\pm0.03$(6) &$-0.14\pm0.02$    \\
$[$Si I/Fe$]$  &$+0.09\pm0.05$(8) &$+0.11\pm0.04$(9) &$+0.12\pm0.03$(12)&$+0.11\pm0.02$    \\
$[$Ca I/Fe$]$  &$+0.02\pm0.04$(10)&$+0.01\pm0.04$(9) &$+0.02\pm0.05$(10) &$+0.02\pm0.03$   \\
$[$Sc I/Fe$]$  &  $\ldots$       &    $\ldots$   &$+0.04\pm0.09$(2)  &$+0.04\pm0.09$       \\
$[$Sc II/Fe$]$ &  $\ldots$       &   $\ldots$   &$+0.10\pm0.01$(2)  &$+0.10\pm0.01$        \\
$[$Ti I/Fe$]$  &$-0.05\pm0.04$(19) &$-0.06\pm0.04$(17)&$-0.01\pm0.05$(17) &$-0.04\pm0.03$  \\
$[$Ti II/Fe$]$ &$-0.05\pm0.04$5) &$-0.05\pm0.04$(8) &$-0.06\pm0.03$(6)   &$-0.05\pm0.02$   \\
$[$V I/Fe$]$   &$-0.03\pm0.04$(13)&$-0.04\pm0.04$(14)&$+0.02\pm0.03$(16)  &$-0.02\pm0.02$  \\
$[$Cr I/Fe$]$  &$-0.05\pm0.04$(13) &$-0.01\pm0.03$(10)&$-0.06\pm0.03$(9) &$-0.04\pm0.02$   \\
$[$Cr II/Fe$]$ &$+0.03\pm0.02$(3)  &$ 0.00\pm0.02$(5) &$+0.02\pm0.04$(4) &$+0.02\pm0.02$   \\
$[$Mn I/Fe$]$  &$ -0.05 $          &$-0.08 $          &    $-0.07 $        &$-0.07$        \\
$[$Fe I/H$]$   &$-0.11\pm0.04$(101)&$-0.13\pm0.04$(102)&$-0.14\pm0.04$(120)&$-0.13\pm0.02$ \\
$[$Fe II/H$]$  &$-0.09\pm0.04$(9) &$-0.12\pm0.03$(11)&$-0.13\pm0.04$(16) &$-0.11\pm0.02$   \\
$[$Co I/Fe$]$  &$-0.07\pm0.04$(6) &$-0.01\pm0.05$(7) &$-0.03\pm0.04$(6)  &$-0.04\pm0.03$   \\
$[$Ni I/Fe$]$  &$-0.08\pm0.04$(25)&$-0.07\pm0.04$(30)&$-0.05\pm0.04$(27) &$-0.07\pm0.02$   \\
$[$Cu I/Fe$]$  &   $-0.27$        &  $-0.25$       &$-0.32$              &$-0.28$          \\
$[$Zn I/Fe$]$  &   $-0.34$        &  $-0.32$       &$-0.49$              &$-0.38$          \\
$[$Rb I/Fe$]$  &   $-0.20$        &  $-0.18$       &  $-0.17$            &$-0.18$          \\
$[$Y II/Fe$]$  &$-0.01\pm0.04$(7) &$-0.02\pm0.03$(5) &$+0.12\pm0.05$(7)  &$+0.03\pm0.02$   \\
$[$Zr I/Fe$]$  &$+0.12\pm0.02$(4) &$+0.11\pm0.05$(5) &$+0.16\pm0.04$(5)  &$+0.13\pm0.02$   \\
$[$Zr II/Fe$]$ & $\ldots$         &$+0.12\pm0.04$(2) &$+0.20\pm0.00$(1)  &$+0.16\pm0.03$   \\
$[$Ba II/Fe$]$ & $\ldots$         &    $\ldots$      &$+0.23$           &$+0.23$           \\
$[$La II/Fe$]$ &$+0.15\pm0.04$(5) &$+0.11\pm0.05$(3) &$+0.12\pm0.03$(6) &$+0.13\pm0.02$    \\
$[$Ce II/Fe$]$ &$+0.32\pm0.01$(4) &$+0.32\pm0.03$(3)  &$+0.31\pm0.03$(5)  &$+0.32\pm0.01$  \\
$[$Nd II/Fe$]$ &$+0.22\pm0.05$(14) &$+0.22\pm0.05$(10) &$+0.22\pm0.04$(14)&$+0.22\pm0.03$  \\
$[$Sm II/Fe$]$ &$+0.22\pm0.05$(6)  &$+0.16\pm0.05$(5)  &$+0.20\pm0.04$(5) &$+0.19\pm0.03$  \\
$[$Eu II/Fe$]$ &  $+0.20 $        & $+0.22$            &$+0.23$           &$+0.22$         \\

\hline
\end{tabular} 
 \end{minipage} 
}
\end{table*}


\begin{table*}   
{\fontsize{7}{7}\selectfont
\caption[Elemental abundance ratios {[X/Fe]} for literature sample of OCs.]{Elemental abundance ratios [X/Fe] for elements Na, Al, Mg, Si, Ca, Ti, Cr and Ni for the literature sample.}
 \label{abu_lit} 
\begin{tabular}{lcccrrrccc} \hline
\multicolumn{1}{l}{Cluster}& \multicolumn{1}{c}{[Na/Fe]}& \multicolumn{1}{c}{[Mg/Fe]}& \multicolumn{1}{c}{[Al/Fe]}& \multicolumn{1}{c}{[Si/Fe]}& \multicolumn{1}{c}{[Ca/Fe]} & \multicolumn{1}{c}{[Ti/Fe]}& \multicolumn{1}{c}{[Cr/Fe]}& \multicolumn{1}{c}{[Fe/H]}& \multicolumn{1}{c}{[Ni/Fe]}  \\
\hline

\multicolumn{10}{c}{\bf Thin disk} \\

NGC 6404 &   $\ldots$    & 0.08$\pm$0.07 & 0.27$\pm$0.07 & 0.18$\pm$0.10 &-0.11$\pm$0.10 &-0.07$\pm$0.07& 0.05$\pm$0.07 & 0.03$\pm$0.10& 0.02$\pm$0.07 \\
NGC 6583 &   $\ldots$    & 0.03$\pm$0.01 & 0.22$\pm$0.01 & 0.01$\pm$0.13 & 0.01$\pm$0.00 &-0.04$\pm$0.02&-0.09$\pm$0.01 & 0.30$\pm$0.01& 0.03$\pm$0.04 \\
NGC 3960 & 0.06$\pm$0.02 &-0.06$\pm$0.04 &-0.02$\pm$0.02 & 0.02$\pm$0.02 & 0.03$\pm$0.03 & 0.01$\pm$0.03& 0.02$\pm$0.02 & 0.12$\pm$0.02&-0.05$\pm$0.02 \\
Collinder 261& 0.08$\pm$0.06& 0.17$\pm$0.07& 0.33$\pm$0.06& 0.24$\pm$0.05 &-0.01$\pm$0.06 &-0.02$\pm$0.09&-0.01$\pm$0.08 & 0.07$\pm$0.06& 0.05$\pm$0.06 \\
NGC 6192 &   $\ldots$    & 0.04$\pm$0.03 & 0.13$\pm$0.03 & 0.11$\pm$0.07 & 0.01$\pm$0.05 & 0.03$\pm$0.09& 0.00$\pm$0.08 & 0.09$\pm$0.09&-0.05$\pm$0.07 \\
 IC 4756 & 0.18$\pm$0.08 & 0.00$\pm$0.03 &   $\ldots$    & 0.00$\pm$0.06 & 0.04$\pm$0.06 &-0.07$\pm$0.05& 0.03$\pm$0.04 & 0.04$\pm$0.04&-0.07$\pm$0.04 \\
NGC 3532 & 0.27$\pm$0.06 & 0.02$\pm$0.01 &   $\ldots$    & 0.02$\pm$0.05 & 0.08$\pm$0.05 &-0.07$\pm$0.05&-0.01$\pm$0.04 & 0.05$\pm$0.04&-0.07$\pm$0.03 \\
NGC 6281 & 0.23$\pm$0.01 & 0.05$\pm$0.02 &   $\ldots$    & 0.07$\pm$0.02 & 0.10$\pm$0.05 &-0.07$\pm$0.04&-0.02$\pm$0.06 & 0.05$\pm$0.05&-0.02$\pm$0.07 \\
NGC 6633 & 0.24$\pm$0.01 & 0.05$\pm$0.00 &   $\ldots$    & 0.06$\pm$0.03 & 0.01$\pm$0.06 &-0.07$\pm$0.08& 0.05$\pm$0.07 & 0.00$\pm$0.07&-0.03$\pm$0.05 \\
 IC 2602 & 0.06$\pm$0.01 &   $\ldots$    &   $\ldots$    & 0.06$\pm$0.04 & 0.08$\pm$0.06 & 0.04$\pm$0.00& 0.06$\pm$0.00 &-0.06$\pm$0.05&-0.01$\pm$0.06 \\
 IC 2391 & 0.03$\pm$0.03 &   $\ldots$    &   $\ldots$    & 0.04$\pm$0.02 & 0.07$\pm$0.03 & 0.10$\pm$0.06& 0.02$\pm$0.00 &-0.02$\pm$0.03& 0.03$\pm$0.02 \\
 NGC 7160 &-0.20$\pm$0.08 &   $\ldots$    &   $\ldots$    & 0.07$\pm$0.09 &-0.02$\pm$0.09 &   $\ldots$   &   $\ldots$    & 0.13$\pm$0.06& 0.01$\pm$0.06 \\
 NGC 5822 & 0.23$\pm$0.04 &   $\ldots$    &-0.01$\pm$0.00 & 0.07$\pm$0.01 & 0.03$\pm$0.08 &-0.02$\pm$0.13& 0.02$\pm$0.11 & 0.01$\pm$0.09&-0.04$\pm$0.07 \\
 NGC 6134 & 0.26$\pm$0.08 &-0.03$\pm$0.05 &   $\ldots$    & 0.02$\pm$0.02 & 0.04$\pm$0.09 & 0.04$\pm$0.02& 0.10$\pm$0.03 & 0.15$\pm$0.04& 0.01$\pm$0.06 \\
 NGC 6253 & 0.29$\pm$0.25 &-0.05$\pm$0.12 & 0.30$\pm$0.16 & 0.23$\pm$0.08 &-0.02$\pm$0.02 & 0.07$\pm$0.08& 0.11$\pm$0.07 & 0.18$\pm$0.01& 0.08$\pm$0.03 \\
 NGC 6791 & 0.15$\pm$0.10 &   $\ldots$    &   $\ldots$    & 0.23$\pm$0.05 & 0.11$\pm$0.12 & 0.26$\pm$0.08& 0.10$\pm$0.06 & 0.24$\pm$0.10& 0.18$\pm$0.03 \\
 NGC 6819 & 0.43$\pm$0.01 &-0.01$\pm$0.01 & 0.15$\pm$0.09 & 0.18$\pm$0.09 & 0.19$\pm$0.08 & 0.07$\pm$0.03& 0.18$\pm$0.03 & 0.04$\pm$0.09&-0.04$\pm$0.01 \\
NGC 2660 & 0.07$\pm$0.03 & 0.00$\pm$0.03 & 0.10$\pm$0.06 & 0.03$\pm$0.01 & 0.02$\pm$0.04 & 0.07$\pm$0.02& 0.02$\pm$0.02 & 0.06$\pm$0.02&-0.04$\pm$0.02 \\
NGC 188 & 0.20$\pm$0.06 & 0.25$\pm$0.04 & 0.25$\pm$0.06 & 0.24$\pm$0.04 & 0.02$\pm$0.02 & 0.07$\pm$0.06&   $\ldots$    & 0.11$\pm$0.07& 0.04$\pm$0.09 \\
NGC 2477 & 0.15$\pm$0.02 &-0.04$\pm$0.03 & 0.12$\pm$0.03 & 0.11$\pm$0.04 & 0.05$\pm$0.03 & 0.01$\pm$0.06& 0.05$\pm$0.06 & 0.07$\pm$0.04&-0.03$\pm$0.05 \\
NGC 6939 & 0.12$\pm$0.02 &   $\ldots$    & 0.27$\pm$0.01 & 0.10$\pm$0.04 & 0.16$\pm$0.03 &   $\ldots$   &   $\ldots$    & 0.07$\pm$0.01& 0.02$\pm$0.02 \\
Collinder 110& 0.13$\pm$0.03& 0.04$\pm$0.01& 0.20$\pm$0.01 & 0.05$\pm$0.03 & 0.03$\pm$0.02 & 0.08$\pm$0.03& 0.08$\pm$0.03 &-0.07$\pm$0.02&-0.03$\pm$0.04 \\
NGC 2099 & 0.19$\pm$0.00 &-0.10$\pm$0.08 & 0.13$\pm$0.07 & 0.02$\pm$0.00 & 0.00$\pm$0.02 &-0.02$\pm$0.04& 0.00$\pm$0.08 &-0.03$\pm$0.02&-0.01$\pm$0.02 \\
NGC 2420 & 0.09$\pm$0.06 &-0.11$\pm$0.04 & 0.11$\pm$0.03 &-0.02$\pm$0.04 & 0.02$\pm$0.01 & 0.12$\pm$0.08& 0.00$\pm$0.08 &-0.06$\pm$0.02& 0.01$\pm$0.01 \\
 IC 2714 & 0.23$\pm$0.03 & 0.04$\pm$0.02 &   $\ldots$    & 0.22$\pm$0.08 & 0.07$\pm$0.05 & 0.01$\pm$0.03&-0.06$\pm$0.07 & 0.08$\pm$0.07&-0.07$\pm$0.04 \\
NGC 7789 & 0.07$\pm$0.05 &-0.09$\pm$0.03 & 0.12$\pm$0.06 &-0.12$\pm$0.03 &-0.05$\pm$0.03 &-0.04$\pm$0.04&-0.01$\pm$0.03 & 0.06$\pm$0.09& 0.03$\pm$0.09 \\
NGC 1245 & 0.18$\pm$0.05 & 0.16$\pm$0.06 &   $\ldots$    & 0.10$\pm$0.04 & 0.09$\pm$0.04 &   $\ldots$   &   $\ldots$    & 0.03$\pm$0.05& 0.00$\pm$0.04 \\
NGC 2194 & 0.16$\pm$0.03 & 0.11$\pm$0.08 &   $\ldots$    & 0.12$\pm$0.02 & 0.01$\pm$0.06 &   $\ldots$   &   $\ldots$    &-0.10$\pm$0.03&-0.03$\pm$0.06 \\
NGC 2355 & 0.21$\pm$0.06 & 0.27$\pm$0.03 & 0.11$\pm$0.00 & 0.19$\pm$0.04 & 0.23$\pm$0.10 &   $\ldots$   &   $\ldots$    &-0.12$\pm$0.01& 0.03$\pm$0.09 \\
NGC 7142 & 0.50$\pm$0.09 &   $\ldots$    & 0.27$\pm$0.12 & 0.26$\pm$0.04 & 0.20$\pm$0.06 &   $\ldots$   &   $\ldots$    &-0.03$\pm$0.11& 0.07$\pm$0.08 \\
NGC 2324 & 0.25$\pm$0.04 & 0.15$\pm$0.04 & 0.04$\pm$0.05 & 0.08$\pm$0.05 & 0.11$\pm$0.07 & 0.02$\pm$0.09& 0.05$\pm$0.03 &-0.12$\pm$0.04&-0.05$\pm$0.03 \\
NGC 2141 & 0.26$\pm$0.12 & 0.08$\pm$0.10 & 0.09$\pm$0.01 & 0.20$\pm$0.16 & 0.09$\pm$0.15 & 0.07$\pm$0.10&   $\ldots$    &-0.14$\pm$0.16&-0.08$\pm$0.11 \\
NGC 2158 & 0.20$\pm$0.08 & 0.09$\pm$0.11 & 0.16$\pm$0.10 & 0.20$\pm$0.22 & 0.29$\pm$0.08 & 0.05$\pm$0.08&   $\ldots$    &-0.15$\pm$0.18& 0.05$\pm$0.21 \\
Berkeley 75& 0.28$\pm$0.04 &   $\ldots$    &   $\ldots$    &   $\ldots$    & 0.16$\pm$0.06 & 0.11$\pm$0.11&-0.10$\pm$0.10 &-0.28$\pm$0.18& 0.04$\pm$0.06 \\
Berkeley 39& 0.14$\pm$0.06 & 0.12$\pm$0.06 & 0.21$\pm$0.03 & 0.20$\pm$0.04 & 0.04$\pm$0.07 & 0.13$\pm$0.02& 0.36$\pm$0.05 &-0.15$\pm$0.09& 0.02$\pm$0.09 \\
NGC 1883 & 0.10$\pm$0.11 & 0.05$\pm$0.05 & 0.06$\pm$0.14 & 0.16$\pm$0.02 &-0.01$\pm$0.03 & 0.01$\pm$0.02& 0.24$\pm$0.07 &-0.06$\pm$0.01&-0.07$\pm$0.04 \\
Ruprecht 4 & 0.10$\pm$0.01 &-0.04$\pm$0.00 & 0.13$\pm$0.00 &-0.01$\pm$0.08 & 0.11$\pm$0.11 & 0.02$\pm$0.06& 0.00$\pm$0.13 &-0.04$\pm$0.05&-0.09$\pm$0.08 \\

\multicolumn{10}{c}{\bf Thin - Thick disk} \\
Melotte 66& 0.10$\pm$0.09 & 0.12$\pm$0.03 & 0.32$\pm$0.04 & 0.20$\pm$0.02 & 0.07$\pm$0.02 & 0.02$\pm$0.02&-0.01$\pm$0.02 &-0.16$\pm$0.01& 0.00$\pm$0.02 \\
NGC 2425  & 0.09$\pm$0.07 &-0.03$\pm$0.07 &   $\ldots$    & 0.02$\pm$0.07 & 0.18$\pm$0.03 &   $\ldots$   &   $\ldots$    & 0.10$\pm$0.01&-0.05$\pm$0.06 \\

 \multicolumn{10}{c}{\bf Thick disk} \\ 
Tombaugh 2 &   $\ldots$    &   $\ldots$    &   $\ldots$    &   $\ldots$    & 0.20$\pm$0.11 &   $\ldots$   &   $\ldots$    &-0.31$\pm$0.10& 0.06$\pm$0.12 \\
Berkeley 73& 0.19$\pm$0.06 & 0.00$\pm$0.00 & 0.17$\pm$0.14 & 0.14$\pm$0.07 &-0.07$\pm$0.07 & 0.03$\pm$0.07& 0.05$\pm$0.13 &-0.23$\pm$0.10& 0.04$\pm$0.14 \\
Berkeley 32&-0.06$\pm$0.04 & 0.13$\pm$0.05 & 0.11$\pm$0.06 & 0.10$\pm$0.04 & 0.07$\pm$0.06 & 0.07$\pm$0.05& 0.01$\pm$0.04 &-0.22$\pm$0.05& 0.01$\pm$0.04 \\
NGC 1193   & 0.14$\pm$0.02 & 0.25$\pm$0.12 & 0.13$\pm$0.14 & 0.17$\pm$0.07 & 0.03$\pm$0.12 &   $\ldots$   &   $\ldots$    &-0.18$\pm$0.10& 0.07$\pm$0.11 \\
Berkeley 22& 0.14$\pm$0.07 & 0.12$\pm$0.05 & 0.41$\pm$0.06 & 0.16$\pm$0.09 &-0.02$\pm$0.03 & 0.17$\pm$0.07&   $\ldots$    &-0.35$\pm$0.02& 0.01$\pm$0.04 \\
Berkeley 18& 0.12$\pm$0.03 & 0.20$\pm$0.06 & 0.30$\pm$0.02 & 0.11$\pm$0.02 & 0.07$\pm$0.02 & 0.12$\pm$0.00&   $\ldots$    &-0.35$\pm$0.00&-0.02$\pm$0.05 \\
NGC 2243   & 0.12$\pm$0.05 & 0.22$\pm$0.04 & 0.40$\pm$0.09 & 0.19$\pm$0.11 & 0.13$\pm$0.03 &   $\ldots$   &   $\ldots$    &-0.42$\pm$0.09& 0.02$\pm$0.09 \\
Berkeley 20& 0.12$\pm$0.05 & 0.26$\pm$0.03 & 0.28$\pm$0.02 & 0.12$\pm$0.10 & 0.05$\pm$0.07 &-0.01$\pm$0.09&-0.03$\pm$0.05 &-0.27$\pm$0.05& 0.03$\pm$0.11 \\
Berkeley 21& 0.18$\pm$0.06 & 0.10$\pm$0.08 & 0.34$\pm$0.02 & 0.18$\pm$0.03 & 0.05$\pm$0.10 & 0.05$\pm$0.09&   $\ldots$    &-0.32$\pm$0.04& 0.03$\pm$0.06 \\
Saurer 1   & 0.21$\pm$0.07 &   $\ldots$    & 0.41$\pm$0.04 & 0.26$\pm$0.01 & 0.25$\pm$0.04 & 0.25$\pm$0.05&   $\ldots$    &-0.39$\pm$0.01& 0.20$\pm$0.01 \\
 
\multicolumn{10}{c}{\bf Thick disk - Halo} \\
Berkeley 33& 0.17$\pm$0.02 &-0.12$\pm$0.10 & 0.33$\pm$0.00 & 0.10$\pm$0.06 & 0.12$\pm$0.13 &-0.10$\pm$0.09& 0.01$\pm$0.12 &-0.34$\pm$0.09&-0.12$\pm$0.08 \\
 \multicolumn{10}{c}{\bf Halo} \\
Berkeley 25& 0.08$\pm$0.14 & 0.10$\pm$0.00 &   $\ldots$    & 0.22$\pm$0.17 & 0.05$\pm$0.10 & 0.20$\pm$0.13& 0.22$\pm$0.14 &-0.30$\pm$0.05&-0.01$\pm$0.21 \\
Berkeley 29& 0.11$\pm$0.05 & 0.18$\pm$0.03 & 0.23$\pm$0.10 & 0.11$\pm$0.06 & 0.06$\pm$0.04 & 0.10$\pm$0.12&-0.09$\pm$0.05 &-0.37$\pm$0.05& 0.00$\pm$0.04 \\
Berkeley 31& 0.14$\pm$0.02 & 0.11$\pm$0.03 & 0.18$\pm$0.05 & 0.16$\pm$0.09 & 0.08$\pm$0.03 &-0.01$\pm$0.05&   $\ldots$    &-0.24$\pm$0.06& 0.01$\pm$0.02 \\

\hline

\end{tabular} }	
\end{table*}

\begin{table*}  
{\fontsize{7}{7}\selectfont
\caption[Elemental abundance ratios {[Y,Zr,Ce/Fe]} for literature sample of OCs.]{Elemental abundance ratios [X/Fe] for elements Y, Zr and Ce for the literature sample.}
\label{lit_heavy} 
\begin{tabular}{lccc|lccc} \hline
\multicolumn{1}{l}{Cluster}& \multicolumn{1}{c}{[Y/Fe]}& \multicolumn{1}{c}{[Zr/Fe]}& \multicolumn{1}{c}{[Ce/Fe]} \vline & \multicolumn{1}{l}{Cluster}& \multicolumn{1}{c}{[Y/Fe]}& \multicolumn{1}{c}{[Zr/Fe]}& \multicolumn{1}{c}{[Ce/Fe]} \\
\hline

\multicolumn{4}{c}{\bf Thin disk} \\

 IC 4756 &  0.06$\pm$0.09 & $\ldots$       & 0.23$\pm$0.06 & \multicolumn{4}{c}{\bf Thick disk} \\
 NGC 3532 &  0.04$\pm$0.11 & $\ldots$      & 0.29$\pm$0.08 & Berkeley 32 &   $\ldots$     & -0.21$\pm$0.07& $\ldots$  \\ 
 NGC 6281 &  0.05$\pm$0.08 & $\ldots$      & 0.26$\pm$0.02 & Berkeley 22 &   $\ldots$     & -0.44$\pm$0.01& $\ldots$  \\
 NGC 6633 &  0.15$\pm$0.11 & $\ldots$      & 0.23$\pm$0.09 & Berkeley 18 &   $\ldots$     & -0.28$\pm$0.02& $\ldots$  \\
 NGC 5822 &  0.02$\pm$0.11 & $\ldots$      & 0.20$\pm$0.08 & Berkeley 21 &   $\ldots$     & -0.26$\pm$0.10& $\ldots$  \\
 NGC 6134 &  0.23$\pm$0.01 & $\ldots$      & $\ldots$      & \multicolumn{4}{c}{\bf Thick disk - Halo} \\
 NGC 6791 &  0.02$\pm$0.03 & $\ldots$      & $\ldots$      & Berkeley 33 & -0.22$\pm$0.05 &   $\ldots$    & $\ldots$   \\
 NGC 188  &    $\ldots$    & 0.08$\pm$0.11 & $\ldots$      &            &                &               &            \\
Collinder 110 &  0.00$\pm$0.02 & $\ldots$  & $\ldots$      &            &                &               &            \\
 NGC 2099 &  0.02$\pm$0.05 & $\ldots$      & $\ldots$      &           &                &               &            \\
 NGC 2420 & -0.04$\pm$0.08 & $\ldots$      & $\ldots$      &           &                &               &            \\
 NGC 7789 &  0.27$\pm$0.02 & $\ldots$      & $\ldots$      &           &                &               &            \\
 NGC 2141 &    $\ldots$    & -0.05$\pm$0.01& $\ldots$      &           &                &               &            \\
 NGC 2158 &    $\ldots$    & -0.12$\pm$0.04& $\ldots$      &           &                &               &            \\
 NGC 1883 &    $\ldots$    & -0.08$\pm$0.05& $\ldots$      &           &                &               &            \\
 Ruprecht 4 & -0.05$\pm$0.08 &  $\ldots$   & $\ldots$      &           &                &               &            \\

\hline

\end{tabular}  }
\end{table*}             



\begin{table*} \vskip-4ex
{\fontsize{7}{8}\selectfont
\caption[The input data of OCs used for membership calculations.]{The input data used in calculating the OC's space coordinates and velocity components. The full sample contains 69 OCs with the parameters extracted from DAML catalogue, while those from different sources in the literature are shown in bold numbers.}
\label{input_kinematics}
\begin{tabular}{lllcrccccrr@{}}  \hline
\multicolumn{1}{l}{Cluster}& \multicolumn{1}{c}{$\alpha$(2000.0)}& \multicolumn{1}{r}{$\delta$(2000.0)}& \multicolumn{1}{c}{l}& 
\multicolumn{1}{c}{b}& \multicolumn{1}{c}{d$_{\odot}$}& \multicolumn{1}{c}{RV} & \multicolumn{1}{c}{$\mu_{\alpha}$ cos $\delta$}& 
\multicolumn{1}{c}{$\mu_{\delta}$}& \multicolumn{1}{c}{Rgc}& \multicolumn{1}{r}{Age} \\
\multicolumn{1}{l}{ }& \multicolumn{1}{c}{hh:mm:ss} & \multicolumn{1}{c}{$^{\circ}$ $^{'}$ $^{''}$}& \multicolumn{1}{c}{(deg.)} &
\multicolumn{1}{c}{(deg.)}& \multicolumn{1}{c}{(kpc)}& \multicolumn{1}{c}{(km sec$^{-1}$)}& \multicolumn{1}{r}{(mas yr$^{-1}$)}& 
\multicolumn{1}{r}{(mas yr$^{-1}$)}& \multicolumn{1}{c}{(kpc)}& \multicolumn{1}{r}{(Myr)} \\
\hline 

 NGC  752 & 01:57:41 & +37:47:06 & 137.125 & $-$23.254 &0.457$\pm$0.091 &   6.30$\pm$0.10 &  7.50$\pm$0.32 &$-$11.50$\pm$0.32 &  8.3 & 1122 \\
 NGC 1817 & 05:12:15 & +16:41:24 & 186.156 & $-$13.096 &1.972$\pm$0.394 &  66.40$\pm$0.11 &  1.02$\pm$1.06 & $-$6.51$\pm$1.06 &  9.9 &  409 \\
 NGC 2360 & 07:17:43 & $-$15:38:30 & 229.807 &  $-$1.426 &1.887$\pm$0.377 &  29.55$\pm$0.30 & $-$3.51$\pm$1.49 &  8.07$\pm$1.57 &  9.3 &  561 \\
 NGC 2506 & 08:00:01 & $-$10:46:12 & 230.564 &   9.935 &3.460$\pm$0.692 &  84.63$\pm$0.18 & $-$2.55$\pm$0.20 &  0.37$\pm$0.14 & 10.5 & 1109 \\
 NGC 2266 & 06:43:19 & +26:58:12 & 187.790 &  10.294 &3.400$\pm$0.680 & $-$29.70$\pm$0.20 & $-$1.98$\pm$0.59 & $-$4.32$\pm$0.59 & 11.3 & 1200 \\
 NGC 2335 & 07:06:49 & $-$10:01:42 & 223.600 &  $-$1.183 &1.417$\pm$0.283 &  $-$3.21$\pm$0.10 & $-$0.91$\pm$1.56 & $-$3.18$\pm$1.25 &  9.1 &  162 \\
 NGC 2482 & 07:55:12 & $-$24:15:30 & 241.626 &   2.035 &1.343$\pm$0.269 &  39.00$\pm$0.20 & $-$4.93$\pm$0.53 &  1.60$\pm$0.53 &  8.7 &  402 \\
 NGC 2251 & 06:34:38 &  +08:22:00 &  203.58  &  +00.10 &1.349$\pm$0.270 &  26.25$\pm$0.14 & $-$0.24$\pm$0.55 & $-$2.60$\pm$0.51& 9.2 & 267  \\
 NGC 2527 & 08:04:58 & $-$28:08:48 & 246.087 &   1.855 &0.601$\pm$0.120 &  40.55$\pm$0.14 & $-$4.10$\pm$0.55 &  6.40$\pm$0.55 &  8.3 &  446 \\
 NGC 2539 & 08:10:37 & $-$12:49:06 & 233.705 &  11.111 &1.363$\pm$0.273 &  29.20$\pm$0.14 & $-$4.07$\pm$0.27 & $-$1.83$\pm$0.27 &  8.9 &  371 \\
 NGC 2682 & 08:51:18 & +11:48:00 & 215.696 &  31.896 &0.908$\pm$0.182 &  34.83$\pm$0.10 & $-$8.62$\pm$0.28 & $-$6.00$\pm$0.28 &  8.7 & 4300 \\
 NGC 1342 & 03:31:38 & +37:22:36 & 154.952 & $-$15.342 & 0.665$\pm$0.133 &  $-$10.67$\pm$0.11 &  $-$1.15$\pm$0.87 &  $-$2.80$\pm$0.87 & 8.6 & 452  \\
 NGC 1662 & 04:48:27 & +10:56:12 & 187.695 & $-$21.114 & 0.437$\pm$0.087 &  $-$13.35$\pm$0.29 &  $-$1.93$\pm$0.28 & $-$2.10$\pm$0.28 & 8.4 & 422 \\
 NGC 2447 & 07:44:30 & $-$23:51:24 & 240.039 & 0.135 & 1.037$\pm$0.207 & 22.08$\pm$0.18 & $-$4.85$\pm$0.33 & 4.47$\pm$0.33  & 8.6 & 387 \\ 
 NGC 2354 & 07:14:10 & $-$25:41:24 & 238.368 & $-$6.792 & 1.445$\pm$0.289 & 34.40$\pm$0.26 & $-$2.44$\pm$0.64 & $-$2.96$\pm$0.64 & 8.8 & 134 \\
 NGC 1912 & 05:28:40 & +35:50:54 & 172.250 & 0.695 & 1.066$\pm$0.213 & 0.18$\pm$0.19 & 0.23$\pm$0.17 & $-$5.44$\pm$0.19 & 9.1 &  290  \\

 NGC 1193 & 03:05:56 & +44:23:00 & 146.812 & $-$12.163 &4.571$\pm$0.914 & $-$88.10$\pm$0.20 &  1.41$\pm$0.54 & $-$4.04$\pm$0.54 & 12.0 & 5012 \\
 NGC 1245 & 03:14:42 & +47:14:12 & 146.647 & $-$8.931 & {\bf 3.000$\pm$0.300}$^{2}$ & {\bf $-$29.70$\pm$1.20}$^{2}$ & $-$2.98$\pm$0.41 & $-$3.05$\pm$0.36 & 10.6 & 1047 \\
 NGC  188 & 04:72:08 & +85:15:18 & 122.843 &  22.384 &1.700$\pm$0.340 & $-$42.36$\pm$0.04 & $-$1.48$\pm$1.25 & $-$0.56$\pm$1.24 &  9.0 & 4285 \\
Berkeley 18 & 05:22:12 & +45:24:00 & 163.632 &   5.017 &5.200$\pm$1.040 &  $-$5.50$\pm$1.10 & $-$2.63$\pm$0.35 & $-$5.95$\pm$0.35 & 13.1 & 4266 \\
NGC 1883 & 05:25:54 & +46:29:24 & 163.083 &   6.159 &4.800$\pm$0.960 & $-$30.80$\pm$0.60 & $-$1.62$\pm$0.39 & $-$1.23$\pm$0.39 & 12.7 & 1000 \\
Berkeley 20 & 05:32:37 & +00:11:18 & 203.483 & $-$17.373 &8.400$\pm$1.680 &  70.00$\pm$13.00 &  1.51$\pm$0.81 & $-$4.11$\pm$0.81 & 15.9 & 6026 \\
Berkeley 21 & 05:51:42 & +21:47:00 & 186.840 &  $-$2.509 &6.200$\pm$1.240 &  $-$1.00$\pm$1.00 & $-$0.65$\pm$0.53 & $-$6.20$\pm$0.53 & 14.2 & 2188 \\
NGC 2099 & 05:52:18 & +32:33:12 & 177.635 &   3.091 &1.383$\pm$0.277 &   8.30$\pm$0.20 &  3.78$\pm$0.29 & $-$7.09$\pm$0.29 &  9.4 &  347 \\
Berkeley 22 & 05:58:27 & +07:45:24 & 199.877 & $-$8.079 &6.000$\pm$1.200 & 95.30$\pm$0.20 & {\bf $-$0.40$\pm$3.87}$^{4}$ & {\bf $-$3.81$\pm$2.90}$^{4}$ & 13.8 & 3311 \\
NGC 2141 & 06:02:55 & +10:26:48 & 198.044 &  $-$5.811 &4.033$\pm$0.807 &  24.10$\pm$0.30 &  1.38$\pm$0.38 & $-$1.78$\pm$0.38 & 11.9 & 1702 \\
NGC 2158 & 06:07:25 & +24:05:48 & 186.634 &   1.781 & {\bf 4.036$\pm$0.125}$^{2}$ &  26.90$\pm$1.90 &  1.43$\pm$0.29 & $-$3.28$\pm$0.20 & 12.0 & 1054 \\
NGC 2194 & 06:13:45 & +12:48:24 & 197.250 &  $-$2.349 &{\bf 1.903$\pm$0.120}$^{2}$ &{\bf 7.80$\pm$0.80}$^{2}$ & $-$0.31$\pm$0.64 & $-$4.39$\pm$0.64 &  9.8 &  327 \\
Berkeley 73 & 06:22:00 & $-$06:21:00 & 215.278 &  $-$9.424 &9.800$\pm$1.960 &  95.70$\pm$0.20 & {\bf +0.54$\pm$2.67}$^{4}$ & {\bf $-$2.60$\pm$2.91}$^{4}$ & 16.9 & 1514 \\
NGC 2243 & 06:29:34 & $-$31:17:00 & 239.478 & $-$18.014 &4.458$\pm$0.892 &  59.84$\pm$0.41 &  2.53$\pm$0.54 &  2.90$\pm$1.30 & 10.9 & 1076 \\
Collinder 110 & 06:38:24 & +02:01:00 & 120.996 &   8.602 &2.184$\pm$0.437 & {\bf 41.00$\pm$3.80}$^{3}$ &  0.59$\pm$0.39 & $-$1.41$\pm$0.39 &  9.3 & 1412 \\
Berkeley 25 & 06:41:00 & $-$16:31:00 & 226.612 & $-$9.700 &11.400$\pm$2.280 & 134.30$\pm$0.20 & {\bf $-$1.27$\pm$4.13}$^{4}$ & {\bf $-$2.53$\pm$3.03}$^{4}$ & 17.8 & 5012 \\
Ruprecht  4 & 06:48:54 & $-$10:32:00 & 222.047 & $-$5.339 &4.700$\pm$0.940 & 47.50$\pm$1.00 & {\bf $-$0.33$\pm$2.34}$^{4}$ & {\bf +0.93$\pm$2.64}$^{4}$ &11.9 &  794 \\

Berkeley 75 & 06:49:01 & $-$23:59:48 & 234.307 & $-$11.188 &9.100$\pm$1.820 &  94.60$\pm$0.35 & {\bf $-$0.61$\pm$3.37}$^{4}$ & {\bf +0.56$\pm$3.36}$^{4}$ & 15.2 & 3981 \\
Berkeley 29 & 06:53:18 & +16:55:00 & 197.983 &   8.025 &14.871$\pm$2.974 &  24.80$\pm$0.10 & $-$0.14$\pm$0.80 & $-$4.75$\pm$0.58 & 22.6 & 1059 \\
Berkeley 31 & 06:57:36 & +08:16:00 & 206.254 &   5.120 &8.272$\pm$1.654 &  55.70$\pm$0.70 & $-$4.30$\pm$0.52 & $-$3.97$\pm$0.52 & 15.8 & 2056 \\
Berkeley 32 & 06:58:06 & +06:26:00 & 207.952 &   4.404 &{\bf 3.420$\pm$0.684}$^{1}$ & 105.20$\pm$0.86 & $-$2.78$\pm$0.88 & $-$3.21$\pm$0.88 & 11.1 & 3388 \\
Tombaugh 2 & 07:03:05 & $-$20:49:00 & 232.832 & $-$6.880 &6.080$\pm$1.216 & 120.90$\pm$0.40 &{\bf $-$2.89$\pm$2.65}$^{4}$ & {\bf 0.95$\pm$2.07}$^{4}$ & 12.6 & 1995 \\
NGC 2324 & 07:04:07 & +01:02:42 & 213.447 &  3.297 &3.800$\pm$0.760 &  41.81$\pm$0.22 & $-$1.70$\pm$1.01 & $-$2.77$\pm$1.01 & 11.4 &  631 \\
NGC 2355 & 07:16:59 & +13:45:00 & 203.390 & 11.803 & {\bf 1.928$\pm$0.130}$^{2}$ &  35.02$\pm$0.16 & $-$2.50$\pm$0.80 & $-$3.00$\pm$1.50 &  9.8 & 162 \\
Saurer 1 & 07:20:56 & +01:48:29 & 214.689 & 7.386 &13.200$\pm$2.640 & 98.00$\pm$9.00 & {\bf +0.48$\pm$2.89}$^{4}$ & {\bf $-$0.69$\pm$3.55}$^{4}$ & 20.3 & 5012 \\
Melotte 66 & 07:26:23 & $-$47:40:00 & 259.559 & $-$14.244 &4.313$\pm$0.863 &  23.00$\pm$6.00 & $-$4.18$\pm$0.61 &  7.67$\pm$1.56 &  9.7 & 2786 \\
NGC 2425 & 07:38:17 & $-$14:52:42 & 231.504 &   3.297 &{\bf 3.062$\pm$0.100}$^{2}$ & 102.90$\pm$1.20 & $-$3.64$\pm$0.87 & $-$0.76$\pm$0.87 & 10.2 & 2188 \\
NGC 2420 & 07:38:23 & +21:34:24 & 198.107 &  19.634 &2.480$\pm$0.496 &  73.57$\pm$0.15 & $-$1.32$\pm$0.42 & $-$4.18$\pm$0.26 & 10.3 & 1995 \\
Berkeley 33 & 06:57:42 & $-$13:13:00 & 225.424 & $-$4.622 & 6500$\pm$1.300 & 76.60$\pm$0.50 & $-$5.73$\pm$0.96 & +3.71$\pm$0.96 & 13.4 & 800 \\
Berkeley 39 & 07:46:42 & $-$04:36:00 & 223.462 &  10.095 &4.780$\pm$0.956 &  55.00$\pm$7.00 & {\bf $-$1.62$\pm$3.61}$^{4}$ & {\bf $-$1.66$\pm$3.07}$^{4}$ & 11.9 & 7943 \\
NGC 2477 & 07:52:10 & $-$38:31:48 & 253.563 &  $-$5.838 &1.300$\pm$0.260 &   7.26$\pm$0.12 & $-$0.73$\pm$0.57 &  1.22$\pm$0.54 & 8.5 &  603 \\
NGC 2660 & 08:42:38 & $-$47:12:00 & 265.929 &  $-$3.013 &2.826$\pm$0.565 &  21.34$\pm$0.46 & $-$5.82$\pm$0.81 &  7.40$\pm$0.83 &  8.7 & 1079 \\
IC 2602 & 10:42:58 & $-$64:24:00 & 289.601 &  $-$4.906 &0.161$\pm$0.032 &  10.10$\pm$0.22 &$-$17.49$\pm$0.22 & 10.10$\pm$0.22 &  7.9 & 32 \\
NGC 3532 & 11:05:39 & $-$58:45:12 & 289.571 &   1.347 &0.492$\pm$0.098 &   4.33$\pm$0.34 &$-$10.95$\pm$0.28 &  4.80$\pm$0.28 &  7.8 &  300 \\
IC 2714 & 11:17:27 & $-$62:44:00 & 198.107 &  19.634 &1.320$\pm$0.264 & $-$14.10$\pm$0.30 & $-$2.56$\pm$0.73 & $-$3.40$\pm$0.61 &  9.2 & 348 \\
NGC 3960 & 11:50:33 & $-$55:40:24 & 294.367 &   6.183 &1.850$\pm$0.370 & $-$22.26$\pm$0.36 & $-$7.01$\pm$0.24 & $-$0.45$\pm$0.33 &  7.4 & 1259 \\
IC 2391  & 08:40:32 & $-$53:02:00 & 270.362 & $-$06.839 & 0.175$\pm$0.035 & 14.40$\pm$0.10 & $-$33.21$\pm$0.30 & $-$5.97$\pm$0.30 & 8.0 & 46 \\
Collinder 261 & 12:37:57 & $-$68:22:00 & 301.684 &  $-$5.528 &2.190$\pm$0.438 & $-$30.00$\pm$9.00 & $-$6.53$\pm$0.72 & $-$1.04$\pm$0.72 &  7.1 & 8912 \\
NGC 5822 & 15:04:21 & $-$54:23:48 & 321.573 &   3.593 &0.933$\pm$0.187 & $-$29.31$\pm$0.18 & $-$7.95$\pm$0.24 & $-$8.20$\pm$0.24 &  7.3 &  891 \\
NGC 6134 & 16:27:46 & $-$49:09:06 & 334.917 &  $-$0.198 &0.913$\pm$0.183 & $-$25.70$\pm$0.19 & $-$0.86$\pm$0.88 & $-$4.60$\pm$0.88 &  7.2 &  929 \\
NGC 6192 & 16:40:23 & $-$43:22:00 & 340.647 &   2.122 &1.547$\pm$0.309 &  $-$7.93$\pm$0.21 &  3.73$\pm$0.83 &  3.18$\pm$1.39 &  6.6 &  135 \\
NGC 6253 & 16:59:05 & $-$52:42:30 & 335.460 &  $-$6.251 &1.510$\pm$0.302 & $-$29.40$\pm$1.30 &{\bf $-$1.55$\pm$2.27}$^{4}$ & {\bf $-$4.64$\pm$2.22}$^{4}$ &  6.7 & 5012 \\
NGC 6281 & 17:04:41 & $-$37:59:06 & 347.731 &   1.972 &0.479$\pm$0.096 &  $-$5.58$\pm$0.26 & $-$3.43$\pm$0.55 & $-$3.60$\pm$0.55 &  7.5 &  314 \\
NGC 6404 & 17:39:37 & $-$33:14:48 & 355.659 & $-$1.177 &2.400$\pm$0.480 &  10.60$\pm$1.10 & {\bf $-$0.67$\pm$2.93}$^{4}$ & {\bf $-$0.55$\pm$3.83}$^{4}$ &  5.6 &  501 \\
NGC 6583 & 18:15:49 & $-$22:08:12 &   9.283 &  $-$2.534 &2.040$\pm$0.408 & 3.00$\pm$0.40 & {\bf $+$0.32$\pm$3.54}$^{4}$ &{\bf $-$1.44$\pm$3.57}$^{4}$ & 6.0 & 1000 \\
NGC 6633 & 18:27:15 & +06:30:30 &  36.011 &   8.328 &0.376$\pm$0.075 & $-$28.95$\pm$0.09 & $-$0.21$\pm$0.31 & $-$1.60$\pm$0.31 &  7.7 &  426 \\
IC 4756 & 18:39:00 & +05:27:00 &  36.381 &   5.242 &0.484$\pm$0.097 & $-$25.80$\pm$0.60 & $-$0.14$\pm$0.23 & $-$3.40$\pm$0.23 &  7.6 &  500 \\
NGC 6791 & 19:20:53 & +37:46:18 &  69.959 &  10.904 &5.853$\pm$1.171 & $-$47.10$\pm$0.70 & $-$0.57$\pm$0.13 & $-$2.45$\pm$0.12 &  8.2 & 4395 \\
NGC 6819 & 19:41:18 & +40:11:12 &  73.978 &   8.481 &2.360$\pm$0.472 &   2.34$\pm$0.05 & $-$3.14$\pm$1.01 & $-$3.34$\pm$1.01 &  7.7 & 1493 \\

NGC 6939 & 20:31:30 & +60:39:42 &  95.903 &  12.305 &1.800$\pm$0.360 & $-$42.00$\pm$10.00 &  0.86$\pm$0.46 & $-$2.16$\pm$0.53 &  8.4 & 1585 \\
NGC 7142 & 21:45:09 & +65:46:30 & 105.347 &   9.485 &2.344$\pm$0.469 & $-$50.30$\pm$0.30 & 1.06$\pm$0.51 & $-$4.43$\pm$0.34 &  8.9 & 6918 \\
NGC 7160 & 21:53:40 & +62:36:12 & 104.012 &   6.457 &0.789$\pm$0.158 & $-$26.28$\pm$3.88 & $-$1.57$\pm$0.51 & $-$1.60$\pm$0.51 & 8.2 &  19 \\
NGC 7789 & 23:57:24 & +56:42:30 & 115.532 &  $-$5.385 &1.795$\pm$0.359 & $-$54.70$\pm$1.30 &  4.08$\pm$0.72 &  0.21$\pm$0.68 &  8.9 & 1412 \\

\hline
\end{tabular} }
\flushleft{ {\bf Source}: 1) Friel et al. (2010), 2) Jacobson et al. (2011), 3) Pancino et al. (2010) and 4) extracted from UCAC4 Catalogue by Zacharias et al. (2013). }
\end{table*}


\begin{table*}  \vspace{-0.4cm}
{\fontsize{6}{6}\selectfont
\caption[]{The present positions and space velocity components of 69 OCs considered in this study along with their membership probabilities.}
\label{membership} 
\begin{tabular}{lcccrrrcl} \hline
\multicolumn{1}{l}{Cluster}& \multicolumn{1}{c}{x}& \multicolumn{1}{c}{y}& \multicolumn{1}{c}{z}& \multicolumn{1}{c}{U$_{\rm GSR}$}& \multicolumn{1}{c}{V$_{\rm GSR}$} & \multicolumn{1}{c}{W$_{\rm GSR}$}& \multicolumn{1}{c}{probability } & \multicolumn{1}{l}{Reference}    \\
\multicolumn{1}{l}{ }& \multicolumn{1}{c}{(kpc)} & \multicolumn{1}{c}{(kpc)} & \multicolumn{1}{c}{(kpc)}& \multicolumn{1}{c}{(km sec$^{-1}$)}& \multicolumn{1}{c}{(km sec$^{-1}$)}& \multicolumn{1}{c}{(km sec$^{-1}$)} & \multicolumn{1}{c}{percent } &  \multicolumn{1}{c}{ } \\
\hline

\multicolumn{8}{c}{\bf Thin disk} \\

NGC  752 &  -8.308$\pm$0.062 &   0.285$\pm$0.057 &  -0.180$\pm$0.036 &   -4.0$\pm$ 2.0 &  207.3$\pm$ 4.4 &  -13.0$\pm$ 3.6 & 98.9$\pm$ 0.2 & Paper {\scs I} \\
NGC 1817 &  -9.909$\pm$0.382 &  -0.208$\pm$0.042 &  -0.447$\pm$0.089 &  -42.5$\pm$ 3.4 &  163.0$\pm$14.7 &  -32.6$\pm$10.9 & 77.8$\pm$22.0 & Paper {\scs I} \\
NGC 2360 &  -9.215$\pm$0.243 &  -1.443$\pm$0.289 &  -0.047$\pm$0.009 &  -69.2$\pm$16.1 &  252.7$\pm$13.4 &   13.2$\pm$13.6 & 97.5$\pm$ 1.8 & Paper {\scs I} \\
NGC 2506 & -10.161$\pm$0.432 &  -2.636$\pm$0.527 &   0.597$\pm$0.119 &  -66.8$\pm$ 5.3 &  172.6$\pm$ 2.9 &   -9.8$\pm$ 7.1 & 93.9$\pm$ 1.5 & Paper {\scs I} \\
NGC 2335 &  -9.024$\pm$0.205 &  -0.979$\pm$0.196 &  -0.029$\pm$0.006 &   23.6$\pm$ 6.6 &  216.0$\pm$ 6.8 &   -8.1$\pm$10.5 & 99.1$\pm$ 0.2 & Paper {\scs II} \\
NGC 2482 &  -8.636$\pm$0.127 &  -1.182$\pm$0.236 &   0.048$\pm$0.010 &  -30.9$\pm$ 5.4 &  201.8$\pm$ 2.7 &  -12.5$\pm$ 5.4 & 98.6$\pm$ 0.3 & Paper {\scs II} \\
NGC 2251 &  -9.236$\pm$0.247 &  -0.541$\pm$0.108 &   0.002$\pm$0.000 &   -8.4$\pm$ 1.8 &  201.7$\pm$ 4.0 &   -1.5$\pm$ 3.9 & 99.1$\pm$ 0.1 & Paper {\scs II} \\
NGC 2527 &  -8.243$\pm$0.049 &  -0.550$\pm$0.110 &   0.019$\pm$0.004 &  -26.0$\pm$ 4.2 &  196.7$\pm$ 1.8 &    8.8$\pm$ 1.6 & 98.7$\pm$ 0.1 & Paper {\scs II} \\
NGC 2539 &  -8.790$\pm$0.158 &  -1.079$\pm$0.216 &   0.263$\pm$0.053 &  -13.2$\pm$ 1.9 &  199.6$\pm$ 1.2 &  -14.9$\pm$ 5.9 & 98.6$\pm$ 0.4 & Paper {\scs II} \\
NGC 2682 &  -8.625$\pm$0.125 &  -0.451$\pm$0.090 &   0.480$\pm$0.096 &  -28.5$\pm$ 3.1 &  187.5$\pm$ 4.2 &  -11.7$\pm$ 7.6 & 98.0$\pm$ 0.6 & Paper {\scs II} \\
NGC 1342 &  -8.581$\pm$0.116 &   0.271$\pm$0.054 &  -0.176$\pm$0.035 &   20.5$\pm$1.3  &  217.8$\pm$2.6  &    1.0$\pm$3.2  & 99.2$\pm$0.0 & This paper \\
NGC 1662 &  -8.404$\pm$0.081 &  -0.055$\pm$0.011 &  -0.157$\pm$0.031 &   24.6$\pm$0.6  &  226.0$\pm$0.6  &    6.5$\pm$1.2  & 99.2$\pm$0.0 & This paper \\
NGC 2447 &  -8.516$\pm$0.103 &  -0.899$\pm$0.180 &   0.002$\pm$0.000 &  -27.9$\pm$5.6  &  221.2$\pm$3.2  &   -2.0$\pm$2.5  & 99.2$\pm$0.0 & This paper \\
NGC 2354 &  -8.750$\pm$0.150 &  -1.223$\pm$0.245 &  -0.171$\pm$0.034 &    2.4$\pm$4.3  &  192.9$\pm$2.4  &  -20.6$\pm$6.5  & 97.8$\pm$0.8 & This paper \\
NGC 1912 &  -9.056$\pm$0.211 &   0.142$\pm$0.028 &   0.013$\pm$0.003 &    6.5$\pm$0.7  &  201.9$\pm$4.8  &   -7.0$\pm$3.0  & 99.0$\pm$0.1 & This paper \\
NGC 1245 & -10.477$\pm$0.495 &   1.627$\pm$0.325 &  -0.466$\pm$0.093 &   48.9$\pm$ 4.4 &  215.2$\pm$ 4.9 &  -47.1$\pm$12.9 & 78.4$\pm$29.7 & Jacobson et al. (2011) \\
NGC  188 &  -8.854$\pm$0.171 &   1.320$\pm$0.264 &   0.648$\pm$0.130 &   40.4$\pm$ 8.9 &  200.4$\pm$ 6.6 &  -13.2$\pm$ 9.3 & 97.4$\pm$ 0.6 & Friel et al. (2010) \\
NGC 2099 &  -9.380$\pm$0.276 &   0.055$\pm$0.011 &   0.075$\pm$0.015 &   -0.3$\pm$ 0.5 &  172.9$\pm$10.7 &    5.6$\pm$ 1.9 & 97.2$\pm$ 1.2  & Pancino et al. (2010) \\
NGC 2141 & -11.813$\pm$0.763 &  -1.248$\pm$0.250 &  -0.408$\pm$0.082 &    0.1$\pm$ 3.5 &  177.0$\pm$10.7 &   11.5$\pm$ 7.3 & 97.4$\pm$ 1.2  & Jacobson et al. (2009) \\
NGC 2158 & -12.006$\pm$0.801 &  -0.471$\pm$0.094 &   0.125$\pm$0.025 &   -8.7$\pm$ 2.6 &  154.3$\pm$14.2 &    1.8$\pm$ 5.4 & 92.6$\pm$ 5.7 & Smiljanic et al. (2009) \\
NGC 2194 &  -9.815$\pm$0.363 &  -0.566$\pm$0.113 &  -0.078$\pm$0.016 &   13.3$\pm$ 2.9 &  191.2$\pm$ 8.4 &  -14.4$\pm$ 7.2 & 98.2$\pm$ 0.6  & Jacobson et al. (2011) \\
Collinder 110 & -9.114$\pm$0.223 & 1.850$\pm$0.370 & 0.327$\pm$0.065 &  -17.6$\pm$ 4.2 &  191.1$\pm$ 4.8 &    4.9$\pm$ 4.0 & 98.6$\pm$ 0.2 & Pancino et al. (2010) \\
NGC 2355 &  -9.731$\pm$0.346 &  -0.752$\pm$0.150 &   0.395$\pm$0.079 &  -21.7$\pm$ 4.6 &  195.0$\pm$12.5 &  -16.9$\pm$10.7 & 98.2$\pm$ 1.1  & Jacobson et al. (2011) \\
NGC 2420 & -10.219$\pm$0.444 &  -0.729$\pm$0.146 &   0.834$\pm$0.167 &  -53.8$\pm$ 2.2 &  162.2$\pm$ 8.8 &    1.6$\pm$ 7.6 & 92.9$\pm$ 3.0  & Pancino et al. (2010) \\
NGC 2477 &  -8.364$\pm$0.073 &  -1.241$\pm$0.248 &  -0.132$\pm$0.026 &   -0.4$\pm$ 3.7 &  220.7$\pm$ 1.1 &    6.5$\pm$ 3.4 & 99.2$\pm$ 0.0 & Bragaglia et al. (2008) \\
NGC 2660 &  -8.195$\pm$0.039 &  -2.815$\pm$0.563 &  -0.149$\pm$0.030 & -117.2$\pm$27.5 &  212.1$\pm$ 1.9 &    7.0$\pm$10.9 & 92.9$\pm$ 6.9 & Bragaglia et al. (2008) \\
NGC 2324 & -11.162$\pm$0.632 &  -2.096$\pm$0.419 &   0.219$\pm$0.044 &  -10.6$\pm$10.6 &  175.0$\pm$16.1 &  -40.0$\pm$20.7 & 79.8$\pm$40.1 & Bragaglia et al. (2008) \\
 IC 2602 &  -7.946$\pm$0.011 &  -0.151$\pm$0.030 &  -0.014$\pm$0.003 &   -1.0$\pm$ 2.9 &  210.4$\pm$ 1.1 &    6.9$\pm$ 0.2 & 99.1$\pm$ 0.0 & D'Orazi et al. (2009) \\
NGC 3532 &  -7.834$\pm$0.033 &  -0.463$\pm$0.093 &   0.012$\pm$0.002 &  -14.7$\pm$ 5.3 &  211.6$\pm$ 1.9 &    7.8$\pm$ 0.7 & 99.1$\pm$ 0.0 & Smiljanic et al. (2009) \\	
 IC 2714 &  -9.181$\pm$0.236 &  -0.388$\pm$0.078 &   0.444$\pm$0.089 &   -2.9$\pm$ 4.4 &  236.0$\pm$ 1.8 &  -17.8$\pm$ 6.4 & 98.9$\pm$ 0.4 & Smiljanic et al. (2009) \\
NGC 3960 &  -7.238$\pm$0.152 &  -1.674$\pm$0.335 &   0.199$\pm$0.040 &  -52.1$\pm$10.7 &  218.8$\pm$ 5.4 &  -12.6$\pm$ 4.5 & 98.6$\pm$ 0.3 & Bragaglia et al. (2008) \\
NGC 1883 & -11.711$\pm$0.742 &  1.124$\pm$0.225  &  0.419$\pm$0.084  &   34.4$\pm$2.5  & 215.2$\pm$ 6.9  & -33.9$\pm$ 10.3 & 95.5$\pm$4.3 & Jacobson et al. (2009) \\                                                            
Collinder 261 & -6.851$\pm$0.230 & -1.852$\pm$0.370 & -0.211$\pm$0.042 & -63.6$\pm$14.0 &  216.0$\pm$11.0 & -3.3$\pm$ 8.0 & 98.6$\pm$ 0.4 & Sestito et al. (2008) \\
NGC 5822 &  -7.269$\pm$0.146 &  -0.577$\pm$0.115 &   0.058$\pm$0.012 &  -42.3$\pm$ 5.9 &  204.5$\pm$ 7.8 &   -8.4$\pm$ 3.0 & 98.7$\pm$ 0.2 & Smiljanic et al. (2009) \\
NGC 6134 &  -7.172$\pm$0.166 &  -0.385$\pm$0.077 &  -0.003$\pm$0.001 &  -20.6$\pm$ 2.2 &  220.6$\pm$ 4.6 &   -3.5$\pm$ 4.4 & 99.2$\pm$ 0.0 & Smiljanic et al. (2009) \\
NGC 6192 &  -6.540$\pm$0.292 &  -0.509$\pm$0.102 &   0.057$\pm$0.011 &   14.2$\pm$ 3.6 &  261.4$\pm$10.6 &    1.8$\pm$ 8.2 & 98.8$\pm$ 0.5 & Magrini et al. (2010) \\
NGC 6281 &  -7.532$\pm$0.094 &  -0.101$\pm$0.020 &   0.016$\pm$0.003 &    2.2$\pm$ 0.6 &  215.4$\pm$ 2.5 &    8.4$\pm$ 1.3 & 99.2$\pm$ 0.0 & Smiljanic et al. (2009) \\
 IC 2391 &  -7.999$\pm$0.000 &  -0.174$\pm$0.035 &  -0.021$\pm$0.004 &   -2.8$\pm$ 2.6 &  213.7$\pm$ 0.6 &   -19.1$\pm$4.9 & 99.2$\pm$ 0.0 & D'Orazi et al. (2009) \\
NGC 6633 &  -7.699$\pm$0.060 &   0.219$\pm$0.044 &   0.054$\pm$0.011 &  -11.4$\pm$ 0.5 &  206.1$\pm$ 0.6 &    2.4$\pm$ 0.6 & 99.1$\pm$ 0.0 & Smiljanic et al. (2009) \\
IC 4756 &  -7.612$\pm$0.078 &   0.286$\pm$0.057 &   0.044$\pm$0.009 &   -6.2$\pm$ 1.1 &  204.3$\pm$ 1.2 &    1.9$\pm$ 0.8 & 99.1$\pm$ 0.0 & Smiljanic et al. (2009) \\
NGC 6404 & -5.607$\pm$0.479 &  -0.176$\pm$0.035 &  -0.049$\pm$0.010 &  20.0$\pm$ 3.5 &  215.1$\pm$40.9 &   10.0$\pm$36.5 & 99.1$\pm$1.0 & Magrini et al. (2010) \\
NGC 6583 & -5.989$\pm$0.402 &   0.329$\pm$0.066 &  -0.090$\pm$0.018 &  14.2$\pm$ 6.1 &  215.0$\pm$34.0 &   -2.3$\pm$34.3 & 99.2$\pm$0.2 & Magrini et al. (2010) \\
NGC 6791 &  -6.650$\pm$0.270 &   3.707$\pm$0.741 &   0.760$\pm$0.152 &   38.9$\pm$ 9.2 &  167.8$\pm$ 3.0 &  -12.4$\pm$ 3.3 & 91.8$\pm$ 2.4 & Boesgaard et al. (2009) \\
NGC 6819 &  -7.357$\pm$0.129 &   2.244$\pm$0.449 &   0.348$\pm$0.070 &   58.0$\pm$14.4 &  212.3$\pm$ 4.6 &   19.9$\pm$11.5 & 97.7$\pm$ 1.6 & Bragaglia et al. (2001) \\
NGC 6939 &  -8.182$\pm$0.036 &   1.749$\pm$0.350 &   0.384$\pm$0.077 &   24.8$\pm$ 4.9 &  189.1$\pm$ 9.8 &  -17.9$\pm$ 5.6 & 97.6$\pm$ 0.9 & Jacobson et al. (2007) \\
NGC 7160 &  -8.190$\pm$0.038 &   0.761$\pm$0.152 &   0.089$\pm$0.018 &   24.5$\pm$ 2.6 &  202.1$\pm$ 3.8 &    3.1$\pm$ 2.0 & 99.0$\pm$ 0.1 & Monroe et al. (2010) \\
NGC 7789 &  -8.772$\pm$0.154 &   1.612$\pm$0.322 &  -0.169$\pm$0.034 &  3.2$\pm$ 8.3 &  160.7$\pm$ 4.3 &  7.1$\pm$ 5.9 & 94.6$\pm$ 1.3 & Pancino et al. (2010) \\
NGC 6253 & -6.633$\pm$0.273 & -0.620$\pm$0.124 &  -0.164$\pm$0.033 & -31.4$\pm$7.5 &  207.5$\pm$15.8 &   -0.6$\pm$16.2 & 99.0$\pm$ 0.2 & Montalto et al. (2012) \\
Ruprecht 4 & -11.470$\pm$0.694 & -3.140$\pm$0.628 & -0.437$\pm$0.087  & -39.8$\pm$39.3 & 209.2$\pm$42.4  &    6.1$\pm$53.4 & 98.9$\pm$1.2 & Carraro et al. (2007) \\
NGC 7142 &  -8.614$\pm$0.123 &   2.229$\pm$0.446 &   0.386$\pm$0.077 &    43.3$\pm$6.3 &  190.8$\pm$ 3.0 &  -45.7$\pm$10.0 & 73.4$\pm$25.9 & Jacobson et al. (2007) \\
Berkeley 39 & -11.411$\pm$0.682 & -3.242$\pm$0.648 &  0.838$\pm$0.168 & -26.4$\pm$ 52.8 & 169.2$\pm$ 52.0&  -31.8$\pm$ 78.4 & 86.3$\pm$89.9 & Friel et al. (2010) \\
Berkeley 75 & -13.196$\pm$1.039 & -7.259$\pm$1.452 & -1.766$\pm$0.353 & -69.4$\pm$119.3 & 170.7$\pm$87.1 & -23.7$\pm$142.8 &  87.2$\pm$120.9 & Carraro et al. (2007) \\

\multicolumn{8}{c}{\bf Thin-Thick disk} \\

Melotte 66  &  -8.750$\pm$0.150 &  -4.112$\pm$0.822 &  -1.062$\pm$0.212 & -169.0$\pm$45.7 &  237.2$\pm$ 9.8 &   -7.9$\pm$17.4 & 56.4$\pm$61.9 (Thin)  & Sestito et al. (2008) \\
            &                   &                   &                   &                  &                 &              & 42.7$\pm$59.6 (Thick) & \\
NGC 2425    &  -9.899$\pm$0.380 &  -2.396$\pm$0.479 &   0.176$\pm$0.035 &  -68.5$\pm$10.4 &  152.0$\pm$ 8.0 &  -37.3$\pm$16.2 & 40.6$\pm$44.3 (Thin)  &  Jacobson et al. (2011) \\
            &                   &                   &                   &                  &                 &               & 58.9$\pm$43.8 (Thick) & \\
\multicolumn{8}{c}{\bf Thick disk} \\

NGC 2266   & -11.314$\pm$0.663 &  -0.458$\pm$0.092 &   0.608$\pm$0.122 &   34.8$\pm$ 2.4 &  179.0$\pm$13.8 &  -56.0$\pm$14.8 & 69.6$\pm$51.2 & Paper {\scs II} \\
Tombaugh 2 & -11.639$\pm$0.728 & -4.816$\pm$0.963 & -0.729$\pm$0.146 & -107.6$\pm$50.9 & 171.9$\pm$40.8 & -67.8$\pm$73.8 & 97.4$\pm$5.9 & Jacobson et al. (2007) \\
Berkeley 73 & -15.884$\pm$1.577 & -5.595$\pm$1.119 & -1.605$\pm$0.321 & 6.5$\pm$81.4  & 76.2$\pm$109.6 & -39.3$\pm$125.0 & 93.8$\pm$28.6 & Carraro et al. (2007) \\
Berkeley 32 & -11.010$\pm$0.602 &  -1.603$\pm$0.321 &   0.263$\pm$0.053 &  -74.9$\pm$ 7.1 &  150.1$\pm$13.6 &  -47.2$\pm$19.1 & 85.7$\pm$31.2 & Yong et al. (2012) \\
NGC 1193    & -11.742$\pm$0.748 &   2.442$\pm$0.488 &  -0.963$\pm$0.193 &   54.3$\pm$ 8.7 &  112.3$\pm$16.5 &  -33.4$\pm$16.4 & 93.9$\pm$ 8.2 & Friel et al. (2010) \\
Berkeley 22 & -13.584$\pm$1.117 & -2.028$\pm$0.406 & -0.844$\pm$0.169 & -39.0$\pm$32.6  & 109.0$\pm$89.9 & -69.3$\pm$107.5 &  96.0$\pm$16.4 & Yong et al. (2012) \\
Berkeley 18 & -12.972$\pm$0.994 &   1.454$\pm$0.291 &   0.455$\pm$0.091 &  -21.1$\pm$ 7.8 &  145.3$\pm$17.7 & -128.5$\pm$28.4 & 85.8$\pm$19.6 & Yong et al. (2012) \\
NGC 2243    & -10.147$\pm$0.429 &  -3.655$\pm$0.731 &  -1.379$\pm$0.276 &  -61.8$\pm$25.2 &  175.7$\pm$11.3 &   58.1$\pm$19.7 & 84.7$\pm$41.7  & Jacobson et al. (2011) \\
Berkeley 20 & -15.348$\pm$1.470 &  -3.205$\pm$0.641 &  -2.509$\pm$0.502 &   25.1$\pm$24.6 &   43.5$\pm$43.2 &  -38.0$\pm$31.5 & 75.2$\pm$43.1 & Sestito et al. (2008) \\
Berkeley 21 & -14.149$\pm$1.230 &  -0.746$\pm$0.149 &  -0.272$\pm$0.054 &   33.0$\pm$ 4.9 &   79.8$\pm$33.0 & -101.9$\pm$26.8 & 71.1$\pm$37.0 & Yong et al. (2012) \\
Saurer 1 & -18.752$\pm$2.150 & -7.466$\pm$1.493 &  1.698$\pm$0.340 & -38.6$\pm$121.3  & 127.4$\pm$178.6 &  27.5$\pm$188.5 &  70.0$\pm$430.0 & Carraro et al. (2004) \\

\multicolumn{8}{c}{\bf Thick disk-halo} \\
Berkeley 33 & -12.540$\pm$0.908 & -4.622$\pm$0.924  &  -0.524$\pm$0.105 &  -142.3$\pm$20.2 &  358.4$\pm$42.4 & -88.5$\pm$36.2 & 56.4$\pm$55.7 (Thick) & Carraro et al. (2007) \\
            &                   &                   &                   &                  &                 &                & 43.6$\pm$55.7 (Halo) & \\
\multicolumn{8}{c}{\bf Halo} \\

Berkeley 29 & -22.000$\pm$2.800 &  -4.566$\pm$0.913 &   2.077$\pm$0.415 &   57.7$\pm$22.6 &  -71.2$\pm$71.1 & -143.6$\pm$61.5 & 100.0$\pm$ 0.0 & Sestito et al. (2008) \\
Berkeley 31 & -15.384$\pm$1.477 &  -3.655$\pm$0.731 &   0.739$\pm$0.148 &  -31.2$\pm$ 9.5 &  134.6$\pm$22.5 & -207.0$\pm$48.4 &  98.5$\pm$ 6.6 & Friel et al. (2010) \\
Berkeley 25 & -15.706$\pm$1.541 & -8.178$\pm$1.636 & -1.922$\pm$0.384   & 0.5$\pm$126.6   & 79.2$\pm$129.4 & -134.3$\pm$211.7 &  66.4$\pm$297.4 & Carraro et al. (2007) \\

\hline
\end{tabular} }
\end{table*}


\begin{thebibliography}{99}

\bibitem{} Alexeeva S. A., Pakhomov Yu. V., Mashonkina L. I., 2014, AstL, 40, 406
\bibitem{} Allende Prieto C., Garc\'{\i}a L\'{o}pez R. J., Lambert D. L., Gustafsson B., 1999, ApJ, 527, 879
\bibitem{} Alonso A., Arribas S., Mart\'{i}nez-Roger C., 1999, A\&AS, 140, 261
\bibitem{} Asplund M., Grevesse N., Sauval A. J., Scott, P., 2009, ARA\&A, 47, 481
\bibitem{} Becker W., Svolopoulos S. N., Fang C., 1976, Astron. Inst. Univ. Basel, 
           Kataloge photographischer und photoelektrischer Helligkeiten von 25 galaktischen Sternhaufen im RGU-und Cap UBV-System 
\bibitem{} Bensby T., Feltzing S., Oey M. S., 2014, A\&A, 562, 71
\bibitem{} Bensby T., Feltzing S., Lundstr\"{o}m I., Ilyin I., 2005, A\&A, 433, 185
\bibitem{} Boesgaard A. M., Jensen E. E. C., Deliyannis C. P., 2009, AJ, 137, 4949
\bibitem{} Bragaglia A., Sestito P., Villanova S., Carretta E., Randich S., Tosi M., 2008, A\&A, 480, 79
\bibitem{} Bragaglia A., Carretta E., Gratton R. G., Tosi M., Bonanno G., Bruno P. et al. 2001, AJ, 121, 327
\bibitem{} Burris D. L., Pilachowski C. A., Armandroff T. E., Sneden C., Cowan J. J., Roe H.,  2000, ApJ, 544, 302
\bibitem{} Carraro G., Geisler D., Villanova S., Frinchaboy P. M., \& Majewski S. R., 2007, A\&A, 476, 217
\bibitem{} Carraro G., Bresolin F., Villanova S., Matteucci F., Patat F., \& Romaniello M., 2004, AJ, 128, 1676
\bibitem{} Carretta E., Bragaglia A., Gratton R. G., 2007, A\&A, 473, 129
\bibitem{} Carretta E., Bragaglia  A., Gratton R. G., Tosi M., 2005, A\&A, 441, 131
\bibitem{} Castelli F. \& Kurucz R. L., 2003, IAU Symposium 210, Modelling of Stellar Atmospheres, Uppsala, Sweden, eds. N.E. Piskunov, W.W. Weiss, and D. F. Gray, 2003, ASP-S210
\bibitem{} Clari\'{a} J. J., Piatti A. E., Lapasset E., \& Parisi M. C., 2005, BaltA, 14, 301
\bibitem{} Clari\'{a} J. J., Mermilliod J.-C., Piatti A. E., 1999, A\&AS, 134, 301
\bibitem{} Cowan J. J., Rose W. K., 1977, ApJ, 212, 149
\bibitem{} Cutri R.M., Skrutskie M.F., Van Dyk S. et al., 2003, Vizier Online Data Catalogue, {\bf II/246}
\bibitem{} De Silva G. M., Freeman K. C., Bland-Hawthorn J., 2009, PASA, 26, 11
\bibitem{} De Silva G. M., Freeman K. C., Asplund M., Bland-Hawthorn, J., Bessell M. S., Collet, R. 2007, AJ, 133, 1161
\bibitem{} De Silva G.M., Sneden C., Paulson D.B., Asplund M., Bland-Hawthorn J., Bessell M.S., Freeman K.C., 2006, AJ, 131, 455.
\bibitem{} Dias W. S., Alessi B. S., Moitinho A., L\'{e}pine J. R. D., 2002, A\&A, 389, 871
\bibitem{} D'Orazi V., Biazzo K., Desidera S., Covino E., Andrievsky S. M., Gratton R. G., 2012, MNRAS, 423, 2789
\bibitem{} D'Orazi V. et al. 2009, ApJ, 693, 31
\bibitem{} D\"{u}rbeck, W., 1960, ZA, 49, 214
\bibitem{} Feltzing S., Gustafsson B., 1998, A\&AS, 129, 237
\bibitem{} Freeman K., \& Bland-Hawthorn J., 2002, ARA\&A, 40, 4875
\bibitem{} Friel E. D., Jacobson H. R., \& Pilachowski C. A. 2010, AJ, 139, 1942
\bibitem{} F\"{u}hr J.R., Wiese W.L., 2006, J. Phys. Chem. Ref. Data, 35, 1669
\bibitem{} Ghez A. M. et al. 2008, ApJ, 689, 1044 
\bibitem{} Glushkova E. V. \& Rastorguev A. S., 1991, SvAL, 17, 13G
\bibitem{} Gorbaneva T. I., Mishenina T. V., Soubiran  C., 2012, KPCB, 28, 121
\bibitem{} Gratton R., 2000, ASPC, 198, 225 
\bibitem{} Hamdani S., North P., Mowlavi N., Raboud D., Mermilliod J. C., 2000, A\&A, 360, 509 
\bibitem{} Heiter U., Soubiran C., Netopil M. \& Paunzen E., 2014, A\&A, 561, 93
\bibitem{} Hinkle K., Wallace L., Valenti J., Harmer D., 2000, \textit{Visible and Near Infrared Atlas of the Arcturus Spectrum 3727-9300 \AA} (San
           Francisco: ASP)
\bibitem{} Hoag A.A., Johnson H.L., Iriarte B., Mitchell R.I., Hallam K.L., Sharpless S., 1961, Publ. Us. Nav. Obs. XVII part VII, 347  
\bibitem{} Jacobson H. R., Friel E. D., 2013, AJ, 145, 107
\bibitem{} Jacobson H. R., Pilachowski C. A., Friel E. D., 2011, AJ, 142, 59
\bibitem{} Jacobson H. R., Friel E. D., Pilachowski C. A., 2009, AJ, 137, 4753
\bibitem{} Jacobson H. R., Friel E. D., Pilachowski C. A., 2007, AJ, 134, 1216
\bibitem{} Karakas A. I. \& Lattanzio J. C., 2014, PASA, 31, 30
\bibitem{} Koch A. \& Edvardsson B., 2002, A\&A, 381, 500
\bibitem{} Kurucz R. L., Furenlid I., Brault J., \& Testerman L. 1984, \textit{Solar Flux Atlas from 296 to 1300 nm}, ed. R. L. Kurucz, I. Furenlid,  
           J. Brault, \& L. Testerman (Sunspot, NM: National Solar Observatory) 
\bibitem{} Lind K., Bergemann M., Asplund M., 2012, MNRAS, 427, 50
\bibitem{} Luck R. E., Heiter U., 2007, AJ, 133, 2464
\bibitem{} Luck R. E., Heiter U., 2006, AJ, 131, 3069
\bibitem{} Luck R. E., \& Heiter U. 2005, AJ, 129, 1063
\bibitem{} Lyng\r{a} G. 1987, Calalog of Open Star Cluster Data (Strasbourg: CDS)
\bibitem{} Magrini L., Randich S., Zoccali M., Jilkova L., Carraro G., Galli D., Maiorca E., Busso M., 2010, A\&A, 523, 11
\bibitem{} Maiorca E., Randich S., Busso M., Magrini L., Palmerini S., 2011, ApJ, 736, 120
\bibitem{} Marigo P., Girardi L., Bressan A., Groenewegen M. A. T., Silva L., Granato G. L., 2008, A\&A, 482, 883 
\bibitem{} Mermilliod J.-C., Mayor M., Udry S., 2008, A\&A, 485, 303
\bibitem{} Mills G. A., 1967, JO, 50, 179
\bibitem{} Mishenina T., Pignatari M., Carraro G., Kovtyukh V., Monaco L., Korotin S., Shereta E., Yegorova I., Herwig F., 2015, MNRAS, 446, 3651
\bibitem{} Mishenina T., Korotin S., Carraro G., Kovtyukh V. V., Yegorova I. A., 2013, MNRAS, 433, 1436
\bibitem{} Mishenina T. V., Soubiran C., Kovtyukh V. V., Korotin S. A., 2004, A\&A, 418, 551
\bibitem{} Montalto M., Santos N. C., Villanova S., Pace G., Piotto G., Desidera S., De Marchi F., Pasquini L., Saviane I., 2012, MNRAS, 423, 3039
\bibitem{} Monroe T. R. \& Pilachowski C. A., 2010, AJ, 140, 2109
\bibitem{} Pancino E., Carrera R., Rossetti E., Gallart C., 2010, A\&A, 511, 56
\bibitem{} Pavani D. B., \& Bica E., 2007, A\&A, 468, 139
\bibitem{} Reddy A.B.S., Giridhar S., Lambert D. L., 2013, MNRAS, 431, 3338 (Paper {\scs II})
\bibitem{} Reddy A.B.S., Giridhar S., Lambert D. L., 2012, MNRAS, 419, 1350 (Paper {\scs I})
\bibitem{} Reddy B. E., Lambert D. L., Allende Prieto C., 2006, MNRAS, 367, 1329
\bibitem{} Sestito P., Bragaglia A., Randich S., Pallavicini R., Andrievsky S. M., \& Korotin S. A. 2008, A\&A, 488, 943
\bibitem{} Smiljanic R., 2012, MNRAS, 422, 1562
\bibitem{} Smiljanic R., Gauderon R., North P., Barbuy B., Charbonnel C., Mowlavi N., 2009, A\&A, 502, 267
\bibitem{} Sneden C., 1973, PhD Thesis, Univ. of Texas, Austin
\bibitem{} Subramaniam A., \& Sagar R. 1999, AJ, 117, 937
\bibitem{} Takeda Y., Sato B., Murata D., 2008, PASJ, 60, 781 
\bibitem{} Tull R.G., MacQueen P.J., Sneden C., Lambert D.L., 1995, PASP, 107, 251
\bibitem{} Twarog B.A., Ashman K.M., Anthony-Twarog B.J., 1997, AJ, 114, 2556 
\bibitem{} Villanova S., Randich S., Geisler D., Carraro G., Costa E., 2010, A\&A, 509, 102
\bibitem{} Yong D., Carney B. W., Friel E. D., 2012, AJ, 144, 95
\bibitem{} Zacharias N., Finch C. T., Girard T. M., et al., 2013, AJ, 145, 44

\end{thebibliography}
\end{document}